\documentclass[aps,prd,eqsecnum,nofootinbib,floatfix,superscriptaddress,preprint,tightenlines,longbibliography]{revtex4-1}

\usepackage{amsfonts}
\usepackage{amsmath}
\usepackage{amssymb}
\usepackage{bm}
\usepackage[pdftex]{color}
\usepackage{graphicx}
\usepackage[sort&compress]{natbib}
\usepackage[colorlinks=true,linkcolor=Blue,filecolor=Blue,urlcolor=Blue,citecolor=Blue,pdftex,plainpages=false]{hyperref}

\definecolor{white}{rgb}{1,1,1}
\definecolor{black}{rgb}{0,0,0}
\definecolor{red}{rgb}{1,0,0}
\definecolor{green}{rgb}{0.317647,0.603922,0.141176}
\definecolor{blue}{rgb}{0,0.2,0.6}
\definecolor{yellow}{rgb}{0.917647,0.686275,0}
\definecolor{brown}{rgb}{0.6,0.4,0}
\definecolor{gray}{rgb}{0.454902,0.454902,0.454902}
\definecolor{violet}{rgb}{0.580392,0,0.827451}
\definecolor{cyan}{rgb}{0,1,1}
\definecolor{magenta}{rgb}{1,0,1}
\definecolor{orange}{rgb}{1,0.6,0}
\definecolor{indigo}{rgb}{0.2,0,0.2}
\definecolor{maroon}{rgb}{0.8,0.2,0}
\definecolor{marlin}{rgb}{0.5,0.8,0.5}
\definecolor{aqua}{rgb}{0.254902,0.607843,0.521569}
\definecolor{azure}{rgb}{0.6,0.8,1}
\definecolor{flesh}{rgb}{1,0.8,0.6}
\definecolor{dkbrown}{rgb}{0.368627,0.176471,0.215686}
\definecolor{straw}{rgb}{1,1,0.8}
\definecolor{gold}{rgb}{0.823529,0.54902,0}
\definecolor{Red}{rgb}{1,0,0}
\definecolor{Green}{rgb}{0,1,0}
\definecolor{Blue}{rgb}{0,0,1}

\renewcommand{\case}[2]{\ensuremath{{\textstyle\frac{#1}{#2}}}}
\newcommand{\half}{\case{1}{2}}

\begin{document}

\title{Lattice QCD and Neutrino-Nucleus Scattering}

\author{Andreas S. Kronfeld}
\email{Editor, \tt ask@fnal.gov}
\affiliation{Theoretical Physics Department, Fermi National Accelerator Laboratory, Batavia, IL 60510}

\author{David G. Richards}
\email{Editor, \tt dgr@jlab.org}
\affiliation{Theory Center, Thomas Jefferson National Accelerator Facility, Newport~News, VA 23606}

\author{William Detmold}
\affiliation{Center for Theoretical Physics, Massachusetts Institute of Technology, Cambridge, MA 02139}

\author{Rajan Gupta}
\affiliation{Group T-2, Los Alamos National Laboratory, Los~Alamos, NM 87545}

\author{Huey-Wen Lin}
\affiliation{Department of Physics and Astronomy, Michigan State University, \\ East~Lansing, MI 48824}

\author{Keh-Fei Liu}
\affiliation{Department of Physics and Astronomy, University of Kentucky, \\ Lexington, KY 40508}

\author{Aaron S. Meyer}
\affiliation{Physics Department, Brookhaven National Laboratory, Upton, NY 11973}

\author{Raza Sufian}
\affiliation{Theory Center, Thomas Jefferson National Accelerator Facility, Newport~News, VA 23606}

\author{Sergey Syritsin}
\affiliation{Department of Physics and Astronomy, Stony Brook University, \\ Stony~Brook, NY 11794}

\collaboration{USQCD Collaboration}
\noaffiliation

\date{\today}

\begin{abstract}
This document is one of a series of whitepapers from the USQCD collaboration.
Here, we discuss opportunities for lattice QCD in neutrino-oscillation physics, which inevitably entails nucleon and nuclear 
structure.
In addition to discussing pertinent lattice-QCD calculations of nucleon and nuclear matrix elements, the interplay with models of 
nuclei is discussed.
This program of lattice-QCD calculations is relevant to current and upcoming neutrino experiments, becoming increasingly important 
on the timescale of LBNF/DUNE and HyperK.
\end{abstract}

\maketitle

\section*{EXECUTIVE SUMMARY}

In 2018, the USQCD collaboration’s Executive Committee organized several subcommittees to recognize future opportunities and
formulate possible goals for lattice field theory calculations in several physics areas.
The conclusions of these studies, along with community input, are presented in seven whitepapers~\cite{Bazavov:2018qcd,Brower:2018qcd,%
Davoudi:2018qcd,Detmold:2018qcd,Joo:2018qcd,Lehner:2018qcd}.
This whitepaper covers the role of lattice QCD in neutrino-nucleus scattering, motivated principally by neutrino
oscillations.

Neutrino-nucleus scattering experiments provide an abundance of information on neutrino masses and flavor mixing, on nucleon and 
nuclear structure, and on non-standard interactions between neutrinos and ordinary matter.
To interpret these experiments cleanly, the key problem is to reconstruct the incident neutrino energy.
The nuclear remnant is not, in these experiments, detected.
It is therefore impossible to reconstruct the neutrino energy without modeling the nucleus in some way.
This problem is complex, because it spans a range of energies---from hundreds of keV to a few GeV---that probe all aspects of the target 
nucleus.

The presence of many energy scales implies that a variety of theoretical techniques must work in concert.
A convenient, organizational framework is nuclear many-body theory, which takes nucleonic properties as inputs.
In this whitepaper, we discuss how these nucleonic properties can be obtained directly from the QCD Lagrangian using numerical
simulations of lattice gauge theory.
Although lattice QCD cannot settle every question in neutrino-nucleus scattering, it is reasonable to demand that our understanding
of these processes be consistent with QCD.
In many cases, the most straightforward route to the needed QCD knowledge is lattice QCD.

In this whitepaper, we discuss several calculations that should, as they mature, be incorporated into nuclear theory and neutrino
event generators.
A very important and very feasible example is the axial form factor of the nucleon.
Lattice QCD has a notable history of calculating this and related observables, and calculations with full control of the systematic
uncertainties are now coming of age.
Here, ``full control of systematic uncertainties'' implies that a complete error budget is provided.
The axial form factor is relatively straightforward: completely analogous calculations of vector form factors are possible with the
same (indeed, overlapping) computational effort.
The vector form factors have been measured in electron-proton and -neutron scattering, so an apt crosscheck is close at hand.
Experience from form factors in meson physics suggests a simple, model-independent way to transmit the output of lattice QCD to
event generators and, thus, analysis of experimental data.

Form factors of nucleons are only the beginning.
Future oscillation experiments span beam energies such that computationally more demanding information is required.
Just at the nucleon level, transition form factors to multibody final states are needed.
For an inclusive data set, the object of interest is the nuclear hadron tensor, which can be obtained by combining the nucleonic
hadron tensor from lattice QCD with a nuclear spectral function.
In the deep inelastic region, new ways of computing parton distribution functions in lattice QCD are an exciting development.
A further emerging component of lattice QCD consists of calculations of the properties of small nuclei---up to ${}^4$He today
and to ${}^6$Li with exascale computing---can be used to test nuclear many-body theory and provide information via chiral effective
theories to pin down the nuclear physics.

Lattice-QCD calculations with nucleon and nuclei are more challenging than the corresponding ones for mesons, because of unavoidable
technical challenges that increase with the number of quark lines.
{Consequently, to perform the requisite calculations}, improvements in methodology, algorithms, and software will be essential.
Even assuming continuing ingenuity on those fronts, much of the work will require exascale computing resources.
As in the past, a combination of high-capability and high-capacity computing will be needed.
The former is needed for timely solution of mature problems, while the latter is necessary for developing feasible techniques for
the challenging calculations, before making the jump to supercomputer centers.

\clearpage

\section{Introduction}
\label{sec:intro}

Along with the first observation of the Higgs boson and the mounting evidence for dark matter, the discovery that neutrinos change
flavor is one of the major advances in particle physics over the past twenty-five years.
The discovery hinged on studies of neutrinos produced at the upper edge of the earth's atmosphere~\cite{Fukuda:1998mi} and also
explained a deficit in electron neutrinos from the sun~\cite{Ahmad:2001an}.
These findings prompted an accelerator-based experimental program in Europe, Japan, and the United States, to make more accurate
measurements of, for example, the squared mass differences.
The increase in precision and sensitivity expected in future experiments raises the question whether the theoretical description of
the relevant experiments must be further refined to exploit the new measurements to the fullest.
In particular, as future, ambitious, long-baseline neutrino-oscillation experiments such as LBNF/DUNE~\cite{Acciarri:2015uup} and
HyperK~\cite{Abe:2018uyc} have come into focus, the quantification of uncertainties from the hadronic and nuclear physics of the
detectors have become increasingly relevant.
To this end, the lattice-QCD community has identified a set of feasible calculations that will be of special relevance.
This program is described in this whitepaper.

An important goal of the experimental neutrino-physics program is to test the three-neutrino paradigm of the Standard Model.
In this context, the Standard Model must be extended to allow for lepton flavor change.
The simplest choice consistent with the standard gauge symmetries is to introduce a set of right-handed neutrino fields.
Then lepton-flavor mixing and neutrino masses arise in the same way as in the quark sector, namely through Yukawa couplings to the
Higgs field with a nonvanishing vacuum expectation value.
To couple to the Higgs and left-handed-lepton doublets, the right-handed neutrino fields have to be gauge singlets.
But then no symmetry principle forbids a mass term connecting neutrinos to themselves (i.e., of the kind first noted by
Majorana~\cite{Majorana:1937vz}), in contrast to the Higgs-generated Dirac mass term, which connects neutrino to antineutrino.
The lack of direct evidence for right-handed neutrinos suggests that in this scenario the Majorana mass $M$ might be very large.
If one supposes that the neutrino Yukawa couplings are not much different from light quarks or charged leptons, the propagating
neutrinos have mass close to $M$ and to $m_\nu\approx y^2v^2/2M$, where $y$ is a Yukawa coupling and $v$ is the vacuum expectation
value of the Higgs field.
This mass hierarchy, known as the see-saw mechanism, provides a possible explanation of the tiny size of neutrino
masses~\cite{Minkowski:1977sc,*Yanagida:1980xy,*GellMann:1980vs}.
For example, if $M$ is a grand-unified mass scale around $10^{15}$~GeV, then $m_\nu\lesssim0.03$~eV (for $y\lesssim1$).

This theoretical framework means that the three-neutrino paradigm can be tested by measuring the neutrino mass-squared differences
and the mixing angles and $CP$~violating phases of the Pontecorvo-Maki-Nakagawa-Sakata (PMNS) mixing
matrix~\cite{Pontecorvo:1957cp,*Pontecorvo:1967fh,*Maki:1962mu}.
Like the Cabibbo-Kobayashi-Maskawa (CKM) quark-mixing matrix~\cite{Cabibbo:1963yz,*Kobayashi:1973fv}, the PMNS has three mixing
angles.
If the Majorana mass term appears, the PMNS matrix has three $CP$-violating phases instead of one as in the CKM matrix.
The mixing angles and the CKM-like $CP$-violating phase can be measured in oscillation experiments, while the extra phases and the
Majorana nature of neutrinos can be probed via the neutrinoless double-beta ($0\nu\beta\beta$) decay of certain nuclei.
For lattice-QCD calculations relevant to $0\nu\beta\beta$, see the companion whitepaper ``The Role of Lattice QCD in Searches for
Violations of Fundamental Symmetries and Signals for New Physics'' \cite{Davoudi:2018qcd}; here, the focus is on lattice-QCD
research that will impact the oscillation experiments.

\begin{figure}[b]
    \includegraphics[width=0.6\textwidth]{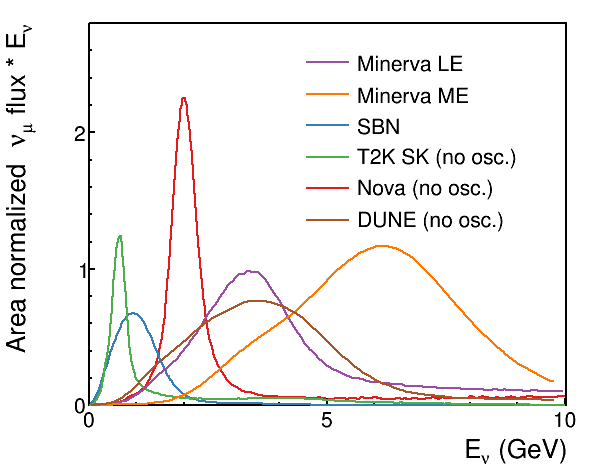}
    \caption{Energy spectrum of the neutrino beam for several experiments.
    In particular, most of DUNE's beam lies in the range $1~\text{GeV}<E_\mu<7$~GeV.
    Courtesy Laura Fields~\cite{Fields:2018enu}.}
    \label{fig:Espect}
\end{figure}

Oscillation experiments measure the energy spectrum of a neutrino beam after it has travelled a certain baseline distance.
Unfortunately, neutrino beams have a wide energy spectrum, as shown in Fig.~\ref{fig:Espect}, so the center-of-mass energy of a
collision is not known.
In contrast, quark-flavor experiments, for which lattice QCD has been crucial, study decays of strange, charmed, or $b$-flavored
hadrons of precisely known mass.
Here, the energy of the incident neutrino must be inferred from measurements of the final state.
The targets in neutrino experiments are medium- to large-sized nuclei, such as ${}^{12}$C, ${}^{16}$O, or ${}^{40}$Ar, the remnants of
which are not, in practice, be detected.
That means that the mapping between final-state measurements and the initial energy inevitably requires theoretical knowledge of the
neutrino interaction with the struck nucleus.

Consistency with QCD is a clearly desirable characteristic of nuclear models used to deduce the connection between final and initial
states.
Thus, it makes sense to incorporate lattice QCD as soon as results with full, reliable error budgets are available.
As discussed in more detail in Ref.~\cite{Alvarez-Ruso:2017oui}, the nuclear models rely in part on properties of the nucleon as
inputs.
Many of these quantities can be calculated in lattice QCD in the near term, with the precision depending on the quantity.
Of course, single-nucleon calculations are not in themselves enough.
Calculations of the properties of multi-nucleon systems must be developed concurrently and, once mature, also incorporated into the
nuclear modeling.

The theory behind neutrino-nucleus collisions is complex because it spans a range of energies that probe all aspects of the target
nucleus.
Nuclear excitation energies are, typically, dozens of keV, while the average binding energy is 8.6~MeV (in $^{40}\text{Ar}$), and
the typical Fermi motion of a nucleon is around 250~MeV.
In the regime relevant to oscillation experiments, the energy transfer to the nucleus ranges between $\sim$200~MeV and the neutrino
energy itself, although much of transferred energy is carried off by nucleons and pions, rather than the nuclear remnant.
Thus, it is a challenge to arrive at a comprehensive approach to the entire problem.
Most approaches start with nuclear many-body theories, in which the nucleus is described by a nuclear wave function of a
collection of interacting nucleons; see, for example, Ref.~\cite{Benhar:2015xga,Carlson:2014vla}.
It is at this point in the analysis that nucleon-level matrix elements enter.
One should bear in mind, however, that single-nucleon physics is not enough: multi-body effects are needed for scattering events
that knock out two (or more) nucleons.
Even in nuclear spectroscopy, three-body potentials improve the agreement with observed nuclear
levels~\cite{Epelbaum:2008ga,Machleidt:2011zz,Carlson:2014vla}.
Often these calculations use phenomenological potentials, but effective field theory (EFT) offers a direct connection to
QCD~\cite{Weinberg:1990rz,vanKolck:1994yi,Kaplan:1998tg,Meissner:2015wva}.
Chiral EFTs are, however, limited to a kinematic range where the momenta are small relative to the chiral symmetry breaking scale
$\Lambda_\chi\sim700~\text{MeV}$.
Even then, the reliability of the application of nuclear EFT to large atomic number systems, such as argon, requires significant
development, testing, and, eventually, verification.
\makebox(0,0){
\phantom{\cite{Juszczak:2005zs,*Zmuda:2015twa}}
\nocite{Hayato:2009zz}
\phantom{\cite{Andreopoulos:2009rq,*Alam:2015nkk}}
\nocite{Gallmeister:2016dnq}
}%
These issues are further intertwined with the constraints of how event
generators~\cite{Juszczak:2005zs,Hayato:2009zz,Casper:2002sd,%
Andreopoulos:2009rq,Gallmeister:2016dnq} and detector simulations are implemented.
Inconsistencies arise in the current approach where, for example, the axial form factor of the nucleon is extracted from $\nu A$
scattering data assuming one nuclear model and then used in event generators employing another.

A central goal of nuclear theory in this arena should therefore be to define a path forward that allows for a quantified nuclear
uncertainty to be presented for experiments such as DUNE and HyperK.
Achieving this is a challenging task and will require input and constraints from lattice QCD in order for it to be successful.
In addition to the single- and few-nucleon amplitudes noted above, it will be valuable to compute directly the properties of small
nuclei.
At present, calculations involving nuclei up to $^4$He are possible.
In addition to being interesting in their own right, such lattice-QCD calculations of few nucleon systems can be used to constrain
low energy constants (LECs) in the EFTs.
This approach has already been applied to static quantities, such as magnetic moments.
A next step will be to work with matrix elements of electroweak currents, to build up effects associated with two- and higher-body
contributions, as well as more complex contributions such as pion production.
In combination with experimental constraints from $eA$ scattering, and neutrino scattering on light nuclear targets,%
\footnote{Indeed, recent discussions of future experiments with deuterium or hydrogen targets~\cite{Morfin:2018int} hinge
on noting the utility of nucleon-level amplitudes in nuclear many-body theory.} %
it is hoped a robust uncertainty can be determined.

To study neutrino oscillations, we are interested in the processes
\begin{equation}
    \nu_\ell A \to \ell^- X, \qquad
    \bar{\nu}_\ell A \to \ell^+ X,
\end{equation}
where $A$ denotes the nucleus and $X$ the combination of all final-state hadrons including the remnant of the nucleus.
The charged weak current responsible for these interactions has the well-known $V-A$ structure.
Properties of the vector current can be inferred from electromagnetic scattering, up to isospin corrections (which are negligible
for the needed precision; see Sec.~\ref{sec:hardest}).
On the other hand, because the weak charge of the proton is so small, $Q_\text{w}^p=0.0719\pm0.0045$~\cite{Androic:2018kni}, at the
energies of interest, only neutron-neutrino (and proton-antineutrino) scattering is sensitive to the axial current.
These circumstances offer the possibility of testing lattice-QCD methodology with the vector current before relying on it for the
axial current.

The quantity needed to describe the strong-interaction side of the scattering depends on the energy transferred.
At the lowest energies, the only possibility is coherent elastic scattering via the weak neutral current, with
$X=A$~\cite{Freedman:1973yd,Brice:2013fwa}.
Coherent neutrino-nucleus interactions have recently been observed for the first time~\cite{Akimov:2017ade}.
As the energy increases slightly, the excitation spectrum of $A$ is traced out: $X=A^*$.
The needed quantities are matrix elements between different nuclear levels.
In lattice QCD, one would have to simulate the whole nucleus directly, which is currently feasible only for nuclei much smaller than
those in the cesium-iodide detector of Ref.~\cite{Akimov:2017ade}.

\begin{figure}[b]
    \centering
    \includegraphics[width=0.48\textwidth,trim={0pt 0pt 48pt 0pt},clip]{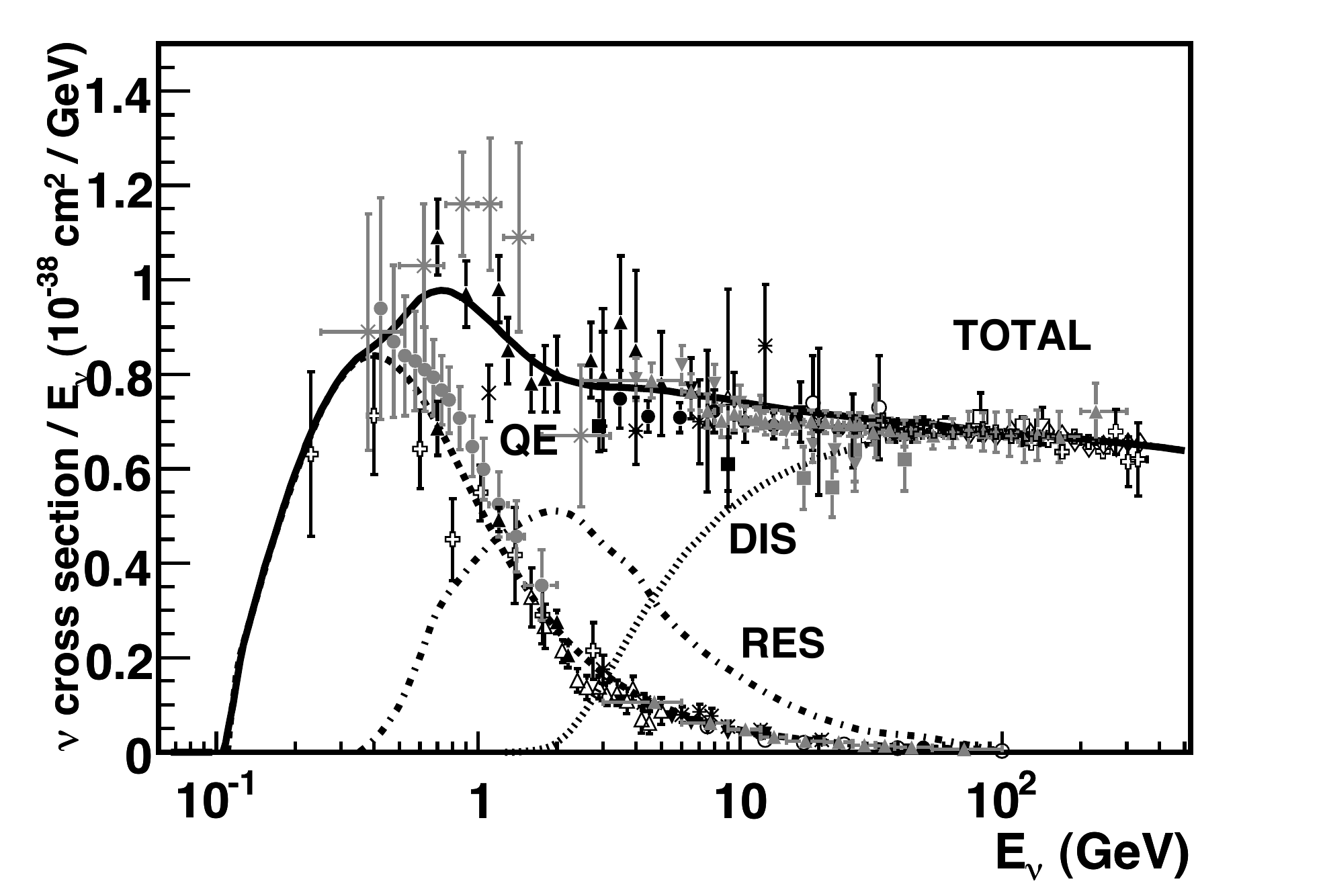} \hfill
    \includegraphics[width=0.48\textwidth,trim={0pt 0pt 48pt 0pt},clip]{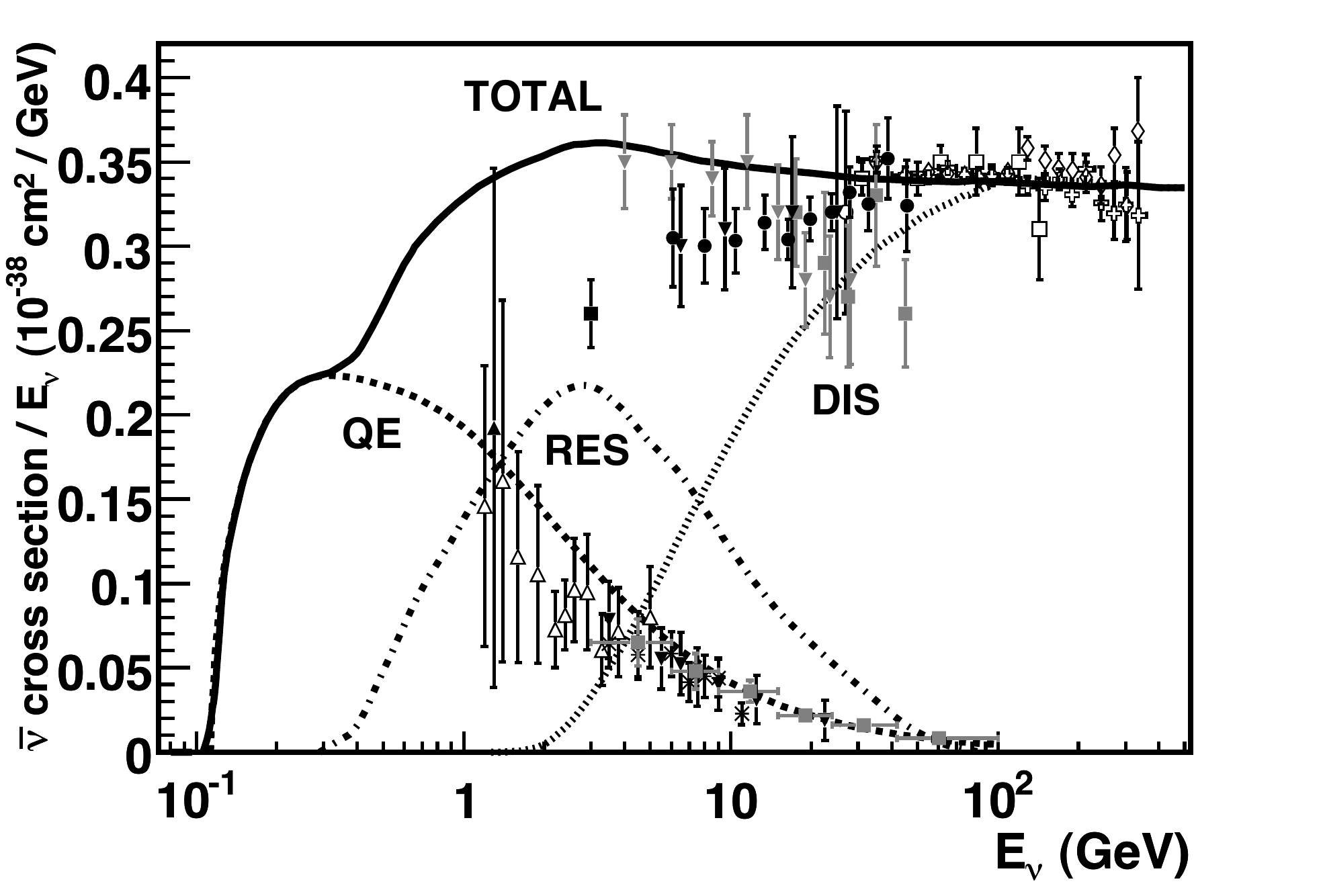}
    \caption[fig:cc_inc]{Cross sections vs neutrino energy (left) or antineutrino energy (right), showing the relative contributions 
        of the underlying processes quasielastic scattering, resonance production, and deep-inelastic
        scattering~\cite{Formaggio:2013kya}.}
    \label{fig:cc_inc}
\end{figure}

At high enough (but still low) energy, a single nucleon can be knocked out.
At its heart, the scattering is
\begin{equation}
    \nu_\ell       n \to l^-p, \qquad
    \bar{\nu}_\ell p \to l^+n,
    \label{eq:ccqe}
\end{equation}
with the initial and final-state nucleons in the nuclear environment.
Such scattering off of a constituent in a bound-state without extra particles is known as quasielastic.
Then nuclear many-body theory requires single-nucleon matrix elements of the form $\langle p(p')|J_\nu|n(p)\rangle$, between a
neutron of momentum $p$ and a proton of momentum~$p'$ (or the $p\to n$ counterpart for antineutrino beams).
These matrix elements are straightforward to calculate in lattice QCD; see Sec.~\ref{sec:easy}.
If pions can be produced, the final state can be a $\Delta(1232)$ resonance, an excited nucleon~$N^*$, or a two-body state~$N\pi$.
In the experiment, these all end up as $N\pi$ so their amplitudes interfere.
In fact, lattice QCD can provide not only the associated transition matrix elements, in the idealization of the resonance as a
stable particle (e.g., $\langle\Delta^{+}|J_\nu|n\rangle$), but also enough information to describe the
{full multi-hadron nature of the final state (at least up to further inelasticities)}; see
Sec.~\ref{sec:hard}.
The quasielastic and resonance regions overlap, because the kinetic energy of Fermi motion is a bit larger than the pion mass.
This overlap is illustrated with experimental data in Fig.~\ref{fig:cc_inc}.
Another contribution in this region arises from many-body nuclear dynamics, for example, when the probe interacts with pairs of
correlated nucleons.
This contribution is described by ``two-body currents'' (see Refs.~\cite{Carlson:1997,Bacca:2014tla} and references therein).
Now a further set of matrix elements is needed, namely of the form $\langle NN|J_\nu|NN\rangle$.
Note that in QCD language, the same current is employed just for two-body initial and final states.

Once the energy is high enough to produce several pions, it is not possible to enumerate every final-state hadron.
In this case, however, lattice QCD can be used to compute nucleon and nuclear structure functions.
In high-energy physics, structure functions are most familiar in deep-inelastic scattering, where the operator-product expansion
(OPE) can be used.
Lattice-QCD calculations can be used to determine the moments of the parton distribution functions (PDFs) that enter in the
deep-inelastic region, and indeed extraction of the full dependence of PDFs on the longitudinal momentum fraction, $x$, is becoming
possible~\cite{Lin:2017snn}.
Moreover, the definition of structure functions is very general.
Lattice QCD may be an ideal way to compute them in the so-called shallow-inelastic region with energy above the resonance region but
insufficient for the OPE; see Sec.~\ref{sec:hard}.

In summary, then, the goals for lattice QCD for neutrino oscillation physics are to calculate matrix elements of the form
\begin{equation}
    \langle f|J_\nu|i\rangle, \qquad
    \langle f|J_\mu^\dagger J_\nu|i\rangle, \qquad
    \langle f|\mathcal{O}|i\rangle,
\end{equation}
where the initial and final states are single nucleons, two nucleons, nucleons with a pion (including resonances), or small nuclei.
In the last case, $\mathcal{O}$ denotes an operator appearing in the OPE, or a bilocal, spatially-separated operator arising in the
calculation of PDFs.
The lattice-QCD calculations of these and related matrix elements have a long history, motivated principally by the desire to
understand nucleon and nuclear structure.
For a broad survey, see our companion whitepaper ``Hadrons and Nuclei''~\cite{Detmold:2018qcd}.

\makebox(0,0){
\nocite{Durr:2010aw}
\phantom{\cite{Bazavov:2012xda,*Bazavov:2017lyh}}
} %
Recall that lattice QCD calculates hadronic correlation functions, which contain information about the masses and matrix elements of
interest; the information is extracted by fitting the behavior of the correlation functions in (Euclidean) time.
Several technical difficulties make baryon calculations more difficult than the corresponding calculations for mesons.
First, statistical errors on baryon correlation functions are larger and more poorly behaved in time
\cite{parisi1983common,lepage1989tasi,Wagman:2017gqi}.
Second, it has proven more difficult, in practice, to disentangle matrix elements of the ground-state baryons from that of their
excitations~\cite{Owen:2012ts}.
Last, the dependence of baryon properties on the light quark mass (used in the simulation) is less well described by the low-energy EFT
of pions and baryons.
All these difficulties can be addressed with more computing.
The signal-to-noise problem can clearly be attacked with higher statistics.
It can also be mitigated by choosing more sophisticated operators to create and annihilate baryon states; this method is also the
way to better filter out the excited states.
Finally, more computing also enables simulations with lighter and even physical quark
masses~\cite{Durr:2010aw,Bazavov:2012xda,Blum:2014tka}.

The rest of this whitepaper is organized as follows.
In Sec.~\ref{sec:easy}, we discuss calculations that are relatively straightforward.
These include nucleon form factors, which are needed to describe quasielastic scattering, and moments of PDFs, which are needed in the deep-inelastic region.
We discuss the form factors in considerable detail, because the time to incorporate these results into event generators is soon or,
arguably, now.
In particular, having the correct slopes for the form factors is crucial to gaining quantitative control of the cross section.
More challenging calculations are covered in Sec.~\ref{sec:hard}.
This class of problems is large and varied: transitions to resonances and multibody states, calculations for shallow- and
deep-inelastic scattering, and the vector and axial matrix elements of small nuclei.
Section~\ref{sec:hardest} turns to calculations that are far enough beyond that state of the art that new ideas or computing
facilities greater than exascale are needed.
Foreseeable computing needs are covered in Sec.~\ref{sec:needs}, noting the separate needs for both capability and capacity
computing.

\section{Straightforward calculations}
\label{sec:easy}

The most straightforward matrix elements to calculate are those with one stable hadron in the initial state, and one or none in the 
final state.
Here we focus on the matrix elements of electroweak currents, $\langle N|J_\mu|N\rangle$, which directly enter neutrino-nucleon 
scattering, and matrix elements of local operators, $\langle N|\mathcal{O}|N\rangle$, where $\mathcal{O}$ appears in the 
operator-product expansion of two $J$ currents, which arise in the analysis of deep-inelastic scattering.

\subsection{Nucleon form factors}

As discussed in Sec.~\ref{sec:intro}, neutrino-nucleon scattering, Eq.~(\ref{eq:ccqe}) is a key process even though the target is a
nucleus.
The $V-A$ charged current of interest is $J_\mu^+=\bar{u}\gamma_\mu(1-\gamma_5)d$.
The matrix element for $n\to p$ can be decomposed into Lorentz covariant combinations of momentum and spin, multiplied by form
factors~\cite{Bhattacharya:2011ah}:
\begin{align}
    \langle p(p')|J_\mu^+|n(p)\rangle &= \bar{u}^{(p)}(p')\left[
        \gamma_\mu F_1^\text{CC} (q^2) +
        i\sigma_{\mu\nu} \frac{q^\nu}{2M_N} F_2^\text{CC} (q^2) +
        \frac{q_\mu}{M_N} F_S^\text{CC} (q^2) \right. \nonumber \\ & \hspace*{2em} + \left.
        \gamma_\mu \gamma_5 F_A^\text{CC} (q^2) +
        \gamma_5 \frac{q_\mu}{M_N} F_P^\text{CC} (q^2) +
        \gamma_5 \frac{(p'+p)_\mu}{M_N} F_T^\text{CC} (q^2) \right] u^{(n)}(p),
    \label{eq:ff}
\end{align}
where $M_N=(M_p+M_n)/2$, $q=p'-p$ and $\bar{u}$ and $u$ are associated spinor factors.
$F_1^\text{CC} (q^2)$, $F_2^\text{CC} (q^2)$, $F_A^\text{CC}(q^2)$, and $F_P^\text{CC}(q^2)$ are known as the Dirac, Pauli, axial,
and induced pseudoscalar form factors, respectively.
The induced scalar and tensor form factors, $F_S^\text{CC}(q^2)$ and $F_T^\text{CC}(q^2)$, are suppressed by $G$~parity violation;
they are known as second-class currents~\cite{Weinberg:1958ut}.
For neutral-current processes, additional form factors $F_i^{\text{EM},N}$ and $F_i^{\text{NC},N}$ are needed: the charged-currents
are all isovector, but the neutral currents contain an isoscalar contribution as well.
Here, $N$ denotes either a proton~$p$ or neutron~$n$.

Because the up- and down-quark masses are so similar, isospin violation can be neglected and, thus, the
charged-current form factors of the vector current (i.e., Dirac and Pauli) can be related to their electromagnetic counterparts, up
to small corrections from isospin violation.
The Dirac and Pauli form factors are usually re-expressed as electric, $G_E(q^2)=F_1(q^2)+q^2F_2(q^2)/(M_n+M_p)^2$, and magnetic,
$G_M(q^2)=F_1(q^2)+F_2(q^2)$, form factors (even for CC and NC).
Expressions relating the differential neutrino-nucleon cross section to the form factors can be found, for example, in
Refs.~\cite{LlewellynSmith:1971uhs,Formaggio:2013kya}.

Most neutrino scattering experiments are performed in a kinematic region of a few~GeV, so tracing out the full $q^2$ dependence is
possible and desirable (see below).
Below 1~GeV it is convenient to focus attention on the intercepts~$F_i(0)$ and (conventionally normalized) slopes
\begin{equation}
    r_E^2 \equiv       6          \left. \frac{dG_E}{dq^2}\right|_{q^2=0}, \quad
    r_M^2 \equiv \frac{6}{G_M(0)} \left. \frac{dG_M}{dq^2}\right|_{q^2=0}, \quad
    r_i^2 \equiv \frac{6}{F_i(0)} \left. \frac{dF_i}{dq^2}\right|_{q^2=0}, \label{eq:rErMri}
\end{equation}
for $i\in\{A,S,T{,P}\}$.
The quantities $r_i$ are usually called ``radii'', although the neutron's $r_E^2$ is negative.

A precise knowledge of the charged-current versions of these quantities is essential for determining the neutrino-nucleon cross
section.
The intercepts and slopes of $G_E^\text{CC}$ and $G_M^\text{CC}$ are well determined from electromagnetic processes and isospin
relations.
Further, the intercept $F_A^\text{CC}(0) = g_A = -1.2723(23)$ is known from neutron $\beta$ decay~\cite{Tanabashi:2018oca}.
{The axial coupling} $g_A$ has been calculated in lattice QCD, although it will be some
time before it can be computed with comparable precision to experiment.
Nevertheless, it is an extremely important benchmark, and once the lattice-QCD precision becomes competitive with experiment, the
result could clear up some puzzles surrounding neutron-decay measurements (see below).

On the other hand, the axial-charge radius-squared $r_A^2$ is less well known.
Historically, the axial form factor has been fit to the so-called ``dipole'' form:
\begin{equation}
    F_A(q^2) = \frac{g_A}{(1-q^2/m_A^2)^2},
    \label{eq:dipole}
\end{equation}
such that $r_A^2=12/m_A^2$.
Experiments report this ``axial mass'', $m_A$, so a comparison of reported values illustrates the current status.
It has been extracted from quasielastic scattering on deuterium targets, finding (e.g.) $m_A=1.02(3)$~GeV~\cite{Bodek:2007ym}, and
from pion electroproduction, finding $m_A=1.08(4)$~GeV~\cite{Liesenfeld:1999mv,Bernard:1992ys}.
More recent experiments find larger values: $m_A=1.20(12)$~GeV at K2K~\cite{Gran:2006jn}, $m_A=1.27(15)$~GeV at
MINOS~\cite{Dorman:2009zz}, and even $m_A=1.35(17)$~GeV at MiniBooNE~\cite{AguilarArevalo:2010zc}, in neutrino charged-current
quasielastic scattering with water, iron, and mineral-oil targets, respectively.
With 2p-2h corrections, however, NOMAD~\cite{Lyubushkin:2008pe}, with a Kevlar target, finds $m_A=1.05(6)$~GeV and

MINERvA~\cite{Fields:2013zhk,*Fiorentini:2013ezn}, with a carbon target, finds the quasielastic cross section to be compatible
with $m_A=0.99$~GeV.
Note that all of these determinations of $m_A$ assume a nuclear model for the target material, which is not the same among the
various collaborations.
Moreover, nuclear modeling uncertainties typically come only from varying parameters of their choice model, not from studying
comparisons among different models.

The uneasy agreement of these results can be removed by switching to a model-independent parametrization of
$F_A(q^2)$~\cite{Bhattacharya:2011ah}.
For example, a reanalysis of 1980s deuterium bubble-chamber data~\cite{Meyer:2016oeg} finds $\sqrt{12/r_A^2}=1.01(24)~\text{GeV}$.
These data are chosen because the nuclear model of the deuteron is under relatively good control.
The main conclusion of Ref.~\cite{Meyer:2016oeg} is that introducing only one free parameter with a qualitatively acceptable but
conceptually incorrect shape, as in Eq.~(\ref{eq:dipole}), leads to gross underestimates of the uncertainty, even when the fit
quality is high.

\begin{figure}[b]
    \centering
    \includegraphics[width=0.5\textwidth,trim={4pt 12pt 12pt 12pt},clip]{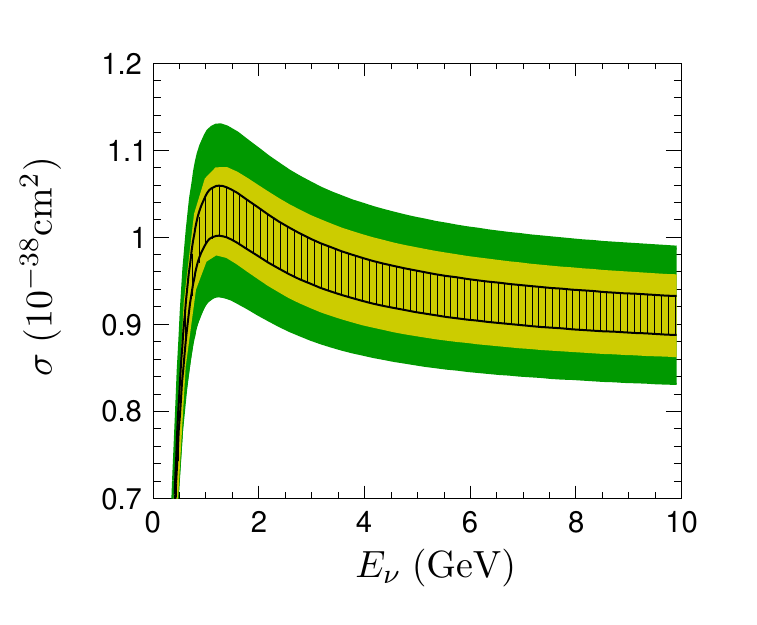}
    \caption{Quasielastic neutrino-neutron cross section assuming current uncertainties for all crucial inputs except $r_A^2$,
    for which a 20\% uncertainty is assumed.
    The hatched (yellow) error band stems from this putative~$r_A^2$ (all other inputs); the green band is the sum in quadrature.
    From Ref.~\cite{Hill:2017wgb}.}
    \label{fig:HKMS}
\end{figure}
\begin{figure} 
    \centering
    \includegraphics[width=0.45\textwidth]{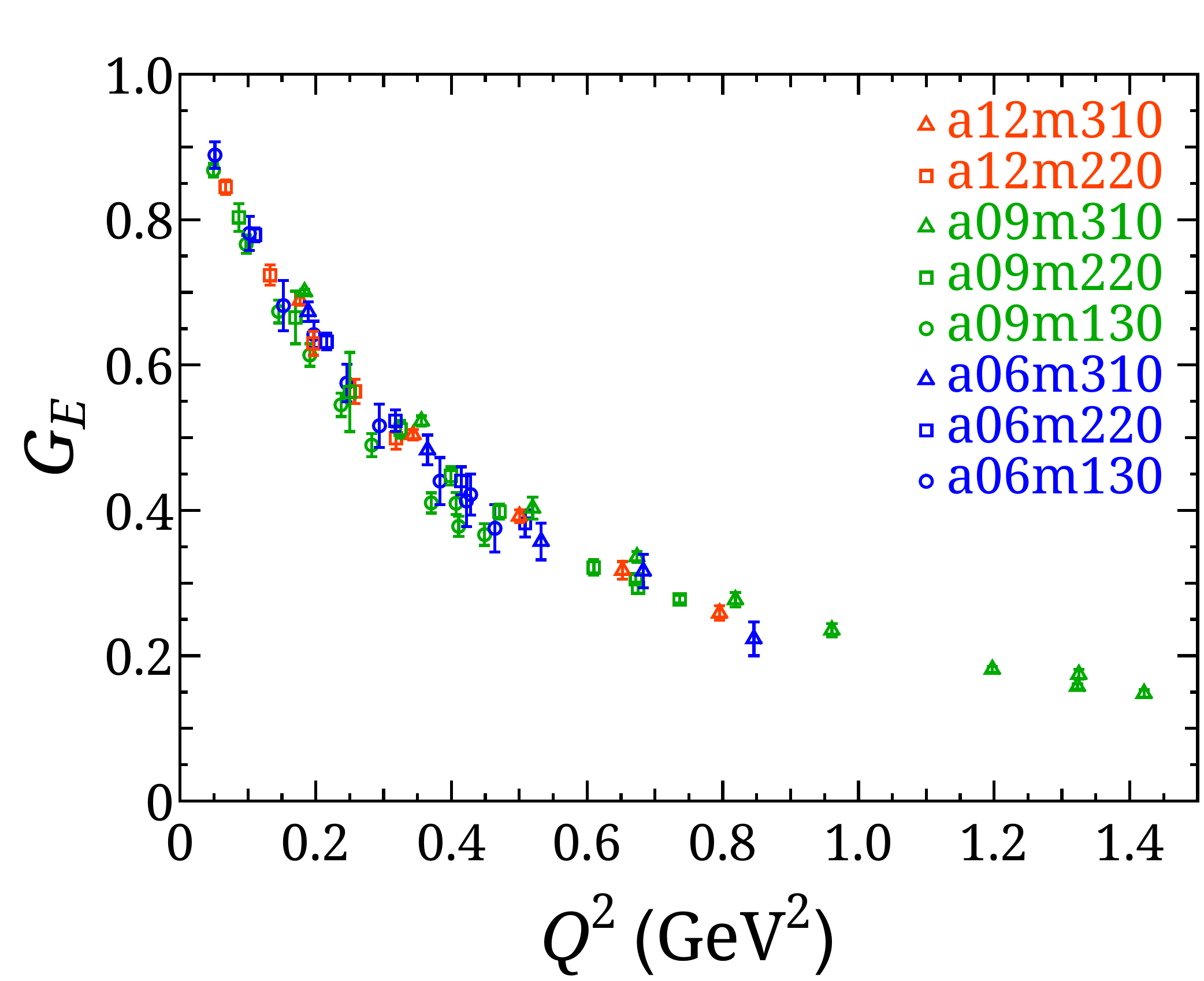} \hfill
    \includegraphics[width=0.45\textwidth]{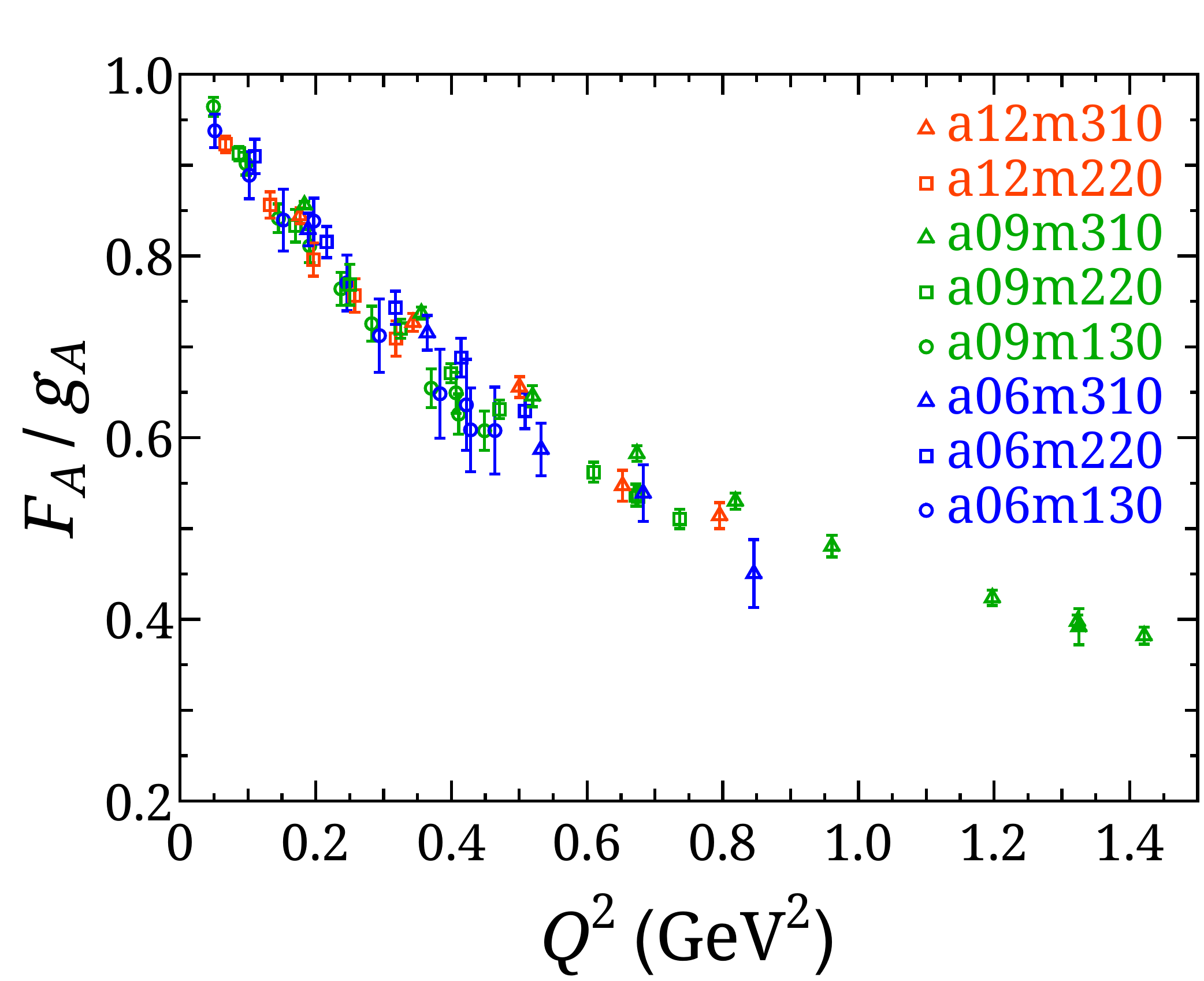}
    \caption{Electric (left) and isovector axial (right) form factors of the nucleon vs $Q^2=-q^2$.
        Data from Ref.~\cite{Jang:2018lup,Jang:2018djx}.}
    \label{fig:EMFF}
\end{figure}

Figure~\ref{fig:EMFF}, from Ref.~\cite{Hill:2017wgb}, shows the dependency of $\nu n$ quasielastic cross section on $E_\nu$,
assuming $r_A^2$ is known with $20$\% uncertainty.
As one can see, this quantity affects the both the normalization and fall-off of the cross section, which are needed, respectively,
to determine the mixing angle and mass difference in an oscillation.
Furthermore, a lattice-QCD calculation with 20\% uncertainty (compared to 50\% in Ref.~\cite{Meyer:2016oeg}) is an important
milestone, because then the $r_A^2$ uncertainty becomes subdominant{, at least until other uncertainties have been
reduced}.

The lattice-QCD community has been pursuing the calculation of the nucleon form factors for a long time.
A representative set of recent work can be found in Refs.~\cite{Bali:2014nma,Bhattacharya:2016zcn,Green:2017keo,Alexandrou:2017hac,%
Capitani:2017qpc,Rajan:2017lxk,Jang:2018lup,Yamanaka:2018uud,Chang:2018uxx,Ishikawa:2018rew,Liang:2018pis,Ottnad:2018fri,%
Bali:2018qus,Shintani:2018ozy,Martha:2018lft} Significant improvements have been made to investigate the quark-mass, finite-volume,
and finite-lattice-spacing dependence, and the effects of excited-state contamination in the correlation functions.
With these technical and algorithmic advances, lattice QCD can calculate not only the isovector contribution but also the
computationally more demanding isoscalar and strange-quark contributions, which are needed for neutral-current processes, discussed
below.

Sample lattice-QCD calculations~\cite{Jang:2018lup,Jang:2018djx} of the nucleon isovector electric and axial form
factors---$G_E$ and $F_A$---are shown in Fig.~\ref{fig:EMFF}.
Eight different $2 + 1 + 1$-flavor HISQ ensembles generated by the MILC collaboration~\cite{Bazavov:2012xda,*Bazavov:2017lyh} with
lattice spacings in the range 0.06--0.12~fm and pion mass in the range 130--310~MeV are employed.
In this calculation, excited-state contamination is controlled via a three-state fit.
The results are in good agreement with the experimental data for the nucleon electromagnetic form factor $G_E(q^2)$
On the other hand, the axial form factor is not as steep as experimental determinations with $m_A\approx1$~GeV~\cite{Bernard:2001rs},
yet is compatible with MiniBooNE's $m_A\approx1.35$~GeV~\cite{AguilarArevalo:2010zc}.
Despite the many laudable aspects of Ref.~\cite{Jang:2018lup}, a full and robust accounting
of all systematics involved in these lattice-QCD calculations has not yet been feasible.
Reliable confrontation with precise experimental data for $G_E$---and, hence, a solid prediction of $F_A$---requires an increase in
computational resources to overcome the technical obstacles to nucleon matrix elements{, discussed in Sec.~\ref{sec:intro}}.

\begin{figure}[b]
    \centering
    \includegraphics[width=0.52\textwidth,trim={0pt 2pt 0pt 12pt},clip]{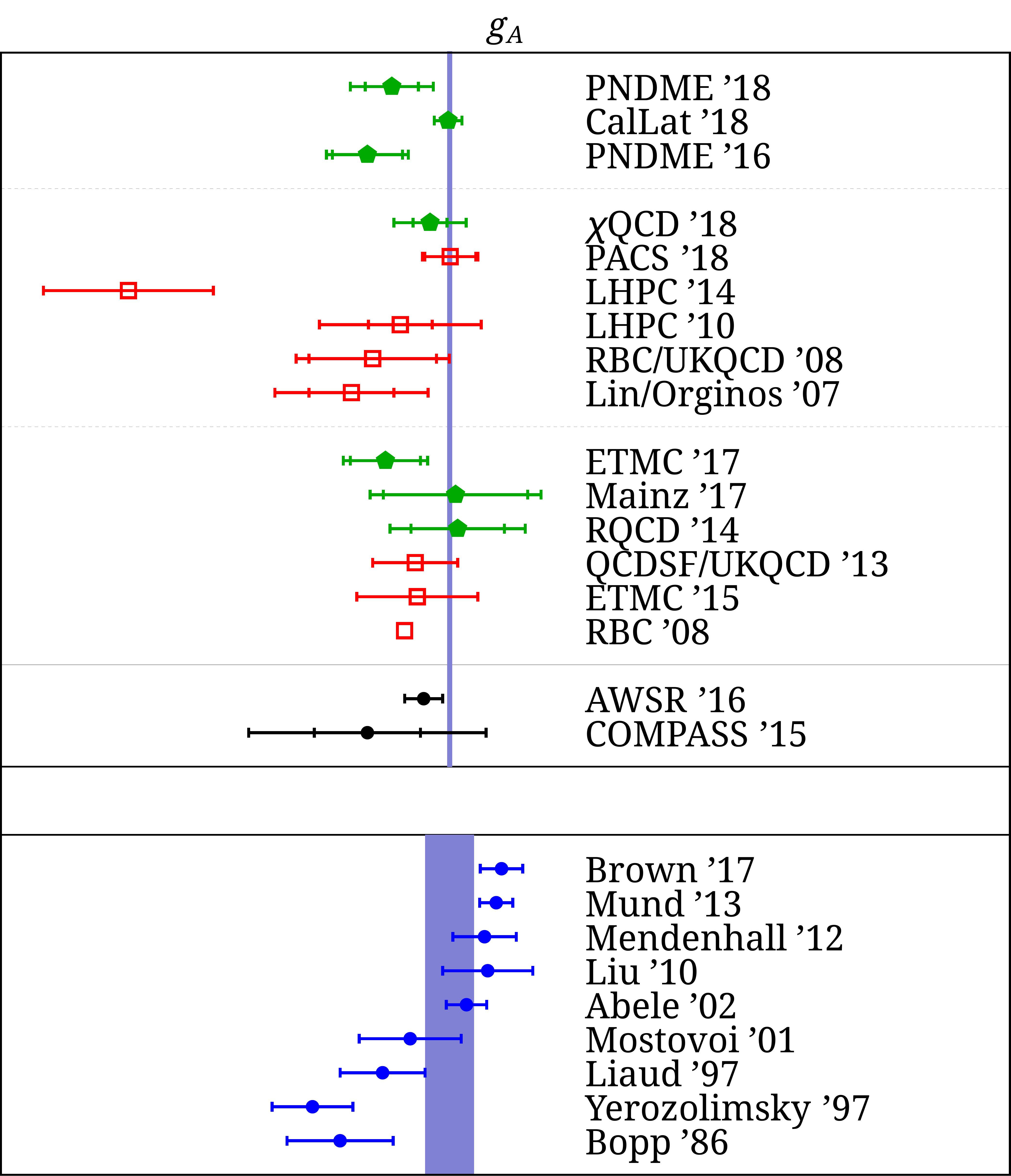} \hfill
    \includegraphics[width=0.47\textwidth,trim={24pt 0pt 18pt 12pt},clip]{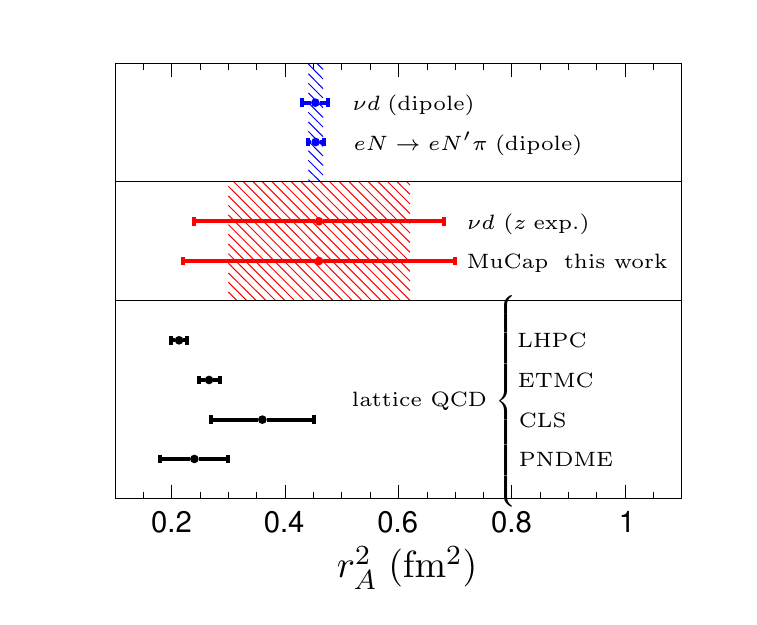}
    \caption{Status of lattice-QCD calculations of $g_A$ (left) and $r_A^2$ (right), together with non-lattice determinations.
        Left: Filled green (unfilled red) lattice-QCD results have (in)complete error budgets.
        The violet line in the upper panel is the PDG average of the results in the bottom panel, in which the scale is blown up by a
        factor of~10.
        Right: As discussed in the text, the error bars on $r_A^2$ from dipole fits are underestimated and the two small lattice-QCD 
        error bars stem from incomplete error analyses (critiqued below).
        The references for $r_A^2$ from top to bottom are as follows:
        ``$\nu d$ and $eN\to eN'\pi$ (dipole)''~\cite{Bodek:2007ym},
        ``$\nu d$ ($z$ exp.)''~\cite{Meyer:2016oeg},
        ``MuCap this work''~\cite{Hill:2017wgb},
        LHPC~\cite{Green:2017keo} (NB: one lattice spacing and $M_\pi=317$~MeV),
        ETMC~\cite{Alexandrou:2017hac} (NB: no strange sea and a small volume such that $M_\pi L<3$),
        CLS~\cite{Capitani:2017qpc},
        PNDME\cite{Rajan:2017lxk}.
        From Refs.~\cite{Gupta:2018qil} (left) and~\cite{Hill:2017wgb} (right, adapted with permission).}
    \label{fig:gArA}
\end{figure}

The status of lattice-QCD calculations of $g_A$ and $r_A^2$ is shown in Fig.~\ref{fig:gArA}.
The left plot~\cite{Gupta:2018qil}, for~$g_A$, shows that lattice-QCD is at this time much less precise than the results from
neutron $\beta$ decay.%
\footnote{The color code here is adapted from the Flavor Lattice Averaging Group~\cite{Aoki:2016frl,*Aoki:2019cca,*FLAGweb}, as
specified in the
Appendix of Ref.~\cite{Bhattacharya:2015wna}.} %
Note, however, that bottle and beam experiments measuring the neutron lifetime yield values of~$g_A$ that differ by $3\sigma$.
For example, a 2015 bottle measurement leads to $g_A=1.2749(11)$~\cite{Arzumanov:2015tea}, while a 2013 beam measurement leads to
$g_A=1.2684(20)$~\cite{Yue:2013qrc}.
It would be interesting to know the answer from lattice QCD.
The precision required depends on whether the (average of several) calculation(s) lands between the two neutron-lifetime values or
outside the interval.
In the latter case, at least percent-level precision is needed, which is likely to be achieved with three years (assuming sustained
computing support).
If lattice QCD lands in the middle, 0.3\% precision is needed.
In this scenario, we would also need 1+1+1(+1)-flavor ensembles, since the isospin symmetry would play an important role at such
precision; it would take 5--10 years to account for full systematics.%
\footnote{Note that the normalization of the matrix element can be blinded with an multiplicative offset~\cite{Meyer:2016kwb}, to
guard against analyst bias.
The results in Fig.~\ref{fig:gArA}~(left) have not, however, employed this technique.}

The right plot~\cite{Hill:2017wgb}, for~$r_A^2$, shows significant problems: the analysis with the $z$
expansion~\cite{Meyer:2016oeg} debunks the uncertainty estimates of determinations predicated on the dipole form.
The model independent results (red; between the horizontal lines) illustrate the best estimate of $r_A^2$ without such strong
assumptions.
One should bear in mind that the ``experimental'' determinations all make assumptions: without new $\nu d$ and $\bar{\nu}p$
experiments~\cite{Morfin:2018int}, it seems nearly impossible to improve the situation via experiment.
On the other hand, lattice gauge theory can provide an \emph{ab initio} result from QCD.
Indeed, lattice QCD is beginning to play a role, but another generation of calculations is needed before fully definitive results
with uncertainties small enough to make an impact on cross section calculations are achieved.

For the full energy range of LBNF/DUNE, it will be necessary to trace out the full $q^2$ dependence of the form factors.
It is imperative to use a model-independent parametrization based on general analytic properties.
In the complex-$q^2$ plane, the vector (axial) form factors have a cut starting at $q^2=t_\text{cut}\equiv4M_\pi^2$
($q^2=t_\text{cut}\equiv9M_\pi^2$) and extending to $\infty$ on the real axis.
The cut lies outside scattering kinematics $q^2<0$ but nevertheless prevents a useful series expansion in $q^2$ around the origin.
A rigorous way to proceed is to introduce a conformal mapping that maps the cut to the unit
circle~\cite{Meiman:1963:zex,Hill:2010yb}:
\begin{equation}
    z(t) = \frac{\sqrt{t_\text{cut}-t} - \sqrt{t_\text{cut}-t_0}}{\sqrt{t_\text{cut}-t} + \sqrt{t_\text{cut}-t_0}},
    \label{eq:zxform}
\end{equation}
where the parameter $t_0$ can be chosen to center the $q^2$ range of interest on $z=0$; in general, spacelike $q^2\to-\infty$ maps
to $z\to1$.
An expansion of the form
\begin{equation}
    F(z) = \sum_k a_k z^k 
    \label{eq:zformfactor}
\end{equation}
thus has an expansion parameter $|z|<1$.
Moreover, unitarity in quantum mechanics ensures that the series is uniformly convergent on this interval.
In fact, unitarity leads to bounds on the coefficients $a_k$ that the dipole form,
 Eq.~(\ref{eq:dipole}), violates~\cite{Bhattacharya:2011ah}.

In practice~\cite{Bhattacharya:2011ah,Hill:2006ub,Bernard:2008dn,Bailey:2008wp}, the $z$~expansion converges after a few terms.
Because on the nonlinear mapping, even an intercept and slope in $z$ give a form factor with a physical shape (i.e., similar to
those shown in Fig.~\ref{fig:EMFF}).
As lattice data improve, more and more terms will become resolvable.
As in CKM physics~\cite{Bernard:2008dn,Bailey:2008wp}, lattice-QCD papers can provide the coefficients, their uncertainties,
and their correlations; several lattice-QCD calculations of $F_A$ do the
same~\cite{Alexandrou:2017hac,Capitani:2017qpc,Green:2017keo,Rajan:2017lxk}.
Finally, code for taking such $z$-expansion input is included in the GENIE event generator~\cite{Andreopoulos:2009rq}
module for the axial form factor, and work is underway to extend this to the vector form factor channel.

Although not crucial to neutrino oscillations, the same experiments study weak neutral-current interactions of the $Z$~boson and,
possibly, non-Standard bosons~\cite{deGouvea:2015ndi,Heeck:2018nzc}.
The corresponding Dirac and Pauli form factors can be obtained from the proton and neutron electromagnetic form factors and the
strange-quark contribution {(accessible in parity-violating elastic electron-scattering experiments \cite{Armstrong:2012bi})} as
\begin{align}
    F_i^\text{NC} &= \left(\half-\sin^2\theta_\text{W}\right) \left(F_i^{\text{em},p}-F_i^{\text{em},n}\right) -
        \sin^2\theta_\text{W}\left(F_i^{\text{em},p}+F_i^{\text{em},n}\right) - \half F_i^s,
    \label{eq:Z} \\
    i &\in \{1,2\}. \nonumber
\end{align}
Using the most recent $z$-expansion fit to nucleon electromagnetic form factors~\cite{Ye:2017gyb} and a new lattice-QCD calculation
of strange-quark form factors~\cite{Sufian:2016vso}, one can see that the strange-quark contribution increases the neutral-current
Pauli form factor, $F_2^\text{NC}(q^2)$, by about 3.1\% and 2.5\% at $q^2=0$ and $q^2=-0.1~\text{GeV}^2$, respectively.
Although the strange-quark contribution is small, the coefficients $(\frac{1}{2}-\sin^2\theta_\text{W})$ and $\sin^2\theta_\text{W}$
suppress the two combinations of nucleon electromagnetic form factors in Eq.~(\ref{eq:Z}), such that the strange-quark sea makes an
important contribution to $F_2^\text{NC}(q^2)$ at low~$q^2$.

Similarly, assuming isospin symmetry and the absence of second-class currents, one can relate the neutral-current axial form factor
to the charged-current axial and strange-quark axial form factors~\cite{Garvey:1993sg,Garvey:1992cg}:
\begin{equation}
    F^\text{NC}_A = \half (-F^\text{CC}_A + F^s_A ).
\end{equation}
It has been shown~\cite{AguilarArevalo:2010cx,Aguilar-Arevalo:2013nkf} that the effect of Pauli blocking becomes very significant
in the region $0<-q^2\lesssim0.2~\text{GeV}^2$.
Therefore, a precise lattice-QCD calculation of $F^\text{NC}_A (q^2)$ is required for a precise estimate of the neutral-current
(anti)neutrino-nucleon scattering cross
section.

Finally, we note that quasielastic neutrino and antineutrino scattering would be sensitive to the presence of the second-class
currents, $F_S$ and $F_T$ in Eq.~(\ref{eq:ff}), characterized by a different $G$-parity to the standard vector and axial currents of
the Standard Model.
The search for such currents has long been pursued in the $\beta$-decay experiments and in muon-capture experiments, but the
measurement of polarization observables in the quasielastic scattering both of nucleons and of hyperons has been shown to be
sensitive both to $G$ invariance and to $T$-invariance~\cite{Fatima:2018tzs}.
Lattice QCD can contribute to these tests through calculations of induced scalar and tensor currents, including calculations
of transition form factors to the rest of the SU(3) baryon octet ($\Lambda$ and $\Sigma$ as well as $p$ and $n$), such as those in
Refs.~\cite{Lin:2008rb,Sasaki:2014osa}.

\subsection{Moments of parton density functions}

Lattice QCD can be used to calculate matrix elements of other operators besides the electroweak currents.
An important class of operators are those the appear in the operator-product expansion of two currents.
Their matrix elements are related to the moments of structure functions in deep-inelastic scattering.
For a full discussion, see the USQCD companion white paper ``Hadrons and Nuclei''~\cite{Detmold:2018qcd}.
Here, {applications to neutrino physics are discussed}.

In 2001, the NuTeV collaboration determined the on-shell weak mixing angle, $\sin^2\theta_W\equiv1-m_W^2/m_Z^2$, to be
$0.2277\pm0.0013_\text{stat}\pm0.0009_\text{syst}$~\cite{Zeller:2001hh} in deep-inelastic neutrino scattering off iron.
This result is $2.7\sigma$ discrepant from the current world average of other experiments, $0.22343\pm0.00007$
\cite{Tanabashi:2018oca}.
This discrepancy, which is known as the ``NuTeV anomaly'', has no universally accepted explanation{, although many possibilities
have been raised \cite{Londergan:2003ij,Ding:2004dv,Gluck:2005xh,Eskola:2006ux,Cloet:2009qs,Bentz:2009yy}}.

One suggestion that may account for part of the anomaly is the strange-antistrange parton
asymmetry~\cite{Davidson:2001ji,Kretzer:2003wy}, $\langle x\rangle_{s_-}=\int dx\, x \left[s(x) - \bar{s}(x)\right]$, where $s(x)$
($\bar{s}(x)$) is the (anti)strange parton distribution function, as a function of parton momentum fraction~$x$.
A~global analysis of several experimental data sets gives $\langle x\rangle_{s_-}\approx0.0018$~\cite{Lai:2007dq}, which is
consistent with a 2006 NuTeV analysis of dimuon production~\cite{Mason:2006qa}.
The global analysis does not, however, find a tight constraint: the authors of Ref.~\cite{Lai:2007dq} present the range
$-0.001<\langle x\rangle_{s_-}<0.005$ at 90\% confidence level.

In view of the uncertain of $\langle x\rangle_{s_-}$ from global fitting, a first-principles lattice-QCD calculation is warranted.
There is, however, no local operator which corresponds to $\langle x\rangle_ {s_- }$.
Instead, one can calculate the third moment from the local operator $\bar{s}\gamma_{\mu}D_{\nu}D_{\lambda}s$ which
corresponds to $\langle x^2\rangle_ {s_-}= \int dx\, x^2 (s(x) - \bar{s}(x))$.
Assuming $s(x) - \bar{s}(x)$ changes sign only once, $\langle x^2\rangle_{s_- }$ should give the same sign as that of $\langle
x\rangle_{s_-}$.
This quantity can also be used to constrain the $x$-dependent distribution, but since it is expected to be small, calculations will
require significant resources.

\section{Challenging calculations}
\label{sec:hard}

In this section, calculations that are computationally more difficult than the form factors in Sec.~\ref{sec:easy} are discussed.
That said, the conceptual formalism underlying these calculations is well established, and pilot calculations provide some idea of
how more complete calculations can be carried out.
More complicated final states in the resonance regions (Sec.~\ref{sec:resonance}), the shallow inelastic region
(Sec.~\ref{sec:shallow}), and the deep inelastic region (Sec.~\ref{sec:deep}) are discussed, as are calculations of the axial
charge, and related quantities, of small nuclei (Sec.~\ref{sec:small-gA}).

\subsection{Transition form factors: resonances and multibody final states}
\label{sec:resonance}

Neutrino scattering above the pion-production threshold constitutes the resonance region, where the scattered nucleon is excited
into resonances, beginning with the $\Delta(1232)$.
To describe the data in this regime thus requires a quantitative knowledge of the $N \to \Delta$ and $N\to N^*$ transitions,
mediated through an external current.
Because these hadrons are unstable, they can also be viewed as a nucleon with one or more pions, which are the only hadrons composed
of the light $u/d$ quarks stable under the strong interaction.

Lattice QCD has a long history of calculations of the transition form factors to the~$\Delta$, treating it as stable.
Both the vector current~\cite{Leinweber:1992pv,Alexandrou:2007dt}, and the axial current~\cite{Alexandrou:2009vqd} have been
studied with unphysically large quark masses, such that $M_\Delta$ at these quark masses lies below the $N \pi$ threshold.
{These calculations} are useful benchmarks for comparisons with non-lattice approaches that neglect the two-body nature of the
resonance.
Although not as rigorous as the methods discussed below, this ``quick and dirty'' approach may be timely, for example, providing
qualitative input to understand better the MiniBooNE low-energy backgrounds from $\Delta\to N\gamma$~\cite{Aguilar-Arevalo:2018gpe}.

Because of the finite volume and Euclidean signature, calculations with two-body states in lattice QCD are conceptually
and computationally more difficult~\cite{Luscher:1986pf,*Luscher:1990ux,Lellouch:2000pv} than the calculations discussed in
Sec.~\ref{sec:easy}.
For example, the L\"uscher method~\cite{Luscher:1986pf,*Luscher:1990ux} relating energy shifts at finite volume to infinite-volume
momentum-dependent phase shifts has been used to study the $\rho$ meson~\cite{Aoki:2007rd,Feng:2010es,Dudek:2012xn,Wilson:2015dqa,%
Alexandrou:2017mpi}, as well as $I=2$ $\pi\pi$ phase shifts~\cite{Beane:2007xs,Feng:2009ij,Beane:2011sc,Dudek:2010ew,Dudek:2012gj}
from first principles.
The theoretical framework for understanding the transition to multihadron states from Euclidean-space lattice QCD calculations have
been further developed over the past several years.
Notably, the formalism has been extended both to inelastic scattering~\cite{Detmold:2004qn,He:2005ey,Hansen:2012tf,Briceno:2012yi,%
Guo:2012hv} with several two-body channels, and to three-body scattering \cite{Kreuzer:2010ti,Briceno:2012rv,Meissner:2014dea,%
Briceno:2017tce,Briceno:2018aml,Doring:2018xxx}, and there have now been several computational applications of these
advances~\cite{Beane:2007es,Dudek:2016cru,Briceno:2017qmb,Woss:2018irj}.

\begin{figure}[b]
    \centering
    \includegraphics[width=0.75\textwidth,trim={4pt 12pt 12pt 6pt},clip]{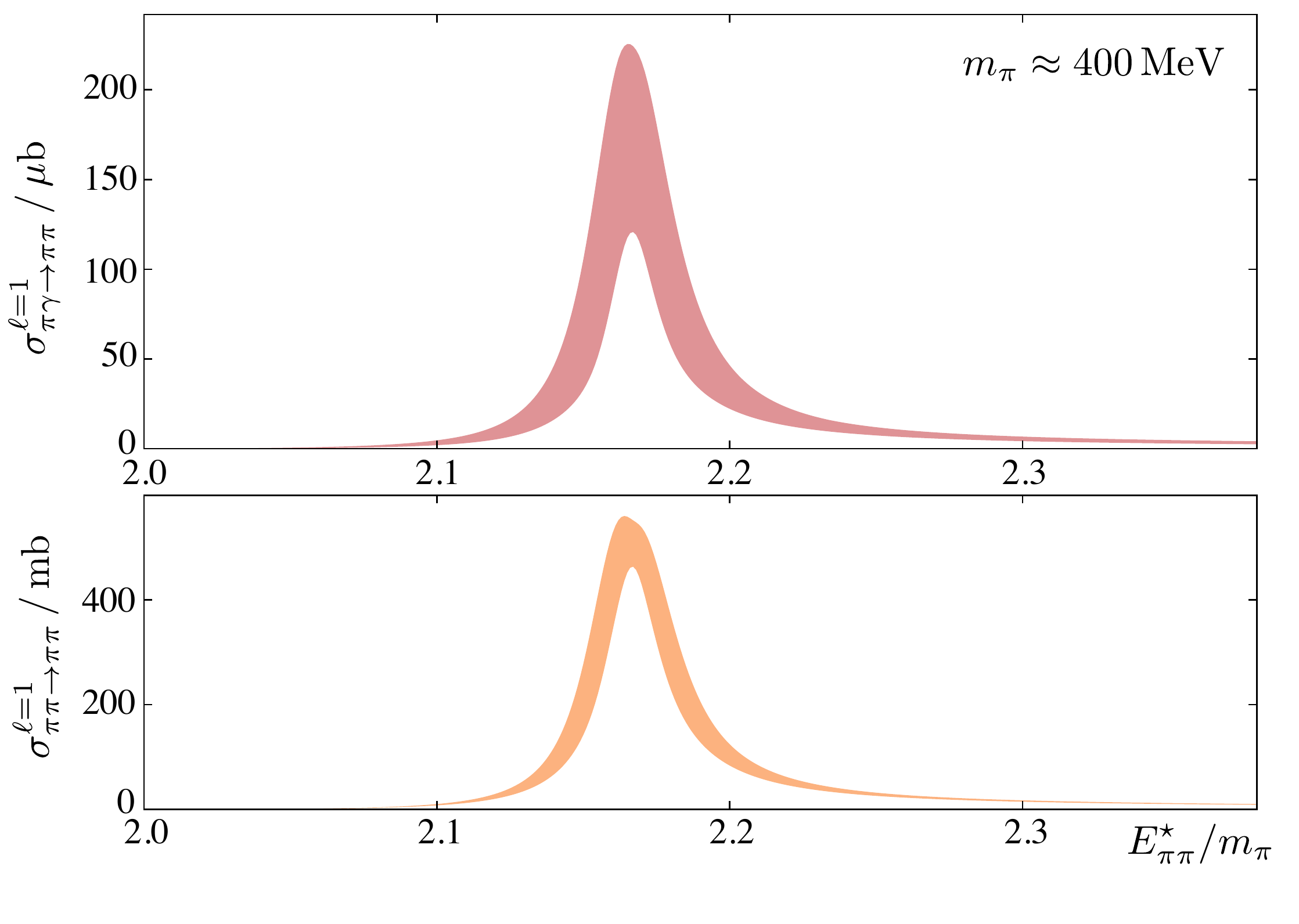}
    \caption{The upper and lower panels show the $\gamma^* \pi^+\to \pi^+ \pi^0$ and $l=1 $ elastic $\pi\pi$ scattering cross sections,
        respectively, as a function of $\pi\pi$ energy, with the $\rho$ resonance clearly visible~\cite{Briceno:2015csa}.}
    \label{fig:resonance}
\end{figure}

A quantitative understanding of resonance production entails extending the formalism to encompassing transitions mediated through
external currents, corresponding here to both vector and axial currents.
The needed formalism to two-body final states, and for arbitrary spin, has now been developed~\cite{Briceno:2015csa}.
The applications have largely focused on the meson sector.
To cite an example bearing some similarity to $W^*n\to \Delta$ in neutrino scattering, the $\gamma^*\pi\to\rho$ transition has been
computed in lattice QCD~\cite{Briceno:2016kkp,Alexandrou:2018jbt}, providing the first rigorous calculation of the transition form
factor to an unstable hadron, illustrated in Fig.~\ref{fig:resonance}
In addition, methods to extract resonance-to-resonance transitions, for example, $\gamma^*\rho\to\rho$, via lattice calculations of
two-to-two transition amplitudes, in this case $\gamma^*\pi\pi\to\pi\pi$, have been developed \cite{Briceno:2015tza,Baroni:2018iau}.
This opens the possibility for calculations of two-body currents, that is, matrix elements of the form 
$\langle NN|J^\mu|NN\rangle$ needed for two-nucleon knockout.

Thus, the theoretical underpinnings for understanding resonance production in $\nu N\to\Delta$, $N^*$, and~$N\pi$, are therefore largely
in place.
Calculations of multihadron states containing baryons are complicated by the extra complexity of the systems relating to the
increased number of quarks, by poorer signal-to-noise ratios, and by the larger number of open channels.
Even so, the first \textit{ab initio} determination of $\Delta(1232)$ resonance parameters appeared in 2017~\cite{Andersen:2017una},
albeit for a simulation with quark masses corresponding to a pion mass of $280~\text{MeV}$, yielding a $\Delta$-$N$-$\pi$
coupling in agreement with phenomenological determinations.
As the invariant mass of the system increases within the resonance regime and the pion mass is decreased to its physical value,
however, inelastic processes and three- and higher-body final states become relevant.
Further theoretical work is needed to encompass transitions to three or more particles, and further development of efficient
algorithms is needed to evaluate the larger number of Wick contractions that increasingly dominate the computational cost of the
calculation.
The interplay between theoretical methods and practical algorithms is, of course, ideally researched on high-capacity computing
facilities.

\subsection{Hadron tensor for shallow and deep inelastic scattering}
\label{sec:shallow}

At higher energies, more and more pions are produced and a full theoretical description of any given final state becomes
impractical.
One can, however, study the sum over all final states via the optical theorem and consider forward matrix elements of the product of
two currents.
Whereas nuclear many-body theory decomposes the low-multiplicity cases into products of nuclear wavefunctions and nucleon (and
$N\pi$, \ldots) form factors, here one can decompose the nuclear hadron tensor, $\langle A|J_\mu^\dagger J_\nu|A\rangle$, into a
spectral function~\cite{Rocco:2015cil} and the nucleon hadron tensor, $\langle N|J_\mu^\dagger J_\nu|N\rangle$.
A recent development in lattice QCD is to calculate this quantity from a four-point correlation function.
This approach is especially appealing in the region, sometimes called shallow inelastic, between the resonances and deep-inelastic
scattering, where no other theoretical tool holds much promise~\cite{Alvarez-Ruso:2017oui}.

The Euclidean hadronic tensor~\cite{Liu:1993cv,Liu:1998um,Liu:1999ak,Aglietti:1998mz,Detmold:2005gg,Liu:2016djw,Chambers:2017dov,%
Hansen:2017mnd} can be decomposed in terms of structure functions that are related to their Minkowski counterparts through a Laplace
transform.
Thus, to obtain the desired structure functions, an inverse Laplace transform is needed, an ill-posed problem that arises in many
fields.
Three approaches to this problem are the maximum entropy method (MEM)~\cite{Liu:1999ak}, the MEM with a prior to stabilize the fit
for Bayesian reconstruction (MEM-BR)~\cite{Liu:2016djw}, and the Backus-Gilbert method~\cite{Hansen:2017mnd}.
These three numerical approaches have been studied recently~\cite{Liang:2017mye}.
The Backus-Gilbert method yields a single broad peak in the energy spectrum from lattice data with 20 points in Euclidean time.
With both the MEM and MEM-BR, the elastic peak and the resonance peak are resolved, with the MEM-BR producing sharper peaks and a
more stable reconstruction.
Given the test lattice spacing of 0.12~fm, there is no excitation spectrum above 2~GeV and, thus, no strength in the spectral weight
above the resonance region.
For a finer lattice with spacing 0.04~fm, the spectral weight up to 5~GeV is accessible.
Even though it may still not be sufficient to resolve the individual resonances, the fact it can cover both the resonance and the
shallow inelastic scattering regions makes the lattice hadronic tensor calculation a promising theoretical tool to address the
total cross section of the neutrino-nucleus scattering over a wide range of energy transfers up to 5~GeV.

The hadron tensor can also be computed for deep-inelastic scattering.
In this case, the calculation needs to be able to access the kinematic region where $Q^2>4~\text{GeV}^2$ and energy transfer
$\nu>5$~GeV where the higher-twist contributions are suppressed.
The Euclidean correlation function can also be analyzed with the OPE, along the lines of the suggestion for calculating the shape
function of the inclusive $B$-meson semileptonic decay rate~\cite{Aglietti:1998mz}.
In addition, using a fictitious heavy-quark propagator between the currents to calculate moments has been
proposed~\cite{Detmold:2005gg}.
A related approach is also discussed in Ref.~\cite{Chambers:2017dov}.
Unlike the approaches discussed in the next subsection, the hadron-tensor approach to deep-inelastic scattering does not need to
match to the infinite-momentum frame.

\subsection{Parton densities for neutrino deep-inelastic scattering}
\label{sec:deep}

The parton distribution functions (PDFs) will be important inputs in the upcoming precision neutrino-physics experiments,
particularly at large Bjorken~$x$ and at the highest energies of the DUNE beam, $\sim4$--5~GeV.
For these kinematics, current global-fit PDFs either suffer greatly from the theoretical uncertainty in their nuclear treatment or
rely mainly on extrapolation from intermediate~$x$.

Direct calculation of the Bjorken~$x$ dependence of hadron structure in lattice QCD has only recently become possible thanks to the
development of the large-momentum effective theory (LaMET)~\cite{Ji:2013dva}, which
introduces a large momentum~$P$ to connect Euclidean lattice~QCD to the desired Minkowski distributions.%
\footnote{See Refs.~\cite{Rossi:2018zkn,Cichy:2018mum,Monahan:2018euv} for critical discussions of this approach.} %
This framework has allowed the first direct lattice-QCD computations of the $x$ dependence of parton distributions~\cite{Lin:2014zya}.
Further developments spurred on by these developments include that of pseudo-PDFs~\cite{Radyushkin:2017cyf}, and that of matrix
elements of gauge-invariant current-current correlators~\cite{Ma:2017pxb}.
In these new approaches, valence- and sea-quark structure can be disentangled, which leads to the possibility of using lattice-QCD
calculations to directly compare with experiments on large-$x$ structure in SoLID at Jefferson Lab, with sea structure in Drell-Yan
experiments at Fermilab or with data from a future electron-ion collider.
In addition to the hadron-tensor method described in the previous subsection, these new approaches and numerical investigations
thereof are described in detail in the companion whitepaper ``Hadrons and Nuclei''~\cite{Detmold:2018qcd}.

The lattice-QCD effort so far has focused on isovector combinations of PDFs, that is, the difference between the up and down
distributions.
A recent joint lattice-QCD and global-fitting community report~\cite{Lin:2017snn}, an effort led by USQCD members, demonstrated that
a calculation of the isovector proton PDF at the 12\% level for $x \in [0.7,0.9]$ can impact our knowledge of the PDF at $x$ near~1
by more than~20\%.
This kinematic region is relevant for neutrino experiments, and such precision should be feasible in the near term.
In addition, as crosscheck of nuclear theory in this region, exploratory calculations of nuclear PDFs will become available; see
Sec.~\ref{sec:small-gA}.
Such precision is already relevant to neutrino-nucleon scattering at 4--5~GeV.
Further, it allows a crosscheck of the nuclear-theory treatment and of the systematic uncertainties of nuclear PDFs.

Neutrino DIS can be important for determining the strange quark and antiquark parton distributions.
Currently, no calculation of the Bjorken-$x$ dependence of the strange PDF has been done with lattice QCD, due to numerical
limitations, but there are USQCD proposals to investigate $s(x)-\bar{s}(x)$.
On the other hand, the nucleon sea flavor asymmetry $\bar{u}(x)-\bar{d}(x)$ has been studied~\cite{Chen:2018fwa}.
Unfortunately, the uncertainties in the quasi-PDF approach are currently much larger than those from experimental/phenomenological
extraction.

Going to larger nucleon boost momenta $P$ with high statistics is key to reducing {several} {systematic
uncertainties} in these quasi-PDF and pseudo-PDF approaches~\cite{Detmold:2018qcd}, especially for the antiquark distribution and
small Bjorken~$x$.
However, this poses several computational challenges.
First, large momentum translates into large $(aP)^n$, and therefore, ensembles with increasingly smaller lattice spacings~$a$ are
needed.
Given the need to keep the spatial size of the box sufficiently large to avoid significant finite-volume effects, which may be
enhanced for some nonlocal matrix elements~\cite{Briceno:2018lfj}, this increases the computational cost.
Second, as the momentum becomes larger, the signal-to-noise ratio degrades, even when using methods such as ``momentum
smearing''\cite{Bali:2016lva}, designed to enhance the contribution of the lowest-lying state in correlation functions at nonzero
three-momentum, thereby increasing the number of measurements that need to be made.
Last, the excited-state contributions themselves become more significant at higher momenta both through the greater number of
contributing states arising from the reduced symmetries at nonzero momentum, and through the relative compression of the energy
spectrum.
This requires either calculations at many source-sink separations, or the use of the variational method with an expanded basis of
operators.
Thus calculations of the precision that the neutrino program demands requires a significant commitment of computational resources.

\subsection{Axial currents in light nuclei}
\label{sec:small-gA}

An important challenge for lattice QCD is to extend the calculations of the axial properties of the nucleon to the more complex
systems of nuclei.
Just as for the nucleon, knowing nuclear matrix elements of axial currents and of quantities relevant in deep-inelastic scattering
on nuclei are high priorities in the lattice-QCD community.
Over the last decade, first studies of a range of nuclear properties have been performed, and calculations of the requisite axial
structure of light nuclei are eminently feasible in a five-year timeframe.

Nuclear effects in neutrino-nucleus scattering are important, extremely nontrivial, and not simply related to those in
electron-nucleus scattering.
For example, Gamow-Teller transitions in nuclei \cite{Buck:1975ae,Krofcheck:1985fg,Chou:1993zz}, which flip spin and isospin of the
nucleus, are poorly described in most simple nuclear models, with deviations of as much as $25\%$ from naive expectations based on
simply scaling from the single-particle $n\to p e \bar{\nu}$ transition.
Once sophisticated nuclear wave functions and many-body axial currents~\cite{Baroni:2015,Krebs:2016rqz} are used, however, agreement
with the data is reached~\cite{Pastore:2017uwc}.
At higher energies, state-of-the-art Green-function Monte-Carlo (an exact many-body method) calculations
\cite{Lovato:2015qka,Lovato:2017cux} show that neutrino response functions describing scattering on nuclei such as $^{12}$C have
effects from two-body currents at the 20\% level, particularly in the transverse response.

In the last few years, USQCD collaboration members have performed lattice-QCD calculations of $A=2,3$ axial transitions in the
forward limit using unphysically heavy quark masses corresponding to $M_\pi\sim800$~MeV
\cite{Savage:2016kon,Chang:2017eiq,Tiburzi:2017iux}.
This work, in which the Gamow-Teller contribution to tritium $\beta$-decay and the rate of the $pp\to d e \bar{\nu}$ fusion
process were extracted, demonstrates that the calculations relevant to constrain neutrino interactions with light nuclei can be
performed.
While nuclear effects in the axial matrix elements of two- and three-body systems are found to be at the few percent level, the
lattice-QCD calculations were able to resolve the relevant effects by isolating the intrinsically two-body contributions.
Pursuing these calculations at the physical quark masses and controlling all sources of systematic uncertainties in them will
require exascale computing resources.
Beyond these forward limit calculations, extensions of this approach to enable calculations of the form factors of light nuclei from
nonforward transition matrix elements are underway.
Calculations involving multiparticle final states are also necessary but challenging: a theoretical understanding of the simplest
inelastic channels is presented in Ref.~\cite{Briceno:2015tza}.
For high-energy neutrino-nucleus scattering in the deep-inelastic regime, lattice-QCD calculations of the moments of the relevant
parton distributions in nuclei will be useful in constraining nuclear effects in a very different kinematic regime.

While these calculations do not directly address the particular nuclear targets for DUNE and other neutrino-scattering experiments,
they are useful in constraining the low-energy constants and meson-exchange currents that enter nuclear-chiral-EFT axial
currents~\cite{Baroni:2015,Krebs:2016rqz}.
Such matchings are already underway for spectroscopy \cite{Barnea:2013uqa,Bansal:2017pwn,Contessi:2017rww} and electromagnetic
interactions \cite{Kirscher:2017fqc} at unphysical values of the quark masses, and the machinery necessary to undertake this at the
physical quark masses is being developed.
As well as studies of currents, lattice-QCD calculations of nuclei and other systems such as three and four neutron systems will
provide input into the three-body forces in nuclear EFTs, particularly those aspects of these forces that are challenging to access
experimentally.

\section{Extremely challenging calculations}
\label{sec:hardest}

Looking further into the future, we can foresee the need for calculations that go far enough beyond the current state of the art 
that it is hard to know when or whether they will be possible.

\subsection{Electromagnetic and isospin-breaking effects}

Going beyond the leading order, calculations of nucleon matrix elements must incorporated the neglected contributions from both QED and
from strong isospin breaking (SIB).
There are two possible approaches for completing these tasks, which can be generally classified as being perturbative and
nonperturbative.
Perturbative calculations make use of existing isospin-symmetric and QCD-only lattice ensembles to compute the desired matrix
elements.
The QED effects are computed with explicit vector current insertions and some scheme for the virtual photon lines.
Similarly, the SIB effects are included via scalar current insertions that allow for expansion in quark mass about the isospin-symmetric
point.
For nucleon matrix elements, perturbative calculations require five-point correlation functions as well as disconnected diagrams and
and pose a significant challenge to pursue.

Nonperturbative calculations make use of gauge ensembles that include explicit sea effects for QED and SIB.
These are $\text{SU}(3)\times\text{U}(1)$ gauge field ensembles with the up and down quark masses tuned to their physical values.
Rather than being restricted to a perturbative expansion of photon fields, these calculations include combined gluon, photon, and
quark loops to all orders.
While these calculations are likely to be cheaper than the perturbative calculations mentioned above, they also have technical
difficulties that must be overcome.
The most challenging of these difficulties is likely to come from finite size effects.
Since photons mediate a long range force, calculations including QED will be sensitive to the size of the lattice.
Many computations on lattice ensembles with different volumes will be necessary to quantify and remove this systematic effect.

\subsection{Axial currents in heavier nuclei}

A holy grail for neutrino-nucleus scattering is controlled QCD calculations of the axial form factors, resonance transition
form factors, and nuclear PDFs for ${}^{40}$Ar, the target nucleus in DUNE and several other experiments.
As yet, exponentially hard challenges must be overcome in order for meaningful lattice-QCD calculations of such heavy nuclei.
Both the factorial growth complexity of many-body contractions and the exponentially degradation of the signal-to-noise ratio are
currently impede to progress on this front.
Since lattice-QCD studies of nuclei are relatively new, it is not unlikely that new algorithms will definitively alter this picture
(algorithms involving machine learning and quantum computation \cite{Roggero:2018hrn} may dramatically improve the situation),%
\footnote{For discussion of these novel approaches, see the companion whitepaper ``Status and Future Perspectives for Lattice Gauge
Theory Calculations to the Exascale and Beyond'' \cite{Joo:2018qcd}.} %
but at present it is realistic to assume that direct lattice-QCD calculations of argon will not occur in a timeframe relevant for
the coming long-baseline experiments.

Perhaps more realistically, significant tests of nuclear EFT frameworks beyond the few-body sector would be enabled by lattice-QCD
calculations of the spectrum and axial structure of an intermediate nucleus such as $^{12}$C.
Aspects of coherent scattering off nuclei will also be addressed by such calculations.
While still challenging, a number of groups are investigating ways to perform the relevant contractions and studying improved ways
to extract signals from noisy multi-baryon data through optimization methods \cite{Detmold:2014hla} or improved estimators
\cite{Beane:2014oea,Wagman:2017gqi,Wagman:2017xfh,Wagman:2016bam,Detmold:2018eqd}.
For carbon targets, experimental scattering data exists and comparison of lattice-QCD calculations with this will help understand
the systematics of the $A$ dependence of nuclear EFT approaches and assess the reliability of the extrapolations to argon.

For light nuclei, adapting the techniques discussed above to address the Bjorken-$x$ dependence on nuclear PDFs will become possible
as computing resources increase.
While challenging, and still in the development stage even for the nucleon, these PDF methods will help constrain the flavor and
spin dependence of nuclear PDFs that are important for high-energy $\nu A$ scattering.

\section{computing needs}
\label{sec:needs}

As we have seen in Secs.~\ref{sec:hard} and~\ref{sec:hardest}, many topics pertaining to lattice QCD for neutrino physics are still
exploratory.
In those cases, computing estimates are impossible because progress depends on innovation and flexible computing, rather than an 
industrial resource.
It is feasible and reasonable, however, to estimate the computing needs of the calculations discussed in Sec.~\ref{sec:easy}.
We do so here, focusing on the example of nucleon form factors.

The methodology for the calculation of the axial and the electromagnetic form factors is well established, and data with control
over statistical errors have been generated by several collaborations worldwide.
Table~\ref{tab:ff} list several of these efforts.
Unfortunately, only a few include several ensembles with strange sea quarks~\cite{Ottnad:2018fri,Rajan:2017lxk,Chang:2018uxx}.
Even those calculations should be pushed to smaller lattice spacing and (in one case) smaller up-down quark mass.
Furthermore, most calculations obtain mean-squared charge radii ($r_A$, $r_E$, and $r_M$) that are smaller than phenomenological
extractions, by about 30\%.
To diagnose where the difference lies, it is crucial to improve control over systematic uncertainties in lattice-QCD calculations 
to obtain a definitive result from~QCD.

Here we base the cost estimates for achieving a 10--15\% result on two ongoing efforts within USQCD.
A convenient starting point is the work of PNDME collaboration that has presented extensive results for the axial form
factors~\cite{Rajan:2017lxk} using the Wilson-clover formulation for the valence quarks on ensembles with 2+1+1 sea quarks with the
staggered formulation.
One way to avoid this ``mixed action'' approach is to use staggered valence quarks~\cite{Meyer:2016kwb}.
Based on current running on institutional clusters at BNL and Fermilab, we estimate 9M GPU-hours%
\footnote{For QCD codes, these 9M GPU hours correspond to 300M conventional (e.g., Intel Skylake) CPU core-hours.}
to carry out a calculation on
eleven ensembles, five at physical sea-quark mass, and six with $m_l=\half(m_u+m_d)=0.2m_s$, with lattice spacing as small as 
0.03~fm.

\begin{table}[b]
    \newcommand{\h}{\phantom{1}}
    \renewcommand{\c}{~\cite{Bazavov:2012xda}}
    \renewcommand{\d}{~\cite{Blum:2014tka}}
    \caption[tab:ff]{Sample of calculations of nucleon form factors going on worldwide.
        In the first column, ``2'', ``2+1'', and ``2+1+1'' all denote two equal-mass quarks for up and down;
        the latter two include strange and charm, respectively.
        The last column indicates work in which USQCD members participate.}
    \label{tab:ff}
    \begin{tabular}{l@{~~}l@{~~}r@{\hfil}ccllc}
    \hline\hline
    Sea quarks      & Valence quarks & $N_\text{ens}$ & $a$ (fm) & $M_\pi$ (MeV) & Collaboration & Ref. & USQCD \\
    \hline
    2 Wilson-clover & same as sea   &  11   &~0.06--0.08~& 150--490 & RQCD          & \cite{Bali:2014nma} &  \\
    2 TM~clover     & same as sea   & \h1   &    0.09    &   130    & ETM           & \cite{Alexandrou:2017hac} &  \\
    2 Wilson-clover & same as sea   &  11   & 0.05--0.08 & 190--470 & Mainz (CLS)   & \cite{Capitani:2017qpc} &  \\
    2+1 overlap     & same as sea   & \h4   &    0.11    & 290--540 & JLQCD         & \cite{Yamanaka:2018uud} &  \\
    2+1 domain wall\d & overlap     & \h3   & 0.08--0.15 & 170--340 & $\chi$QCD     & \cite{Liang:2018pis} & \checkmark \\
    2+1 Wilson-clover & same as sea & \h1   &    0.085   & 146, 135 & PACS          & \cite{Shintani:2018ozy} &  \\
    2+1 Wilson-clover & same as sea &  11   & 0.05--0.09 & 200--350 & Mainz (CLS)   & \cite{Ottnad:2018fri} &  \\
    2+1+1 HISQ\c    & Wilson-clover & \h8   & 0.06--0.12 & 135--210 & PNDME         & \cite{Rajan:2017lxk} & \checkmark \\
    2+1+1 HISQ\c    & domain wall   &  16   & 0.09--0.15 & 130--400 & CalLat        & \cite{Chang:2018uxx} & \checkmark \\
    2+1+1 TM~clover & same as sea   & \h3   & 0.09--0.15 &  140     & ETM           & \cite{Martha:2018lft} & \checkmark \\
    2+1+1 HISQ      & same as sea   & \h3   & 0.09--0.15 &  135     & Fermilab/MILC & \cite{Meyer:2016kwb} & \checkmark \\
    \hline\hline
    \end{tabular}
\end{table}

\pagebreak
Similarly, a significant subset of USQCD plans on generating a suite of ensembles with Wilson-clover sea quarks.
To generate eight such ensembles, with light sea-quark masses corresponding to pion masses of 170~MeV and 270~MeV (four ensembles 
each), with lattice spacing as small as 0.05~fm.
The estimate to finish generating these ensembles is 8M GPU-hours (assuming the GPUs on Summit at ORNL).
This chunk of computing will be shared with many other projects, particularly those described in the companion white paper 
``Hadrons and Nuclei'' \cite{Detmold:2018qcd}.
The computation of the needed nucleon correlation functions is estimated to require 15M GPU-hours.

These estimates set the scale for a modern calculation of the simplest quantity needed for neutrino physics.
At the same time, comparably demanding work with small nuclei, but not yet physical pion mass, will be needed.
Such work is necessary to understand the technical issues facing more realistic calculations and to find better methods and
algorithms.
Even assuming gains from innovation, it is hard to imagine that nuclear form factors will end up below 10M GPU-hours.
The same line of reasoning can be applied to other calculations discussed in Sec.~\ref{sec:hard}.

\acknowledgments We would like to thank Ra\'ul Brice\~no, Maxwell Hansen, Richard Hill, Ciaran Hughes, William Marciano, Saori
Pastore, Noemi Rocco, and Michael Wagman for useful input.
This material is based upon work supported by the U.S.\ Department of Energy, Office of Science, Office of Nuclear Physics under
contract No.\ DE-AC05-06OR23177.
Fermilab is operated by Fermi Research Alliance, LLC, under Contract No.\ DE-AC02-07CH11359 with the United States Department of
Energy, Office of Science, Office of High Energy Physics.

\bibliography{nu,usqcd-wp}

\begin{thebibliography}{200}%
\makeatletter
\providecommand \@ifxundefined [1]{%
 \@ifx{#1\undefined}
}%
\providecommand \@ifnum [1]{%
 \ifnum #1\expandafter \@firstoftwo
 \else \expandafter \@secondoftwo
 \fi
}%
\providecommand \@ifx [1]{%
 \ifx #1\expandafter \@firstoftwo
 \else \expandafter \@secondoftwo
 \fi
}%
\providecommand \natexlab [1]{#1}%
\providecommand \enquote  [1]{``#1''}%
\providecommand \bibnamefont  [1]{#1}%
\providecommand \bibfnamefont [1]{#1}%
\providecommand \citenamefont [1]{#1}%
\providecommand \href@noop [0]{\@secondoftwo}%
\providecommand \href [0]{\begingroup \@sanitize@url \@href}%
\providecommand \@href[1]{\@@startlink{#1}\@@href}%
\providecommand \@@href[1]{\endgroup#1\@@endlink}%
\providecommand \@sanitize@url [0]{\catcode `\\12\catcode `\$12\catcode
  `\&12\catcode `\#12\catcode `\^12\catcode `\_12\catcode `\%12\relax}%
\providecommand \@@startlink[1]{}%
\providecommand \@@endlink[0]{}%
\providecommand \url  [0]{\begingroup\@sanitize@url \@url }%
\providecommand \@url [1]{\endgroup\@href {#1}{\urlprefix }}%
\providecommand \urlprefix  [0]{URL }%
\providecommand \Eprint [0]{\href }%
\providecommand \doibase [0]{http://dx.doi.org/}%
\providecommand \selectlanguage [0]{\@gobble}%
\providecommand \bibinfo  [0]{\@secondoftwo}%
\providecommand \bibfield  [0]{\@secondoftwo}%
\providecommand \translation [1]{[#1]}%
\providecommand \BibitemOpen [0]{}%
\providecommand \bibitemStop [0]{}%
\providecommand \bibitemNoStop [0]{.\EOS\space}%
\providecommand \EOS [0]{\spacefactor3000\relax}%
\providecommand \BibitemShut  [1]{\csname bibitem#1\endcsname}%
\let\auto@bib@innerbib\@empty
\bibitem [{\citenamefont {Bazavov}\ \emph {et~al.}(2019)\citenamefont
  {Bazavov}, \citenamefont {Karsch}, \citenamefont {Mukherjee},\ and\
  \citenamefont {Petreczky}}]{Bazavov:2018qcd}%
  \BibitemOpen
  \bibfield  {author} {\bibinfo {author} {\bibfnamefont {Alexei}\ \bibnamefont
  {Bazavov}}, \bibinfo {author} {\bibfnamefont {Frithjof}\ \bibnamefont
  {Karsch}}, \bibinfo {author} {\bibfnamefont {Swagato}\ \bibnamefont
  {Mukherjee}}, \ and\ \bibinfo {author} {\bibfnamefont {Peter}\ \bibnamefont
  {Petreczky}} (\bibinfo {collaboration} {USQCD}),\ }\bibfield  {title}
  {\enquote {\bibinfo {title} {Hot-dense lattice {QCD}},}\ }\href@noop {} {\
  (\bibinfo {year} {2019})}\BibitemShut {NoStop}%
\bibitem [{\citenamefont {Brower}\ \emph {et~al.}(2019)\citenamefont {Brower},
  \citenamefont {Hasenfratz}, \citenamefont {Neil} \emph
  {et~al.}}]{Brower:2018qcd}%
  \BibitemOpen
  \bibfield  {author} {\bibinfo {author} {\bibfnamefont {Richard}\ \bibnamefont
  {Brower}}, \bibinfo {author} {\bibfnamefont {Anna}\ \bibnamefont
  {Hasenfratz}}, \bibinfo {author} {\bibfnamefont {Ethan~T.}\ \bibnamefont
  {Neil}},  \emph {et~al.} (\bibinfo {collaboration} {USQCD}),\ }\bibfield
  {title} {\enquote {\bibinfo {title} {Lattice gauge theory for physics beyond
  the {Standard Model}},}\ }\href@noop {} {\  (\bibinfo {year}
  {2019})}\BibitemShut {NoStop}%
\bibitem [{\citenamefont {Cirigliano}\ \emph {et~al.}(2019)\citenamefont
  {Cirigliano}, \citenamefont {Davoudi} \emph {et~al.}}]{Davoudi:2018qcd}%
  \BibitemOpen
  \bibfield  {author} {\bibinfo {author} {\bibfnamefont {Vincenzo}\
  \bibnamefont {Cirigliano}}, \bibinfo {author} {\bibfnamefont {Zohreh}\
  \bibnamefont {Davoudi}},  \emph {et~al.} (\bibinfo {collaboration} {USQCD}),\
  }\bibfield  {title} {\enquote {\bibinfo {title} {The role of lattice {QCD} in
  searches for violations of fundamental symmetries and signals for new
  physics},}\ }\href@noop {} {\  (\bibinfo {year} {2019})}\BibitemShut
  {NoStop}%
\bibitem [{\citenamefont {Detmold}\ \emph {et~al.}(2019)\citenamefont
  {Detmold}, \citenamefont {Edwards} \emph {et~al.}}]{Detmold:2018qcd}%
  \BibitemOpen
  \bibfield  {author} {\bibinfo {author} {\bibfnamefont {William}\ \bibnamefont
  {Detmold}}, \bibinfo {author} {\bibfnamefont {Robert~G.}\ \bibnamefont
  {Edwards}},  \emph {et~al.} (\bibinfo {collaboration} {USQCD}),\ }\bibfield
  {title} {\enquote {\bibinfo {title} {Hadrons and nuclei},}\ }\href@noop {} {\
   (\bibinfo {year} {2019})}\BibitemShut {NoStop}%
\bibitem [{\citenamefont {{Jo\'o}}\ \emph {et~al.}(2019)\citenamefont
  {{Jo\'o}}, \citenamefont {Jung} \emph {et~al.}}]{Joo:2018qcd}%
  \BibitemOpen
  \bibfield  {author} {\bibinfo {author} {\bibfnamefont {{B\'alint}}\
  \bibnamefont {{Jo\'o}}}, \bibinfo {author} {\bibfnamefont {Chulwoo}\
  \bibnamefont {Jung}},  \emph {et~al.} (\bibinfo {collaboration} {USQCD}),\
  }\bibfield  {title} {\enquote {\bibinfo {title} {Status and future
  perspectives for lattice gauge theory calculations to the exascale and
  beyond},}\ }\href@noop {} {\  (\bibinfo {year} {2019})}\BibitemShut {NoStop}%
\bibitem [{\citenamefont {Lehner}\ \emph {et~al.}(2019)\citenamefont {Lehner},
  \citenamefont {Meinel} \emph {et~al.}}]{Lehner:2018qcd}%
  \BibitemOpen
  \bibfield  {author} {\bibinfo {author} {\bibfnamefont {Christoph}\
  \bibnamefont {Lehner}}, \bibinfo {author} {\bibfnamefont {Stefan}\
  \bibnamefont {Meinel}},  \emph {et~al.} (\bibinfo {collaboration} {USQCD}),\
  }\bibfield  {title} {\enquote {\bibinfo {title} {Opportunities for lattice
  {QCD} in quark and lepton flavor physics},}\ }\href@noop {} {\  (\bibinfo
  {year} {2019})}\BibitemShut {NoStop}%
\bibitem [{\citenamefont {Fukuda}\ \emph {et~al.}(1998)\citenamefont {Fukuda}
  \emph {et~al.}}]{Fukuda:1998mi}%
  \BibitemOpen
  \bibfield  {author} {\bibinfo {author} {\bibfnamefont {Y.}~\bibnamefont
  {Fukuda}} \emph {et~al.} (\bibinfo {collaboration} {Super-Kamiokande}),\
  }\bibfield  {title} {\enquote {\bibinfo {title} {{Evidence for oscillation of
  atmospheric neutrinos}},}\ }\href {\doibase 10.1103/PhysRevLett.81.1562}
  {\bibfield  {journal} {\bibinfo  {journal} {Phys. Rev. Lett.}\ }\textbf
  {\bibinfo {volume} {81}},\ \bibinfo {pages} {1562--1567} (\bibinfo {year}
  {1998})},\ \Eprint {http://arxiv.org/abs/hep-ex/9807003}
  {arXiv:hep-ex/9807003 [hep-ex]} \BibitemShut {NoStop}%
\bibitem [{\citenamefont {Ahmad}\ \emph {et~al.}(2001)\citenamefont {Ahmad}
  \emph {et~al.}}]{Ahmad:2001an}%
  \BibitemOpen
  \bibfield  {author} {\bibinfo {author} {\bibfnamefont {Q.~R.}\ \bibnamefont
  {Ahmad}} \emph {et~al.} (\bibinfo {collaboration} {SNO}),\ }\bibfield
  {title} {\enquote {\bibinfo {title} {Measurement of the rate of $\nu_e+d \to
  p+p+e^-$ interactions produced by {$^8$B} solar neutrinos at the {Sudbury
  Neutrino Observatory}},}\ }\href {\doibase 10.1103/PhysRevLett.87.071301}
  {\bibfield  {journal} {\bibinfo  {journal} {Phys. Rev. Lett.}\ }\textbf
  {\bibinfo {volume} {87}},\ \bibinfo {pages} {071301} (\bibinfo {year}
  {2001})},\ \Eprint {http://arxiv.org/abs/nucl-ex/0106015}
  {arXiv:nucl-ex/0106015 [nucl-ex]} \BibitemShut {NoStop}%
\bibitem [{\citenamefont {Acciarri}\ \emph {et~al.}(2015)\citenamefont
  {Acciarri} \emph {et~al.}}]{Acciarri:2015uup}%
  \BibitemOpen
  \bibfield  {author} {\bibinfo {author} {\bibfnamefont {R.}~\bibnamefont
  {Acciarri}} \emph {et~al.} (\bibinfo {collaboration} {DUNE}),\ }\bibfield
  {title} {\enquote {\bibinfo {title} {{Long-Baseline Neutrino Facility (LBNF)
  and Deep Underground Neutrino Experiment (DUNE)}},}\ }\href@noop {} {\
  (\bibinfo {year} {2015})},\ \Eprint {http://arxiv.org/abs/1512.06148}
  {arXiv:1512.06148 [physics.ins-det]} \BibitemShut {NoStop}%
\bibitem [{\citenamefont {Abe}\ \emph {et~al.}(2018)\citenamefont {Abe} \emph
  {et~al.}}]{Abe:2018uyc}%
  \BibitemOpen
  \bibfield  {author} {\bibinfo {author} {\bibfnamefont {K.}~\bibnamefont
  {Abe}} \emph {et~al.} (\bibinfo {collaboration} {Hyper-Kamiokande}),\
  }\bibfield  {title} {\enquote {\bibinfo {title} {{Hyper-Kamiokande} design
  report},}\ }\href@noop {} {\  (\bibinfo {year} {2018})},\ \Eprint
  {http://arxiv.org/abs/1805.04163} {arXiv:1805.04163 [physics.ins-det]}
  \BibitemShut {NoStop}%
\bibitem [{\citenamefont {Majorana}(1937)}]{Majorana:1937vz}%
  \BibitemOpen
  \bibfield  {author} {\bibinfo {author} {\bibfnamefont {Ettore}\ \bibnamefont
  {Majorana}},\ }\bibfield  {title} {\enquote {\bibinfo {title} {Teoria
  simmetrica dell’elettrone e del positrone},}\ }\href {\doibase
  10.1007/BF02961314} {\bibfield  {journal} {\bibinfo  {journal} {Nuovo Cim.}\
  }\textbf {\bibinfo {volume} {14}},\ \bibinfo {pages} {171--184} (\bibinfo
  {year} {1937})}\BibitemShut {NoStop}%
\bibitem [{\citenamefont {Minkowski}(1977)}]{Minkowski:1977sc}%
  \BibitemOpen
  \bibfield  {author} {\bibinfo {author} {\bibfnamefont {Peter}\ \bibnamefont
  {Minkowski}},\ }\bibfield  {title} {\enquote {\bibinfo {title} {$\mu \to
  e\gamma$ at a rate of one out of $10^{9}$ muon decays?}}\ }\href {\doibase
  10.1016/0370-2693(77)90435-X} {\bibfield  {journal} {\bibinfo  {journal}
  {Phys. Lett.}\ }\textbf {\bibinfo {volume} {67B}},\ \bibinfo {pages}
  {421--428} (\bibinfo {year} {1977})}\BibitemShut {NoStop}%
\bibitem [{\citenamefont {Yanagida}(1980)}]{Yanagida:1980xy}%
  \BibitemOpen
  \bibfield  {author} {\bibinfo {author} {\bibfnamefont {Tsutomu}\ \bibnamefont
  {Yanagida}},\ }\bibfield  {title} {\enquote {\bibinfo {title} {Horizontal
  symmetry and masses of neutrinos},}\ }\href {\doibase 10.1143/PTP.64.1103}
  {\bibfield  {journal} {\bibinfo  {journal} {Prog. Theor. Phys.}\ }\textbf
  {\bibinfo {volume} {64}},\ \bibinfo {pages} {1103} (\bibinfo {year}
  {1980})}\BibitemShut {NoStop}%
\bibitem [{\citenamefont {Gell-Mann}\ \emph {et~al.}(1979)\citenamefont
  {Gell-Mann}, \citenamefont {Ramond},\ and\ \citenamefont
  {Slansky}}]{GellMann:1980vs}%
  \BibitemOpen
  \bibfield  {author} {\bibinfo {author} {\bibfnamefont {Murray}\ \bibnamefont
  {Gell-Mann}}, \bibinfo {author} {\bibfnamefont {Pierre}\ \bibnamefont
  {Ramond}}, \ and\ \bibinfo {author} {\bibfnamefont {Richard}\ \bibnamefont
  {Slansky}},\ }\enquote {\bibinfo {title} {Complex spinors and unified
  theories},}\ in\ \href@noop {} {\emph {\bibinfo {booktitle}
  {Supergravity}}},\ \bibinfo {editor} {edited by\ \bibinfo {editor}
  {\bibfnamefont {P.}~\bibnamefont {{van Nieuwenhuizen}}}}\ (\bibinfo
  {publisher} {North-Holland},\ \bibinfo {address} {Amsterdam},\ \bibinfo
  {year} {1979})\ pp.\ \bibinfo {pages} {315--321},\ \Eprint
  {http://arxiv.org/abs/1306.4669} {arXiv:1306.4669 [hep-th]} \BibitemShut
  {NoStop}%
\bibitem [{\citenamefont {Pontecorvo}(1957)}]{Pontecorvo:1957cp}%
  \BibitemOpen
  \bibfield  {author} {\bibinfo {author} {\bibfnamefont {B.}~\bibnamefont
  {Pontecorvo}},\ }\bibfield  {title} {\enquote {\bibinfo {title} {Mesonium and
  antimesonium},}\ }\href@noop {} {\bibfield  {journal} {\bibinfo  {journal}
  {Sov. Phys. JETP}\ }\textbf {\bibinfo {volume} {6}},\ \bibinfo {pages} {429}
  (\bibinfo {year} {1957})},\ \bibinfo {note}
  {[\href{http://www.jetp.ac.ru/cgi-bin/e/index/e/6/2/p429?a=list}{Zh. Eksp.
  Teor. Fiz.~\textbf{33}, 549 (1957)}]}\BibitemShut {NoStop}%
\bibitem [{\citenamefont {Pontecorvo}(1968)}]{Pontecorvo:1967fh}%
  \BibitemOpen
  \bibfield  {author} {\bibinfo {author} {\bibfnamefont {B.}~\bibnamefont
  {Pontecorvo}},\ }\bibfield  {title} {\enquote {\bibinfo {title} {Neutrino
  experiments and the problem of conservation of leptonic charge},}\
  }\href@noop {} {\bibfield  {journal} {\bibinfo  {journal} {Sov. Phys. JETP}\
  }\textbf {\bibinfo {volume} {26}},\ \bibinfo {pages} {984--988} (\bibinfo
  {year} {1968})},\ \bibinfo {note}
  {[\href{http://www.jetp.ac.ru/cgi-bin/e/index/e/26/5/p984?a=list}{Zh. Eksp.
  Teor. Fiz.~\textbf{53}, 1717 (1967)}]}\BibitemShut {NoStop}%
\bibitem [{\citenamefont {Maki}\ \emph {et~al.}(1962)\citenamefont {Maki},
  \citenamefont {Nakagawa},\ and\ \citenamefont {Sakata}}]{Maki:1962mu}%
  \BibitemOpen
  \bibfield  {author} {\bibinfo {author} {\bibfnamefont {Ziro}\ \bibnamefont
  {Maki}}, \bibinfo {author} {\bibfnamefont {Masami}\ \bibnamefont {Nakagawa}},
  \ and\ \bibinfo {author} {\bibfnamefont {Shoichi}\ \bibnamefont {Sakata}},\
  }\bibfield  {title} {\enquote {\bibinfo {title} {{Remarks on the unified
  model of elementary particles}},}\ }\href {\doibase 10.1143/PTP.28.870}
  {\bibfield  {journal} {\bibinfo  {journal} {Prog. Theor. Phys.}\ }\textbf
  {\bibinfo {volume} {28}},\ \bibinfo {pages} {870--880} (\bibinfo {year}
  {1962})}\BibitemShut {NoStop}%
\bibitem [{\citenamefont {Cabibbo}(1963)}]{Cabibbo:1963yz}%
  \BibitemOpen
  \bibfield  {author} {\bibinfo {author} {\bibfnamefont {Nicola}\ \bibnamefont
  {Cabibbo}},\ }\bibfield  {title} {\enquote {\bibinfo {title} {Unitary
  symmetry and leptonic decays},}\ }\href {\doibase 10.1103/PhysRevLett.10.531}
  {\bibfield  {journal} {\bibinfo  {journal} {Phys. Rev. Lett.}\ }\textbf
  {\bibinfo {volume} {10}},\ \bibinfo {pages} {531--533} (\bibinfo {year}
  {1963})}\BibitemShut {NoStop}%
\bibitem [{\citenamefont {Kobayashi}\ and\ \citenamefont
  {Maskawa}(1973)}]{Kobayashi:1973fv}%
  \BibitemOpen
  \bibfield  {author} {\bibinfo {author} {\bibfnamefont {Makoto}\ \bibnamefont
  {Kobayashi}}\ and\ \bibinfo {author} {\bibfnamefont {Toshihide}\ \bibnamefont
  {Maskawa}},\ }\bibfield  {title} {\enquote {\bibinfo {title} {{$CP$}
  violation in the renormalizable theory of weak interaction},}\ }\href
  {\doibase 10.1143/PTP.49.652} {\bibfield  {journal} {\bibinfo  {journal}
  {Prog. Theor. Phys.}\ }\textbf {\bibinfo {volume} {49}},\ \bibinfo {pages}
  {652--657} (\bibinfo {year} {1973})}\BibitemShut {NoStop}%
\bibitem [{\citenamefont {Fields}(2018)}]{Fields:2018enu}%
  \BibitemOpen
  \bibfield  {author} {\bibinfo {author} {\bibfnamefont {Laura}\ \bibnamefont
  {Fields}},\ }\href@noop {} {}\bibinfo {howpublished} {private communication}
  (\bibinfo {year} {2018})\BibitemShut {NoStop}%
\bibitem [{\citenamefont {Alvarez-Ruso}\ \emph {et~al.}(2018)\citenamefont
  {Alvarez-Ruso} \emph {et~al.}}]{Alvarez-Ruso:2017oui}%
  \BibitemOpen
  \bibfield  {author} {\bibinfo {author} {\bibfnamefont {L.}~\bibnamefont
  {Alvarez-Ruso}} \emph {et~al.} (\bibinfo {collaboration} {NuSTEC}),\
  }\bibfield  {title} {\enquote {\bibinfo {title} {{NuSTEC} white paper: Status
  and challenges of neutrino-nucleus scattering},}\ }\href@noop {} {\bibfield
  {journal} {\bibinfo  {journal} {Prog. Part. Nucl. Phys.}\ }\textbf {\bibinfo
  {volume} {100}},\ \bibinfo {pages} {1--68} (\bibinfo {year} {2018})},\
  \Eprint {http://arxiv.org/abs/1706.03621} {arXiv:1706.03621 [hep-ph]}
  \BibitemShut {NoStop}%
\bibitem [{\citenamefont {Benhar}\ and\ \citenamefont
  {Lovato}(2015)}]{Benhar:2015xga}%
  \BibitemOpen
  \bibfield  {author} {\bibinfo {author} {\bibfnamefont {Omar}\ \bibnamefont
  {Benhar}}\ and\ \bibinfo {author} {\bibfnamefont {Alessandro}\ \bibnamefont
  {Lovato}},\ }\bibfield  {title} {\enquote {\bibinfo {title} {Towards a
  unified description of the electroweak nuclear response},}\ }\href {\doibase
  10.1142/S0218301315300064} {\bibfield  {journal} {\bibinfo  {journal} {Int.
  J. Mod. Phys.}\ }\textbf {\bibinfo {volume} {E24}},\ \bibinfo {pages}
  {1530006} (\bibinfo {year} {2015})},\ \Eprint
  {http://arxiv.org/abs/1506.05225} {arXiv:1506.05225 [nucl-th]} \BibitemShut
  {NoStop}%
\bibitem [{\citenamefont {Carlson}\ \emph {et~al.}(2015)\citenamefont
  {Carlson}, \citenamefont {Gandolfi}, \citenamefont {Pederiva}, \citenamefont
  {Pieper}, \citenamefont {Schiavilla}, \citenamefont {Schmidt},\ and\
  \citenamefont {Wiringa}}]{Carlson:2014vla}%
  \BibitemOpen
  \bibfield  {author} {\bibinfo {author} {\bibfnamefont {J.}~\bibnamefont
  {Carlson}}, \bibinfo {author} {\bibfnamefont {S.}~\bibnamefont {Gandolfi}},
  \bibinfo {author} {\bibfnamefont {F.}~\bibnamefont {Pederiva}}, \bibinfo
  {author} {\bibfnamefont {Steven~C.}\ \bibnamefont {Pieper}}, \bibinfo
  {author} {\bibfnamefont {R.}~\bibnamefont {Schiavilla}}, \bibinfo {author}
  {\bibfnamefont {K.~E.}\ \bibnamefont {Schmidt}}, \ and\ \bibinfo {author}
  {\bibfnamefont {R.~B.}\ \bibnamefont {Wiringa}},\ }\bibfield  {title}
  {\enquote {\bibinfo {title} {Quantum {Monte Carlo} methods for nuclear
  physics},}\ }\href {\doibase 10.1103/RevModPhys.87.1067} {\bibfield
  {journal} {\bibinfo  {journal} {Rev. Mod. Phys.}\ }\textbf {\bibinfo {volume}
  {87}},\ \bibinfo {pages} {1067} (\bibinfo {year} {2015})},\ \bibinfo {note}
  {and references within},\ \Eprint {http://arxiv.org/abs/1412.3081}
  {arXiv:1412.3081 [nucl-th]} \BibitemShut {NoStop}%
\bibitem [{\citenamefont {Epelbaum}\ \emph {et~al.}(2009)\citenamefont
  {Epelbaum}, \citenamefont {Hammer},\ and\ \citenamefont
  {Meissner}}]{Epelbaum:2008ga}%
  \BibitemOpen
  \bibfield  {author} {\bibinfo {author} {\bibfnamefont {Evgeny}\ \bibnamefont
  {Epelbaum}}, \bibinfo {author} {\bibfnamefont {Hans-Werner}\ \bibnamefont
  {Hammer}}, \ and\ \bibinfo {author} {\bibfnamefont {Ulf-G.}\ \bibnamefont
  {Meissner}},\ }\bibfield  {title} {\enquote {\bibinfo {title} {Modern theory
  of nuclear forces},}\ }\href {\doibase 10.1103/RevModPhys.81.1773} {\bibfield
   {journal} {\bibinfo  {journal} {Rev. Mod. Phys.}\ }\textbf {\bibinfo
  {volume} {81}},\ \bibinfo {pages} {1773--1825} (\bibinfo {year} {2009})},\
  \Eprint {http://arxiv.org/abs/0811.1338} {arXiv:0811.1338 [nucl-th]}
  \BibitemShut {NoStop}%
\bibitem [{\citenamefont {Machleidt}\ and\ \citenamefont
  {Entem}(2011)}]{Machleidt:2011zz}%
  \BibitemOpen
  \bibfield  {author} {\bibinfo {author} {\bibfnamefont {R.}~\bibnamefont
  {Machleidt}}\ and\ \bibinfo {author} {\bibfnamefont {D.~R.}\ \bibnamefont
  {Entem}},\ }\bibfield  {title} {\enquote {\bibinfo {title} {{Chiral effective
  field theory and nuclear forces}},}\ }\href {\doibase
  10.1016/j.physrep.2011.02.001} {\bibfield  {journal} {\bibinfo  {journal}
  {Phys. Rept.}\ }\textbf {\bibinfo {volume} {503}},\ \bibinfo {pages} {1--75}
  (\bibinfo {year} {2011})},\ \Eprint {http://arxiv.org/abs/1105.2919}
  {arXiv:1105.2919 [nucl-th]} \BibitemShut {NoStop}%
\bibitem [{\citenamefont {Weinberg}(1990)}]{Weinberg:1990rz}%
  \BibitemOpen
  \bibfield  {author} {\bibinfo {author} {\bibfnamefont {Steven}\ \bibnamefont
  {Weinberg}},\ }\bibfield  {title} {\enquote {\bibinfo {title} {Nuclear forces
  from chiral {Lagrangians}},}\ }\href {\doibase 10.1016/0370-2693(90)90938-3}
  {\bibfield  {journal} {\bibinfo  {journal} {Phys. Lett.}\ }\textbf {\bibinfo
  {volume} {B251}},\ \bibinfo {pages} {288--292} (\bibinfo {year}
  {1990})}\BibitemShut {NoStop}%
\bibitem [{\citenamefont {van Kolck}(1994)}]{vanKolck:1994yi}%
  \BibitemOpen
  \bibfield  {author} {\bibinfo {author} {\bibfnamefont {U.}~\bibnamefont {van
  Kolck}},\ }\bibfield  {title} {\enquote {\bibinfo {title} {Few nucleon forces
  from chiral {Lagrangians}},}\ }\href {\doibase 10.1103/PhysRevC.49.2932}
  {\bibfield  {journal} {\bibinfo  {journal} {Phys. Rev.}\ }\textbf {\bibinfo
  {volume} {C49}},\ \bibinfo {pages} {2932--2941} (\bibinfo {year}
  {1994})}\BibitemShut {NoStop}%
\bibitem [{\citenamefont {Kaplan}\ \emph {et~al.}(1998)\citenamefont {Kaplan},
  \citenamefont {Savage},\ and\ \citenamefont {Wise}}]{Kaplan:1998tg}%
  \BibitemOpen
  \bibfield  {author} {\bibinfo {author} {\bibfnamefont {David~B.}\
  \bibnamefont {Kaplan}}, \bibinfo {author} {\bibfnamefont {Martin~J.}\
  \bibnamefont {Savage}}, \ and\ \bibinfo {author} {\bibfnamefont {Mark~B.}\
  \bibnamefont {Wise}},\ }\bibfield  {title} {\enquote {\bibinfo {title} {A new
  expansion for nucleon-nucleon interactions},}\ }\href {\doibase
  10.1016/S0370-2693(98)00210-X} {\bibfield  {journal} {\bibinfo  {journal}
  {Phys. Lett.}\ }\textbf {\bibinfo {volume} {B424}},\ \bibinfo {pages}
  {390--396} (\bibinfo {year} {1998})},\ \Eprint
  {http://arxiv.org/abs/nucl-th/9801034} {arXiv:nucl-th/9801034 [nucl-th]}
  \BibitemShut {NoStop}%
\bibitem [{\citenamefont {Meißner}(2016)}]{Meissner:2015wva}%
  \BibitemOpen
  \bibfield  {author} {\bibinfo {author} {\bibfnamefont {Ulf-G}\ \bibnamefont
  {Meißner}},\ }\bibfield  {title} {\enquote {\bibinfo {title} {{The long and
  winding road from chiral effective Lagrangians to nuclear structure}},}\
  }\href {\doibase 10.1088/0031-8949/91/3/033005} {\bibfield  {journal}
  {\bibinfo  {journal} {Phys. Scripta}\ }\textbf {\bibinfo {volume} {91}},\
  \bibinfo {pages} {033005} (\bibinfo {year} {2016})},\ \Eprint
  {http://arxiv.org/abs/1510.03230} {arXiv:1510.03230 [nucl-th]} \BibitemShut
  {NoStop}%
\bibitem [{\citenamefont {Juszczak}\ \emph {et~al.}(2006)\citenamefont
  {Juszczak}, \citenamefont {Nowak},\ and\ \citenamefont
  {Sobczyk}}]{Juszczak:2005zs}%
  \BibitemOpen
  \bibfield  {author} {\bibinfo {author} {\bibfnamefont {Cezary}\ \bibnamefont
  {Juszczak}}, \bibinfo {author} {\bibfnamefont {Jaroslaw~A.}\ \bibnamefont
  {Nowak}}, \ and\ \bibinfo {author} {\bibfnamefont {Jan~T.}\ \bibnamefont
  {Sobczyk}},\ }\bibfield  {title} {\enquote {\bibinfo {title} {{Simulations
  from a new neutrino event generator}},}\ }\href {\doibase
  10.1016/j.nuclphysbps.2006.08.069} {\bibfield  {journal} {\bibinfo  {journal}
  {Nucl. Phys. Proc. Suppl.}\ }\textbf {\bibinfo {volume} {159}},\ \bibinfo
  {pages} {211--216} (\bibinfo {year} {2006})},\ \Eprint
  {http://arxiv.org/abs/hep-ph/0512365} {arXiv:hep-ph/0512365 [hep-ph]}
  \BibitemShut {NoStop}%
\bibitem [{\citenamefont {Żmuda}\ \emph {et~al.}(2015)\citenamefont {Żmuda},
  \citenamefont {Graczyk}, \citenamefont {Juszczak},\ and\ \citenamefont
  {Sobczyk}}]{Zmuda:2015twa}%
  \BibitemOpen
  \bibfield  {author} {\bibinfo {author} {\bibfnamefont {Jakub}\ \bibnamefont
  {Żmuda}}, \bibinfo {author} {\bibfnamefont {Krzysztof~M.}\ \bibnamefont
  {Graczyk}}, \bibinfo {author} {\bibfnamefont {Cezary}\ \bibnamefont
  {Juszczak}}, \ and\ \bibinfo {author} {\bibfnamefont {Jan~T.}\ \bibnamefont
  {Sobczyk}},\ }\bibfield  {title} {\enquote {\bibinfo {title} {{NuWro Monte
  Carlo generator of neutrino interactions---first electron scattering
  results}},}\ }\href {\doibase 10.5506/APhysPolB.46.2329} {\bibfield
  {journal} {\bibinfo  {journal} {Acta Phys. Polon.}\ }\textbf {\bibinfo
  {volume} {B46}},\ \bibinfo {pages} {2329} (\bibinfo {year} {2015})},\ \Eprint
  {http://arxiv.org/abs/1510.03268} {arXiv:1510.03268 [hep-ph]} \BibitemShut
  {NoStop}%
\bibitem [{\citenamefont {Hayato}(2009)}]{Hayato:2009zz}%
  \BibitemOpen
  \bibfield  {author} {\bibinfo {author} {\bibfnamefont {Yoshinari}\
  \bibnamefont {Hayato}},\ }\bibfield  {title} {\enquote {\bibinfo {title} {A
  neutrino interaction simulation program library {NEUT}},}\ }\href@noop {}
  {\bibfield  {journal} {\bibinfo  {journal} {Acta Phys. Polon.}\ }\textbf
  {\bibinfo {volume} {B40}},\ \bibinfo {pages} {2477--2489} (\bibinfo {year}
  {2009})}\BibitemShut {NoStop}%
\bibitem [{\citenamefont {Andreopoulos}\ \emph {et~al.}(2010)\citenamefont
  {Andreopoulos} \emph {et~al.}}]{Andreopoulos:2009rq}%
  \BibitemOpen
  \bibfield  {author} {\bibinfo {author} {\bibfnamefont {C.}~\bibnamefont
  {Andreopoulos}} \emph {et~al.},\ }\bibfield  {title} {\enquote {\bibinfo
  {title} {The {GENIE} neutrino {Monte Carlo} generator},}\ }\href {\doibase
  10.1016/j.nima.2009.12.009} {\bibfield  {journal} {\bibinfo  {journal} {Nucl.
  Instrum. Meth.}\ }\textbf {\bibinfo {volume} {A614}},\ \bibinfo {pages}
  {87--104} (\bibinfo {year} {2010})},\ \Eprint
  {http://arxiv.org/abs/0905.2517} {arXiv:0905.2517 [hep-ph]} \BibitemShut
  {NoStop}%
\bibitem [{\citenamefont {Alam}\ \emph {et~al.}(2015)\citenamefont {Alam} \emph
  {et~al.}}]{Alam:2015nkk}%
  \BibitemOpen
  \bibfield  {author} {\bibinfo {author} {\bibfnamefont {M.}~\bibnamefont
  {Alam}} \emph {et~al.},\ }\bibfield  {title} {\enquote {\bibinfo {title}
  {{GENIE} production release 2.10.0},}\ }\href@noop {} {\  (\bibinfo {year}
  {2015})},\ \Eprint {http://arxiv.org/abs/1512.06882} {arXiv:1512.06882
  [hep-ph]} \BibitemShut {NoStop}%
\bibitem [{\citenamefont {Gallmeister}\ \emph {et~al.}(2016)\citenamefont
  {Gallmeister}, \citenamefont {Mosel},\ and\ \citenamefont
  {Weil}}]{Gallmeister:2016dnq}%
  \BibitemOpen
  \bibfield  {author} {\bibinfo {author} {\bibfnamefont {K.}~\bibnamefont
  {Gallmeister}}, \bibinfo {author} {\bibfnamefont {U.}~\bibnamefont {Mosel}},
  \ and\ \bibinfo {author} {\bibfnamefont {J.}~\bibnamefont {Weil}},\
  }\bibfield  {title} {\enquote {\bibinfo {title} {Neutrino-induced reactions
  on nuclei},}\ }\href {\doibase 10.1103/PhysRevC.94.035502} {\bibfield
  {journal} {\bibinfo  {journal} {Phys. Rev.}\ }\textbf {\bibinfo {volume}
  {C94}},\ \bibinfo {pages} {035502} (\bibinfo {year} {2016})},\ \Eprint
  {http://arxiv.org/abs/1605.09391} {arXiv:1605.09391 [nucl-th]} \BibitemShut
  {NoStop}%
\bibitem [{\citenamefont {Casper}(2002)}]{Casper:2002sd}%
  \BibitemOpen
  \bibfield  {author} {\bibinfo {author} {\bibfnamefont {D.}~\bibnamefont
  {Casper}},\ }\bibfield  {title} {\enquote {\bibinfo {title} {The {Nuance}
  neutrino physics simulation, and the future},}\ }\href {\doibase
  10.1016/S0920-5632(02)01756-5} {\bibfield  {journal} {\bibinfo  {journal}
  {Nucl. Phys. Proc. Suppl.}\ }\textbf {\bibinfo {volume} {112}},\ \bibinfo
  {pages} {161--170} (\bibinfo {year} {2002})},\ \Eprint
  {http://arxiv.org/abs/hep-ph/0208030} {arXiv:hep-ph/0208030 [hep-ph]}
  \BibitemShut {NoStop}%
\bibitem [{\citenamefont {{Morf\'in}}(2018)}]{Morfin:2018int}%
  \BibitemOpen
  \bibfield  {author} {\bibinfo {author} {\bibfnamefont {Jorge}\ \bibnamefont
  {{Morf\'in}}},\ }\href@noop {} {\enquote {\bibinfo {title} {Past and future
  of $\nu$/$\bar\nu$ deuterium/hydrogen experiments},}\ }\bibinfo
  {howpublished}
  {\href{http://www.int.washington.edu/talks/WorkShops/int_18_2a/People/Morfin_J/Morfin.pdf}{talk
  at INT Seattle}} (\bibinfo {year} {2018})\BibitemShut {NoStop}%
\bibitem [{\citenamefont {Androić}\ \emph {et~al.}(2018)\citenamefont
  {Androić} \emph {et~al.}}]{Androic:2018kni}%
  \BibitemOpen
  \bibfield  {author} {\bibinfo {author} {\bibfnamefont {D.}~\bibnamefont
  {Androić}} \emph {et~al.} (\bibinfo {collaboration} {Qweak}),\ }\bibfield
  {title} {\enquote {\bibinfo {title} {{Precision measurement of the weak
  charge of the proton}},}\ }\href {\doibase 10.1038/s41586-018-0096-0}
  {\bibfield  {journal} {\bibinfo  {journal} {Nature}\ }\textbf {\bibinfo
  {volume} {557}},\ \bibinfo {pages} {207--211} (\bibinfo {year}
  {2018})}\BibitemShut {NoStop}%
\bibitem [{\citenamefont {Freedman}(1974)}]{Freedman:1973yd}%
  \BibitemOpen
  \bibfield  {author} {\bibinfo {author} {\bibfnamefont {Daniel~Z.}\
  \bibnamefont {Freedman}},\ }\bibfield  {title} {\enquote {\bibinfo {title}
  {Coherent neutrino nucleus scattering as a probe of the weak neutral
  current},}\ }\href {\doibase 10.1103/PhysRevD.9.1389} {\bibfield  {journal}
  {\bibinfo  {journal} {Phys. Rev.}\ }\textbf {\bibinfo {volume} {D9}},\
  \bibinfo {pages} {1389--1392} (\bibinfo {year} {1974})}\BibitemShut {NoStop}%
\bibitem [{\citenamefont {Brice}\ \emph {et~al.}(2014)\citenamefont {Brice}
  \emph {et~al.}}]{Brice:2013fwa}%
  \BibitemOpen
  \bibfield  {author} {\bibinfo {author} {\bibfnamefont {S.~J.}\ \bibnamefont
  {Brice}} \emph {et~al.},\ }\bibfield  {title} {\enquote {\bibinfo {title} {A
  method for measuring coherent elastic neutrino-nucleus scattering at a far
  off-axis high-energy neutrino beam target},}\ }\href {\doibase
  10.1103/PhysRevD.89.072004} {\bibfield  {journal} {\bibinfo  {journal} {Phys.
  Rev.}\ }\textbf {\bibinfo {volume} {D89}},\ \bibinfo {pages} {072004}
  (\bibinfo {year} {2014})},\ \Eprint {http://arxiv.org/abs/1311.5958}
  {arXiv:1311.5958 [physics.ins-det]} \BibitemShut {NoStop}%
\bibitem [{\citenamefont {Akimov}\ \emph {et~al.}(2017)\citenamefont {Akimov}
  \emph {et~al.}}]{Akimov:2017ade}%
  \BibitemOpen
  \bibfield  {author} {\bibinfo {author} {\bibfnamefont {D.}~\bibnamefont
  {Akimov}} \emph {et~al.} (\bibinfo {collaboration} {COHERENT}),\ }\bibfield
  {title} {\enquote {\bibinfo {title} {Observation of coherent elastic
  neutrino-nucleus scattering},}\ }\href {\doibase 10.1126/science.aao0990}
  {\bibfield  {journal} {\bibinfo  {journal} {Science}\ }\textbf {\bibinfo
  {volume} {357}},\ \bibinfo {pages} {1123--1126} (\bibinfo {year} {2017})},\
  \Eprint {http://arxiv.org/abs/1708.01294} {arXiv:1708.01294 [nucl-ex]}
  \BibitemShut {NoStop}%
\bibitem [{\citenamefont {Formaggio}\ and\ \citenamefont
  {Zeller}(2012)}]{Formaggio:2013kya}%
  \BibitemOpen
  \bibfield  {author} {\bibinfo {author} {\bibfnamefont {J.~A.}\ \bibnamefont
  {Formaggio}}\ and\ \bibinfo {author} {\bibfnamefont {G.~P.}\ \bibnamefont
  {Zeller}},\ }\bibfield  {title} {\enquote {\bibinfo {title} {From {eV} to
  {EeV}: Neutrino cross sections across energy scales},}\ }\href {\doibase
  10.1103/RevModPhys.84.1307} {\bibfield  {journal} {\bibinfo  {journal} {Rev.
  Mod. Phys.}\ }\textbf {\bibinfo {volume} {84}},\ \bibinfo {pages}
  {1307--1341} (\bibinfo {year} {2012})},\ \Eprint
  {http://arxiv.org/abs/1305.7513} {arXiv:1305.7513 [hep-ex]} \BibitemShut
  {NoStop}%
\bibitem [{\citenamefont {Carlson}\ and\ \citenamefont
  {Schiavilla}(1998)}]{Carlson:1997}%
  \BibitemOpen
  \bibfield  {author} {\bibinfo {author} {\bibfnamefont {J.}~\bibnamefont
  {Carlson}}\ and\ \bibinfo {author} {\bibfnamefont {R.}~\bibnamefont
  {Schiavilla}},\ }\bibfield  {title} {\enquote {\bibinfo {title} {{Structure
  and dynamics of few nucleon systems}},}\ }\href {\doibase
  10.1103/RevModPhys.70.743} {\bibfield  {journal} {\bibinfo  {journal} {Rev.
  Mod. Phys.}\ }\textbf {\bibinfo {volume} {70}},\ \bibinfo {pages} {743--842}
  (\bibinfo {year} {1998})}\BibitemShut {NoStop}%
\bibitem [{\citenamefont {Bacca}\ and\ \citenamefont
  {Pastore}(2014)}]{Bacca:2014tla}%
  \BibitemOpen
  \bibfield  {author} {\bibinfo {author} {\bibfnamefont {Sonia}\ \bibnamefont
  {Bacca}}\ and\ \bibinfo {author} {\bibfnamefont {Saori}\ \bibnamefont
  {Pastore}},\ }\bibfield  {title} {\enquote {\bibinfo {title}
  {{Electromagnetic reactions on light nuclei}},}\ }\href {\doibase
  10.1088/0954-3899/41/12/123002} {\bibfield  {journal} {\bibinfo  {journal}
  {J. Phys.}\ }\textbf {\bibinfo {volume} {G41}},\ \bibinfo {pages} {123002}
  (\bibinfo {year} {2014})},\ \Eprint {http://arxiv.org/abs/1407.3490}
  {arXiv:1407.3490 [nucl-th]} \BibitemShut {NoStop}%
\bibitem [{\citenamefont {Lin}\ \emph {et~al.}(2018)\citenamefont {Lin} \emph
  {et~al.}}]{Lin:2017snn}%
  \BibitemOpen
  \bibfield  {author} {\bibinfo {author} {\bibfnamefont {Huey-Wen}\
  \bibnamefont {Lin}} \emph {et~al.},\ }\bibfield  {title} {\enquote {\bibinfo
  {title} {Parton distributions and lattice {QCD} calculations: A~community
  white paper},}\ }\href {\doibase 10.1016/j.ppnp.2018.01.007} {\bibfield
  {journal} {\bibinfo  {journal} {Prog. Part. Nucl. Phys.}\ }\textbf {\bibinfo
  {volume} {100}},\ \bibinfo {pages} {107--160} (\bibinfo {year} {2018})},\
  \Eprint {http://arxiv.org/abs/1711.07916} {arXiv:1711.07916 [hep-ph]}
  \BibitemShut {NoStop}%
\bibitem [{\citenamefont {{D\"urr}}\ \emph {et~al.}(2011)\citenamefont
  {{D\"urr}} \emph {et~al.}}]{Durr:2010aw}%
  \BibitemOpen
  \bibfield  {author} {\bibinfo {author} {\bibfnamefont {S.}~\bibnamefont
  {{D\"urr}}} \emph {et~al.} (\bibinfo {collaboration}
  {Budapest-Marseille-Wuppertal}),\ }\bibfield  {title} {\enquote {\bibinfo
  {title} {Lattice {QCD} at the physical point: Simulation and analysis
  details},}\ }\href {\doibase 10.1007/JHEP08(2011)148} {\bibfield  {journal}
  {\bibinfo  {journal} {JHEP}\ }\textbf {\bibinfo {volume} {08}},\ \bibinfo
  {pages} {148} (\bibinfo {year} {2011})},\ \Eprint
  {http://arxiv.org/abs/1011.2711} {arXiv:1011.2711 [hep-lat]} \BibitemShut
  {NoStop}%
\bibitem [{\citenamefont {Bazavov}\ \emph {et~al.}(2013)\citenamefont {Bazavov}
  \emph {et~al.}}]{Bazavov:2012xda}%
  \BibitemOpen
  \bibfield  {author} {\bibinfo {author} {\bibfnamefont {A.}~\bibnamefont
  {Bazavov}} \emph {et~al.} (\bibinfo {collaboration} {MILC}),\ }\bibfield
  {title} {\enquote {\bibinfo {title} {Lattice {QCD} ensembles with four
  flavors of highly improved staggered quarks},}\ }\href {\doibase
  10.1103/PhysRevD.87.054505} {\bibfield  {journal} {\bibinfo  {journal} {Phys.
  Rev.}\ }\textbf {\bibinfo {volume} {D87}},\ \bibinfo {pages} {054505}
  (\bibinfo {year} {2013})},\ \Eprint {http://arxiv.org/abs/1212.4768}
  {arXiv:1212.4768 [hep-lat]} \BibitemShut {NoStop}%
\bibitem [{\citenamefont {Bazavov}\ \emph {et~al.}(2018)\citenamefont {Bazavov}
  \emph {et~al.}}]{Bazavov:2017lyh}%
  \BibitemOpen
  \bibfield  {author} {\bibinfo {author} {\bibfnamefont {A.}~\bibnamefont
  {Bazavov}} \emph {et~al.} (\bibinfo {collaboration} {Fermilab Lattice,
  MILC}),\ }\bibfield  {title} {\enquote {\bibinfo {title} {{$B$}- and
  {$D$}-meson leptonic decay constants from four-flavor lattice {QCD}},}\
  }\href {\doibase 10.1103/PhysRevD.98.074512} {\bibfield  {journal} {\bibinfo
  {journal} {Phys. Rev.}\ }\textbf {\bibinfo {volume} {D98}},\ \bibinfo {pages}
  {074512} (\bibinfo {year} {2018})},\ \Eprint
  {http://arxiv.org/abs/1712.09262} {arXiv:1712.09262 [hep-lat]} \BibitemShut
  {NoStop}%
\bibitem [{\citenamefont {Parisi}(1984)}]{parisi1983common}%
  \BibitemOpen
  \bibfield  {author} {\bibinfo {author} {\bibfnamefont {G.}~\bibnamefont
  {Parisi}},\ }\bibfield  {title} {\enquote {\bibinfo {title} {The strategy for
  computing the hadronic mass spectrum},}\ }\href {\doibase
  10.1016/0370-1573(84)90081-4} {\bibfield  {journal} {\bibinfo  {journal}
  {Phys. Rept.}\ }\textbf {\bibinfo {volume} {103}},\ \bibinfo {pages}
  {203--211} (\bibinfo {year} {1984})}\BibitemShut {NoStop}%
\bibitem [{\citenamefont {Lepage}(1989)}]{lepage1989tasi}%
  \BibitemOpen
  \bibfield  {author} {\bibinfo {author} {\bibfnamefont {G.~Peter}\
  \bibnamefont {Lepage}},\ }\bibfield  {title} {\enquote {\bibinfo {title} {The
  analysis of algorithms for lattice field theory},}\ }in\ \href@noop {} {\emph
  {\bibinfo {booktitle} {From Actions to Answers}}},\ \bibinfo {editor} {edited
  by\ \bibinfo {editor} {\bibfnamefont {T.}~\bibnamefont {DeGrand}}\ and\
  \bibinfo {editor} {\bibfnamefont {W.~D.}\ \bibnamefont {Toussaint}}}\
  (\bibinfo  {publisher} {World Scientific},\ \bibinfo {address} {Singapore},\
  \bibinfo {year} {1989})\ pp.\ \bibinfo {pages} {97--120}\BibitemShut
  {NoStop}%
\bibitem [{\citenamefont {Wagman}(2017)}]{Wagman:2017gqi}%
  \BibitemOpen
  \bibfield  {author} {\bibinfo {author} {\bibfnamefont {Michael~L.}\
  \bibnamefont {Wagman}},\ }\emph {\bibinfo {title} {Statistical Angles on the
  Lattice {QCD} Signal-to-Noise Problem}},\ \href@noop {} {Ph.D. thesis},\
  \bibinfo  {school} {University of Washington} (\bibinfo {year} {2017}),\
  \Eprint {http://arxiv.org/abs/1711.00062} {arXiv:1711.00062 [hep-lat]}
  \BibitemShut {NoStop}%
\bibitem [{\citenamefont {Owen}\ \emph {et~al.}(2013)\citenamefont {Owen},
  \citenamefont {Dragos}, \citenamefont {Kamleh}, \citenamefont {Leinweber},
  \citenamefont {Mahbub}, \citenamefont {Menadue},\ and\ \citenamefont
  {Zanotti}}]{Owen:2012ts}%
  \BibitemOpen
  \bibfield  {author} {\bibinfo {author} {\bibfnamefont {Benjamin~J.}\
  \bibnamefont {Owen}}, \bibinfo {author} {\bibfnamefont {Jack}\ \bibnamefont
  {Dragos}}, \bibinfo {author} {\bibfnamefont {Waseem}\ \bibnamefont {Kamleh}},
  \bibinfo {author} {\bibfnamefont {Derek~B.}\ \bibnamefont {Leinweber}},
  \bibinfo {author} {\bibfnamefont {M.~Selim}\ \bibnamefont {Mahbub}}, \bibinfo
  {author} {\bibfnamefont {Benjamin~J.}\ \bibnamefont {Menadue}}, \ and\
  \bibinfo {author} {\bibfnamefont {James~M.}\ \bibnamefont {Zanotti}},\
  }\bibfield  {title} {\enquote {\bibinfo {title} {Variational approach to the
  calculation of {$g_A$}},}\ }\href {\doibase 10.1016/j.physletb.2013.04.063}
  {\bibfield  {journal} {\bibinfo  {journal} {Phys. Lett.}\ }\textbf {\bibinfo
  {volume} {B723}},\ \bibinfo {pages} {217--223} (\bibinfo {year} {2013})},\
  \Eprint {http://arxiv.org/abs/1212.4668} {arXiv:1212.4668 [hep-lat]}
  \BibitemShut {NoStop}%
\bibitem [{\citenamefont {Blum}\ \emph {et~al.}(2016)\citenamefont {Blum} \emph
  {et~al.}}]{Blum:2014tka}%
  \BibitemOpen
  \bibfield  {author} {\bibinfo {author} {\bibfnamefont {T.}~\bibnamefont
  {Blum}} \emph {et~al.} (\bibinfo {collaboration} {RBC, UKQCD}),\ }\bibfield
  {title} {\enquote {\bibinfo {title} {{Domain wall QCD with physical quark
  masses}},}\ }\href {\doibase 10.1103/PhysRevD.93.074505} {\bibfield
  {journal} {\bibinfo  {journal} {Phys. Rev.}\ }\textbf {\bibinfo {volume}
  {D93}},\ \bibinfo {pages} {074505} (\bibinfo {year} {2016})},\ \Eprint
  {http://arxiv.org/abs/1411.7017} {arXiv:1411.7017 [hep-lat]} \BibitemShut
  {NoStop}%
\bibitem [{\citenamefont {Bhattacharya}\ \emph {et~al.}(2011)\citenamefont
  {Bhattacharya}, \citenamefont {Hill},\ and\ \citenamefont
  {Paz}}]{Bhattacharya:2011ah}%
  \BibitemOpen
  \bibfield  {author} {\bibinfo {author} {\bibfnamefont {Bhubanjyoti}\
  \bibnamefont {Bhattacharya}}, \bibinfo {author} {\bibfnamefont {Richard~J.}\
  \bibnamefont {Hill}}, \ and\ \bibinfo {author} {\bibfnamefont {Gil}\
  \bibnamefont {Paz}},\ }\bibfield  {title} {\enquote {\bibinfo {title} {Model
  independent determination of the axial mass parameter in quasielastic
  neutrino-nucleon scattering},}\ }\href {\doibase 10.1103/PhysRevD.84.073006}
  {\bibfield  {journal} {\bibinfo  {journal} {Phys. Rev.}\ }\textbf {\bibinfo
  {volume} {D84}},\ \bibinfo {pages} {073006} (\bibinfo {year} {2011})},\
  \Eprint {http://arxiv.org/abs/1108.0423} {arXiv:1108.0423 [hep-ph]}
  \BibitemShut {NoStop}%
\bibitem [{\citenamefont {Weinberg}(1958)}]{Weinberg:1958ut}%
  \BibitemOpen
  \bibfield  {author} {\bibinfo {author} {\bibfnamefont {Steven}\ \bibnamefont
  {Weinberg}},\ }\bibfield  {title} {\enquote {\bibinfo {title} {Charge
  symmetry of weak interactions},}\ }\href {\doibase 10.1103/PhysRev.112.1375}
  {\bibfield  {journal} {\bibinfo  {journal} {Phys. Rev.}\ }\textbf {\bibinfo
  {volume} {112}},\ \bibinfo {pages} {1375--1379} (\bibinfo {year}
  {1958})}\BibitemShut {NoStop}%
\bibitem [{\citenamefont {Llewellyn~Smith}(1972)}]{LlewellynSmith:1971uhs}%
  \BibitemOpen
  \bibfield  {author} {\bibinfo {author} {\bibfnamefont {C.~H.}\ \bibnamefont
  {Llewellyn~Smith}},\ }\bibfield  {title} {\enquote {\bibinfo {title}
  {Neutrino reactions at accelerator energies},}\ }\href {\doibase
  10.1016/0370-1573(72)90010-5} {\bibfield  {journal} {\bibinfo  {journal}
  {Phys. Rept.}\ }\textbf {\bibinfo {volume} {3}},\ \bibinfo {pages} {261--379}
  (\bibinfo {year} {1972})}\BibitemShut {NoStop}%
\bibitem [{\citenamefont {Tanabashi}\ \emph {et~al.}(2018)\citenamefont
  {Tanabashi} \emph {et~al.}}]{Tanabashi:2018oca}%
  \BibitemOpen
  \bibfield  {author} {\bibinfo {author} {\bibfnamefont {M.}~\bibnamefont
  {Tanabashi}} \emph {et~al.} (\bibinfo {collaboration} {Particle Data
  Group}),\ }\bibfield  {title} {\enquote {\bibinfo {title} {Review of particle
  physics},}\ }\href {\doibase 10.1103/PhysRevD.98.030001} {\bibfield
  {journal} {\bibinfo  {journal} {Phys. Rev.}\ }\textbf {\bibinfo {volume}
  {D98}},\ \bibinfo {pages} {030001} (\bibinfo {year} {2018})}\BibitemShut
  {NoStop}%
\bibitem [{\citenamefont {Bodek}\ \emph {et~al.}(2008)\citenamefont {Bodek},
  \citenamefont {Avvakumov}, \citenamefont {Bradford},\ and\ \citenamefont
  {Budd}}]{Bodek:2007ym}%
  \BibitemOpen
  \bibfield  {author} {\bibinfo {author} {\bibfnamefont {A.}~\bibnamefont
  {Bodek}}, \bibinfo {author} {\bibfnamefont {S.}~\bibnamefont {Avvakumov}},
  \bibinfo {author} {\bibfnamefont {R.}~\bibnamefont {Bradford}}, \ and\
  \bibinfo {author} {\bibfnamefont {Howard~S.}\ \bibnamefont {Budd}},\
  }\bibfield  {title} {\enquote {\bibinfo {title} {Vector and axial nucleon
  form factors: A duality constrained parameterization},}\ }\href {\doibase
  10.1140/epjc/s10052-007-0491-4} {\bibfield  {journal} {\bibinfo  {journal}
  {Eur. Phys. J.}\ }\textbf {\bibinfo {volume} {C53}},\ \bibinfo {pages}
  {349--354} (\bibinfo {year} {2008})},\ \Eprint
  {http://arxiv.org/abs/0708.1946} {arXiv:0708.1946 [hep-ex]} \BibitemShut
  {NoStop}%
\bibitem [{\citenamefont {Liesenfeld}\ \emph {et~al.}(1999)\citenamefont
  {Liesenfeld} \emph {et~al.}}]{Liesenfeld:1999mv}%
  \BibitemOpen
  \bibfield  {author} {\bibinfo {author} {\bibfnamefont {A.}~\bibnamefont
  {Liesenfeld}} \emph {et~al.} (\bibinfo {collaboration} {A1}),\ }\bibfield
  {title} {\enquote {\bibinfo {title} {A measurement of the axial form-factor
  of the nucleon by the $p(e,e'\pi^+)n$ reaction at {$W=1125~\text{MeV}$}},}\
  }\href {\doibase 10.1016/S0370-2693(99)01204-6} {\bibfield  {journal}
  {\bibinfo  {journal} {Phys. Lett.}\ }\textbf {\bibinfo {volume} {B468}},\
  \bibinfo {pages} {20} (\bibinfo {year} {1999})},\ \Eprint
  {http://arxiv.org/abs/nucl-ex/9911003} {arXiv:nucl-ex/9911003 [nucl-ex]}
  \BibitemShut {NoStop}%
\bibitem [{\citenamefont {Bernard}\ \emph {et~al.}(1992)\citenamefont
  {Bernard}, \citenamefont {Kaiser},\ and\ \citenamefont
  {Meissner}}]{Bernard:1992ys}%
  \BibitemOpen
  \bibfield  {author} {\bibinfo {author} {\bibfnamefont {Veronique}\
  \bibnamefont {Bernard}}, \bibinfo {author} {\bibfnamefont {Norbert}\
  \bibnamefont {Kaiser}}, \ and\ \bibinfo {author} {\bibfnamefont {Ulf~G.}\
  \bibnamefont {Meissner}},\ }\bibfield  {title} {\enquote {\bibinfo {title}
  {Measuring the axial radius of the nucleon in pion electroproduction},}\
  }\href {\doibase 10.1103/PhysRevLett.69.1877} {\bibfield  {journal} {\bibinfo
   {journal} {Phys. Rev. Lett.}\ }\textbf {\bibinfo {volume} {69}},\ \bibinfo
  {pages} {1877--1879} (\bibinfo {year} {1992})}\BibitemShut {NoStop}%
\bibitem [{\citenamefont {Gran}\ \emph {et~al.}(2006)\citenamefont {Gran} \emph
  {et~al.}}]{Gran:2006jn}%
  \BibitemOpen
  \bibfield  {author} {\bibinfo {author} {\bibfnamefont {R.}~\bibnamefont
  {Gran}} \emph {et~al.} (\bibinfo {collaboration} {K2K}),\ }\bibfield  {title}
  {\enquote {\bibinfo {title} {Measurement of the quasielastic axial vector
  mass in neutrino-oxygen interactions},}\ }\href {\doibase
  10.1103/PhysRevD.74.052002} {\bibfield  {journal} {\bibinfo  {journal} {Phys.
  Rev.}\ }\textbf {\bibinfo {volume} {D74}},\ \bibinfo {pages} {052002}
  (\bibinfo {year} {2006})},\ \Eprint {http://arxiv.org/abs/hep-ex/0603034}
  {arXiv:hep-ex/0603034 [hep-ex]} \BibitemShut {NoStop}%
\bibitem [{\citenamefont {Dorman}(2009)}]{Dorman:2009zz}%
  \BibitemOpen
  \bibfield  {author} {\bibinfo {author} {\bibfnamefont {M.}~\bibnamefont
  {Dorman}} (\bibinfo {collaboration} {MINOS}),\ }\bibfield  {title} {\enquote
  {\bibinfo {title} {Preliminary results for {CCQE} scattering with the {MINOS}
  near detector},}\ }\href {\doibase 10.1063/1.3274143} {\bibfield  {journal}
  {\bibinfo  {journal} {AIP Conf. Proc.}\ }\textbf {\bibinfo {volume} {1189}},\
  \bibinfo {pages} {133--138} (\bibinfo {year} {2009})}\BibitemShut {NoStop}%
\bibitem [{\citenamefont {Aguilar-Arevalo}\ \emph
  {et~al.}(2010{\natexlab{a}})\citenamefont {Aguilar-Arevalo} \emph
  {et~al.}}]{AguilarArevalo:2010zc}%
  \BibitemOpen
  \bibfield  {author} {\bibinfo {author} {\bibfnamefont {A.~A.}\ \bibnamefont
  {Aguilar-Arevalo}} \emph {et~al.} (\bibinfo {collaboration} {MiniBooNE}),\
  }\bibfield  {title} {\enquote {\bibinfo {title} {First measurement of the
  muon neutrino charged current quasielastic double differential cross
  section},}\ }\href {\doibase 10.1103/PhysRevD.81.092005} {\bibfield
  {journal} {\bibinfo  {journal} {Phys. Rev.}\ }\textbf {\bibinfo {volume}
  {D81}},\ \bibinfo {pages} {092005} (\bibinfo {year} {2010}{\natexlab{a}})},\
  \Eprint {http://arxiv.org/abs/1002.2680} {arXiv:1002.2680 [hep-ex]}
  \BibitemShut {NoStop}%
\bibitem [{\citenamefont {Lyubushkin}\ \emph {et~al.}(2009)\citenamefont
  {Lyubushkin} \emph {et~al.}}]{Lyubushkin:2008pe}%
  \BibitemOpen
  \bibfield  {author} {\bibinfo {author} {\bibfnamefont {V.}~\bibnamefont
  {Lyubushkin}} \emph {et~al.} (\bibinfo {collaboration} {NOMAD}),\ }\bibfield
  {title} {\enquote {\bibinfo {title} {A study of quasielastic muon neutrino
  and antineutrino scattering in the {NOMAD} experiment},}\ }\href {\doibase
  10.1140/epjc/s10052-009-1113-0} {\bibfield  {journal} {\bibinfo  {journal}
  {Eur. Phys. J.}\ }\textbf {\bibinfo {volume} {C63}},\ \bibinfo {pages}
  {355--381} (\bibinfo {year} {2009})},\ \Eprint
  {http://arxiv.org/abs/0812.4543} {arXiv:0812.4543 [hep-ex]} \BibitemShut
  {NoStop}%
\bibitem [{\citenamefont {Fields}\ \emph {et~al.}(2013)\citenamefont {Fields}
  \emph {et~al.}}]{Fields:2013zhk}%
  \BibitemOpen
  \bibfield  {author} {\bibinfo {author} {\bibfnamefont {L.}~\bibnamefont
  {Fields}} \emph {et~al.} (\bibinfo {collaboration} {MINERvA}),\ }\bibfield
  {title} {\enquote {\bibinfo {title} {Measurement of muon antineutrino
  quasielastic scattering on a hydrocarbon target at
  $e_\nu\sim3.5~\text{GeV}$},}\ }\href {\doibase
  10.1103/PhysRevLett.111.022501} {\bibfield  {journal} {\bibinfo  {journal}
  {Phys. Rev. Lett.}\ }\textbf {\bibinfo {volume} {111}},\ \bibinfo {pages}
  {022501} (\bibinfo {year} {2013})},\ \Eprint {http://arxiv.org/abs/1305.2234}
  {arXiv:1305.2234 [hep-ex]} \BibitemShut {NoStop}%
\bibitem [{\citenamefont {Fiorentini}\ \emph {et~al.}(2013)\citenamefont
  {Fiorentini} \emph {et~al.}}]{Fiorentini:2013ezn}%
  \BibitemOpen
  \bibfield  {author} {\bibinfo {author} {\bibfnamefont {G.~A.}\ \bibnamefont
  {Fiorentini}} \emph {et~al.} (\bibinfo {collaboration} {MINERvA}),\
  }\bibfield  {title} {\enquote {\bibinfo {title} {Measurement of muon neutrino
  quasielastic scattering on a hydrocarbon target at
  $e_\nu\sim3.5~\text{GeV}$},}\ }\href {\doibase
  10.1103/PhysRevLett.111.022502} {\bibfield  {journal} {\bibinfo  {journal}
  {Phys. Rev. Lett.}\ }\textbf {\bibinfo {volume} {111}},\ \bibinfo {pages}
  {022502} (\bibinfo {year} {2013})},\ \Eprint {http://arxiv.org/abs/1305.2243}
  {arXiv:1305.2243 [hep-ex]} \BibitemShut {NoStop}%
\bibitem [{\citenamefont {Meyer}\ \emph
  {et~al.}(2016{\natexlab{a}})\citenamefont {Meyer}, \citenamefont
  {Betancourt}, \citenamefont {Gran},\ and\ \citenamefont
  {Hill}}]{Meyer:2016oeg}%
  \BibitemOpen
  \bibfield  {author} {\bibinfo {author} {\bibfnamefont {Aaron~S.}\
  \bibnamefont {Meyer}}, \bibinfo {author} {\bibfnamefont {Minerba}\
  \bibnamefont {Betancourt}}, \bibinfo {author} {\bibfnamefont {Richard}\
  \bibnamefont {Gran}}, \ and\ \bibinfo {author} {\bibfnamefont {Richard~J.}\
  \bibnamefont {Hill}},\ }\bibfield  {title} {\enquote {\bibinfo {title}
  {{Deuterium target data for precision neutrino-nucleus cross sections}},}\
  }\href {\doibase 10.1103/PhysRevD.93.113015} {\bibfield  {journal} {\bibinfo
  {journal} {Phys. Rev.}\ }\textbf {\bibinfo {volume} {D93}},\ \bibinfo {pages}
  {113015} (\bibinfo {year} {2016}{\natexlab{a}})},\ \Eprint
  {http://arxiv.org/abs/1603.03048} {arXiv:1603.03048 [hep-ph]} \BibitemShut
  {NoStop}%
\bibitem [{\citenamefont {Hill}\ \emph {et~al.}(2018)\citenamefont {Hill},
  \citenamefont {Kammel}, \citenamefont {Marciano},\ and\ \citenamefont
  {Sirlin}}]{Hill:2017wgb}%
  \BibitemOpen
  \bibfield  {author} {\bibinfo {author} {\bibfnamefont {Richard~J.}\
  \bibnamefont {Hill}}, \bibinfo {author} {\bibfnamefont {Peter}\ \bibnamefont
  {Kammel}}, \bibinfo {author} {\bibfnamefont {William~J.}\ \bibnamefont
  {Marciano}}, \ and\ \bibinfo {author} {\bibfnamefont {Alberto}\ \bibnamefont
  {Sirlin}},\ }\bibfield  {title} {\enquote {\bibinfo {title} {Nucleon axial
  radius and muonic hydrogen: A new analysis and review},}\ }\href {\doibase
  10.1088/1361-6633/aac190} {\bibfield  {journal} {\bibinfo  {journal} {Rept.
  Prog. Phys.}\ }\textbf {\bibinfo {volume} {81}},\ \bibinfo {pages} {096301}
  (\bibinfo {year} {2018})},\ \Eprint {http://arxiv.org/abs/1708.08462}
  {arXiv:1708.08462 [hep-ph]} \BibitemShut {NoStop}%
\bibitem [{\citenamefont {Bali}\ \emph {et~al.}(2015)\citenamefont {Bali},
  \citenamefont {Collins}, \citenamefont {Glä\ss{}le}, \citenamefont
  {Göckeler}, \citenamefont {Najjar}, \citenamefont {Rödl}, \citenamefont
  {Schäfer}, \citenamefont {Schiel}, \citenamefont {Söldner},\ and\
  \citenamefont {Sternbeck}}]{Bali:2014nma}%
  \BibitemOpen
  \bibfield  {author} {\bibinfo {author} {\bibfnamefont {Gunnar~S.}\
  \bibnamefont {Bali}}, \bibinfo {author} {\bibfnamefont {Sara}\ \bibnamefont
  {Collins}}, \bibinfo {author} {\bibfnamefont {Benjamin}\ \bibnamefont
  {Glä\ss{}le}}, \bibinfo {author} {\bibfnamefont {Meinulf}\ \bibnamefont
  {Göckeler}}, \bibinfo {author} {\bibfnamefont {Johannes}\ \bibnamefont
  {Najjar}}, \bibinfo {author} {\bibfnamefont {Rudolf~H.}\ \bibnamefont
  {Rödl}}, \bibinfo {author} {\bibfnamefont {Andreas}\ \bibnamefont
  {Schäfer}}, \bibinfo {author} {\bibfnamefont {Rainer~W.}\ \bibnamefont
  {Schiel}}, \bibinfo {author} {\bibfnamefont {Wolfgang}\ \bibnamefont
  {Söldner}}, \ and\ \bibinfo {author} {\bibfnamefont {André}\ \bibnamefont
  {Sternbeck}},\ }\bibfield  {title} {\enquote {\bibinfo {title} {Nucleon
  isovector couplings from {$N_f=2$} lattice {QCD}},}\ }\href {\doibase
  10.1103/PhysRevD.91.054501} {\bibfield  {journal} {\bibinfo  {journal} {Phys.
  Rev.}\ }\textbf {\bibinfo {volume} {D91}},\ \bibinfo {pages} {054501}
  (\bibinfo {year} {2015})},\ \Eprint {http://arxiv.org/abs/1412.7336}
  {arXiv:1412.7336 [hep-lat]} \BibitemShut {NoStop}%
\bibitem [{\citenamefont {Bhattacharya}\ \emph {et~al.}(2016)\citenamefont
  {Bhattacharya}, \citenamefont {Cirigliano}, \citenamefont {Cohen},
  \citenamefont {Gupta}, \citenamefont {Lin},\ and\ \citenamefont
  {Yoon}}]{Bhattacharya:2016zcn}%
  \BibitemOpen
  \bibfield  {author} {\bibinfo {author} {\bibfnamefont {Tanmoy}\ \bibnamefont
  {Bhattacharya}}, \bibinfo {author} {\bibfnamefont {Vincenzo}\ \bibnamefont
  {Cirigliano}}, \bibinfo {author} {\bibfnamefont {Saul}\ \bibnamefont
  {Cohen}}, \bibinfo {author} {\bibfnamefont {Rajan}\ \bibnamefont {Gupta}},
  \bibinfo {author} {\bibfnamefont {Huey-Wen}\ \bibnamefont {Lin}}, \ and\
  \bibinfo {author} {\bibfnamefont {Boram}\ \bibnamefont {Yoon}},\ }\bibfield
  {title} {\enquote {\bibinfo {title} {Axial, scalar and tensor charges of the
  nucleon from 2+1+1-flavor lattice {QCD}},}\ }\href {\doibase
  10.1103/PhysRevD.94.054508} {\bibfield  {journal} {\bibinfo  {journal} {Phys.
  Rev.}\ }\textbf {\bibinfo {volume} {D94}},\ \bibinfo {pages} {054508}
  (\bibinfo {year} {2016})},\ \Eprint {http://arxiv.org/abs/1606.07049}
  {arXiv:1606.07049 [hep-lat]} \BibitemShut {NoStop}%
\bibitem [{\citenamefont {Green}\ \emph {et~al.}(2017)\citenamefont {Green},
  \citenamefont {Hasan}, \citenamefont {Meinel}, \citenamefont {Engelhardt},
  \citenamefont {Krieg}, \citenamefont {Laeuchli}, \citenamefont {Negele},
  \citenamefont {Orginos}, \citenamefont {Pochinsky},\ and\ \citenamefont
  {Syritsyn}}]{Green:2017keo}%
  \BibitemOpen
  \bibfield  {author} {\bibinfo {author} {\bibfnamefont {Jeremy}\ \bibnamefont
  {Green}}, \bibinfo {author} {\bibfnamefont {Nesreen}\ \bibnamefont {Hasan}},
  \bibinfo {author} {\bibfnamefont {Stefan}\ \bibnamefont {Meinel}}, \bibinfo
  {author} {\bibfnamefont {Michael}\ \bibnamefont {Engelhardt}}, \bibinfo
  {author} {\bibfnamefont {Stefan}\ \bibnamefont {Krieg}}, \bibinfo {author}
  {\bibfnamefont {Jesse}\ \bibnamefont {Laeuchli}}, \bibinfo {author}
  {\bibfnamefont {John}\ \bibnamefont {Negele}}, \bibinfo {author}
  {\bibfnamefont {Kostas}\ \bibnamefont {Orginos}}, \bibinfo {author}
  {\bibfnamefont {Andrew}\ \bibnamefont {Pochinsky}}, \ and\ \bibinfo {author}
  {\bibfnamefont {Sergey}\ \bibnamefont {Syritsyn}},\ }\bibfield  {title}
  {\enquote {\bibinfo {title} {Up, down, and strange nucleon axial form factors
  from lattice {QCD}},}\ }\href {\doibase 10.1103/PhysRevD.95.114502}
  {\bibfield  {journal} {\bibinfo  {journal} {Phys. Rev.}\ }\textbf {\bibinfo
  {volume} {D95}},\ \bibinfo {pages} {114502} (\bibinfo {year} {2017})},\
  \Eprint {http://arxiv.org/abs/1703.06703} {arXiv:1703.06703 [hep-lat]}
  \BibitemShut {NoStop}%
\bibitem [{\citenamefont {Alexandrou}\ \emph
  {et~al.}(2017{\natexlab{a}})\citenamefont {Alexandrou}, \citenamefont
  {Constantinou}, \citenamefont {Hadjiyiannakou}, \citenamefont {Jansen},
  \citenamefont {Kallidonis}, \citenamefont {Koutsou},\ and\ \citenamefont
  {Vaquero Aviles-Casco}}]{Alexandrou:2017hac}%
  \BibitemOpen
  \bibfield  {author} {\bibinfo {author} {\bibfnamefont {Constantia}\
  \bibnamefont {Alexandrou}}, \bibinfo {author} {\bibfnamefont {Martha}\
  \bibnamefont {Constantinou}}, \bibinfo {author} {\bibfnamefont {Kyriakos}\
  \bibnamefont {Hadjiyiannakou}}, \bibinfo {author} {\bibfnamefont {Karl}\
  \bibnamefont {Jansen}}, \bibinfo {author} {\bibfnamefont {Christos}\
  \bibnamefont {Kallidonis}}, \bibinfo {author} {\bibfnamefont {Giannis}\
  \bibnamefont {Koutsou}}, \ and\ \bibinfo {author} {\bibfnamefont {Alejandro}\
  \bibnamefont {Vaquero Aviles-Casco}},\ }\bibfield  {title} {\enquote
  {\bibinfo {title} {Nucleon axial form factors using {$N_f=2$} twisted mass
  fermions with a physical value of the pion mass},}\ }\href {\doibase
  10.1103/PhysRevD.96.054507} {\bibfield  {journal} {\bibinfo  {journal} {Phys.
  Rev.}\ }\textbf {\bibinfo {volume} {D96}},\ \bibinfo {pages} {054507}
  (\bibinfo {year} {2017}{\natexlab{a}})},\ \Eprint
  {http://arxiv.org/abs/1705.03399} {arXiv:1705.03399 [hep-lat]} \BibitemShut
  {NoStop}%
\bibitem [{\citenamefont {Capitani}\ \emph {et~al.}(2019)\citenamefont
  {Capitani}, \citenamefont {Della~Morte}, \citenamefont {Djukanovic},
  \citenamefont {von Hippel}, \citenamefont {Hua}, \citenamefont {Jäger},
  \citenamefont {Junnarkar}, \citenamefont {Meyer}, \citenamefont {Rae},\ and\
  \citenamefont {Wittig}}]{Capitani:2017qpc}%
  \BibitemOpen
  \bibfield  {author} {\bibinfo {author} {\bibfnamefont {Stefano}\ \bibnamefont
  {Capitani}}, \bibinfo {author} {\bibfnamefont {Michele}\ \bibnamefont
  {Della~Morte}}, \bibinfo {author} {\bibfnamefont {Dalibor}\ \bibnamefont
  {Djukanovic}}, \bibinfo {author} {\bibfnamefont {Georg~M.}\ \bibnamefont {von
  Hippel}}, \bibinfo {author} {\bibfnamefont {Jiayu}\ \bibnamefont {Hua}},
  \bibinfo {author} {\bibfnamefont {Benjamin}\ \bibnamefont {Jäger}}, \bibinfo
  {author} {\bibfnamefont {Parikshit~M.}\ \bibnamefont {Junnarkar}}, \bibinfo
  {author} {\bibfnamefont {Harvey~B.}\ \bibnamefont {Meyer}}, \bibinfo {author}
  {\bibfnamefont {Thomas~D.}\ \bibnamefont {Rae}}, \ and\ \bibinfo {author}
  {\bibfnamefont {Hartmut}\ \bibnamefont {Wittig}},\ }\bibfield  {title}
  {\enquote {\bibinfo {title} {Isovector axial form factors of the nucleon in
  two-flavor lattice {QCD}},}\ }\href {\doibase 10.1142/S0217751X1950009X}
  {\bibfield  {journal} {\bibinfo  {journal} {Int. J. Mod. Phys.}\ }\textbf
  {\bibinfo {volume} {A34}},\ \bibinfo {pages} {1950009} (\bibinfo {year}
  {2019})},\ \Eprint {http://arxiv.org/abs/1705.06186} {arXiv:1705.06186
  [hep-lat]} \BibitemShut {NoStop}%
\bibitem [{\citenamefont {Gupta}\ \emph {et~al.}(2017)\citenamefont {Gupta},
  \citenamefont {Jang}, \citenamefont {Lin}, \citenamefont {Yoon},\ and\
  \citenamefont {Bhattacharya}}]{Rajan:2017lxk}%
  \BibitemOpen
  \bibfield  {author} {\bibinfo {author} {\bibfnamefont {Rajan}\ \bibnamefont
  {Gupta}}, \bibinfo {author} {\bibfnamefont {Yong-Chull}\ \bibnamefont
  {Jang}}, \bibinfo {author} {\bibfnamefont {Huey-Wen}\ \bibnamefont {Lin}},
  \bibinfo {author} {\bibfnamefont {Boram}\ \bibnamefont {Yoon}}, \ and\
  \bibinfo {author} {\bibfnamefont {Tanmoy}\ \bibnamefont {Bhattacharya}}
  (\bibinfo {collaboration} {PNDME}),\ }\bibfield  {title} {\enquote {\bibinfo
  {title} {Axial vector form factors of the nucleon from lattice {QCD}},}\
  }\href {\doibase 10.1103/PhysRevD.96.114503} {\bibfield  {journal} {\bibinfo
  {journal} {Phys. Rev.}\ }\textbf {\bibinfo {volume} {D96}},\ \bibinfo {pages}
  {114503} (\bibinfo {year} {2017})},\ \Eprint
  {http://arxiv.org/abs/1705.06834} {arXiv:1705.06834 [hep-lat]} \BibitemShut
  {NoStop}%
\bibitem [{\citenamefont {Jang}\ \emph
  {et~al.}(2018{\natexlab{a}})\citenamefont {Jang}, \citenamefont
  {Bhattacharya}, \citenamefont {Gupta}, \citenamefont {Lin},\ and\
  \citenamefont {Yoon}}]{Jang:2018lup}%
  \BibitemOpen
  \bibfield  {author} {\bibinfo {author} {\bibfnamefont {Yong-Chull}\
  \bibnamefont {Jang}}, \bibinfo {author} {\bibfnamefont {Tanmoy}\ \bibnamefont
  {Bhattacharya}}, \bibinfo {author} {\bibfnamefont {Rajan}\ \bibnamefont
  {Gupta}}, \bibinfo {author} {\bibfnamefont {Huey-Wen}\ \bibnamefont {Lin}}, \
  and\ \bibinfo {author} {\bibfnamefont {Boram}\ \bibnamefont {Yoon}},\
  }\bibfield  {title} {\enquote {\bibinfo {title} {Nucleon axial and
  electromagnetic form factors},}\ }\href {\doibase
  10.1051/epjconf/201817506033} {\bibfield  {journal} {\bibinfo  {journal} {EPJ
  Web Conf.}\ }\textbf {\bibinfo {volume} {175}},\ \bibinfo {pages} {06033}
  (\bibinfo {year} {2018}{\natexlab{a}})},\ \Eprint
  {http://arxiv.org/abs/1801.01635} {arXiv:1801.01635 [hep-lat]} \BibitemShut
  {NoStop}%
\bibitem [{\citenamefont {Yamanaka}\ \emph {et~al.}(2018)\citenamefont
  {Yamanaka}, \citenamefont {Hashimoto}, \citenamefont {Kaneko},\ and\
  \citenamefont {Ohki}}]{Yamanaka:2018uud}%
  \BibitemOpen
  \bibfield  {author} {\bibinfo {author} {\bibfnamefont {Nodoka}\ \bibnamefont
  {Yamanaka}}, \bibinfo {author} {\bibfnamefont {Shoji}\ \bibnamefont
  {Hashimoto}}, \bibinfo {author} {\bibfnamefont {Takashi}\ \bibnamefont
  {Kaneko}}, \ and\ \bibinfo {author} {\bibfnamefont {Hiroshi}\ \bibnamefont
  {Ohki}} (\bibinfo {collaboration} {JLQCD}),\ }\bibfield  {title} {\enquote
  {\bibinfo {title} {Nucleon charges with dynamical overlap fermions},}\ }\href
  {\doibase 10.1103/PhysRevD.98.054516} {\bibfield  {journal} {\bibinfo
  {journal} {Phys. Rev.}\ }\textbf {\bibinfo {volume} {D98}},\ \bibinfo {pages}
  {054516} (\bibinfo {year} {2018})},\ \Eprint
  {http://arxiv.org/abs/1805.10507} {arXiv:1805.10507 [hep-lat]} \BibitemShut
  {NoStop}%
\bibitem [{\citenamefont {Chang}\ \emph
  {et~al.}(2018{\natexlab{a}})\citenamefont {Chang} \emph
  {et~al.}}]{Chang:2018uxx}%
  \BibitemOpen
  \bibfield  {author} {\bibinfo {author} {\bibfnamefont {C.~C.}\ \bibnamefont
  {Chang}} \emph {et~al.} (\bibinfo {collaboration} {CalLat}),\ }\bibfield
  {title} {\enquote {\bibinfo {title} {A per-cent-level determination of the
  nucleon axial coupling from quantum chromodynamics},}\ }\href {\doibase
  10.1038/s41586-018-0161-8} {\bibfield  {journal} {\bibinfo  {journal}
  {Nature}\ }\textbf {\bibinfo {volume} {558}},\ \bibinfo {pages} {91--94}
  (\bibinfo {year} {2018}{\natexlab{a}})},\ \Eprint
  {http://arxiv.org/abs/1805.12130} {arXiv:1805.12130 [hep-lat]} \BibitemShut
  {NoStop}%
\bibitem [{\citenamefont {Ishikawa}\ \emph {et~al.}(2018)\citenamefont
  {Ishikawa}, \citenamefont {Kuramashi}, \citenamefont {Sasaki}, \citenamefont
  {Tsukamoto}, \citenamefont {Ukawa},\ and\ \citenamefont
  {Yamazaki}}]{Ishikawa:2018rew}%
  \BibitemOpen
  \bibfield  {author} {\bibinfo {author} {\bibfnamefont {Ken-Ichi}\
  \bibnamefont {Ishikawa}}, \bibinfo {author} {\bibfnamefont {Yoshinobu}\
  \bibnamefont {Kuramashi}}, \bibinfo {author} {\bibfnamefont {Shoichi}\
  \bibnamefont {Sasaki}}, \bibinfo {author} {\bibfnamefont {Natsuki}\
  \bibnamefont {Tsukamoto}}, \bibinfo {author} {\bibfnamefont {Akira}\
  \bibnamefont {Ukawa}}, \ and\ \bibinfo {author} {\bibfnamefont {Takeshi}\
  \bibnamefont {Yamazaki}} (\bibinfo {collaboration} {PACS}),\ }\bibfield
  {title} {\enquote {\bibinfo {title} {Nucleon form factors on a large volume
  lattice near the physical point in 2+1 flavor {QCD}},}\ }\href {\doibase
  10.1103/PhysRevD.98.074510} {\bibfield  {journal} {\bibinfo  {journal} {Phys.
  Rev.}\ }\textbf {\bibinfo {volume} {D98}},\ \bibinfo {pages} {074510}
  (\bibinfo {year} {2018})},\ \Eprint {http://arxiv.org/abs/1807.03974}
  {arXiv:1807.03974 [hep-lat]} \BibitemShut {NoStop}%
\bibitem [{\citenamefont {Liang}\ \emph
  {et~al.}(2018{\natexlab{a}})\citenamefont {Liang}, \citenamefont {Yang},
  \citenamefont {Draper}, \citenamefont {Gong},\ and\ \citenamefont
  {Liu}}]{Liang:2018pis}%
  \BibitemOpen
  \bibfield  {author} {\bibinfo {author} {\bibfnamefont {Jian}\ \bibnamefont
  {Liang}}, \bibinfo {author} {\bibfnamefont {Yi-Bo}\ \bibnamefont {Yang}},
  \bibinfo {author} {\bibfnamefont {Terrence}\ \bibnamefont {Draper}}, \bibinfo
  {author} {\bibfnamefont {Ming}\ \bibnamefont {Gong}}, \ and\ \bibinfo
  {author} {\bibfnamefont {Keh-Fei}\ \bibnamefont {Liu}} (\bibinfo
  {collaboration} {$\chi$QCD}),\ }\bibfield  {title} {\enquote {\bibinfo
  {title} {{Quark spins and anomalous Ward identity}},}\ }\href {\doibase
  10.1103/PhysRevD.98.074505} {\bibfield  {journal} {\bibinfo  {journal} {Phys.
  Rev.}\ }\textbf {\bibinfo {volume} {D98}},\ \bibinfo {pages} {074505}
  (\bibinfo {year} {2018}{\natexlab{a}})},\ \Eprint
  {http://arxiv.org/abs/1806.08366} {arXiv:1806.08366 [hep-ph]} \BibitemShut
  {NoStop}%
\bibitem [{\citenamefont {Ottnad}\ \emph {et~al.}(2018)\citenamefont {Ottnad},
  \citenamefont {Harris}, \citenamefont {Meyer}, \citenamefont {von Hippel},
  \citenamefont {Wilhelm},\ and\ \citenamefont {Wittig}}]{Ottnad:2018fri}%
  \BibitemOpen
  \bibfield  {author} {\bibinfo {author} {\bibfnamefont {Konstantin}\
  \bibnamefont {Ottnad}}, \bibinfo {author} {\bibfnamefont {Tim}\ \bibnamefont
  {Harris}}, \bibinfo {author} {\bibfnamefont {Harvey}\ \bibnamefont {Meyer}},
  \bibinfo {author} {\bibfnamefont {Georg}\ \bibnamefont {von Hippel}},
  \bibinfo {author} {\bibfnamefont {Jonas}\ \bibnamefont {Wilhelm}}, \ and\
  \bibinfo {author} {\bibfnamefont {Hartmut}\ \bibnamefont {Wittig}},\
  }\bibfield  {title} {\enquote {\bibinfo {title} {Nucleon charges and quark
  momentum fraction with {$N_f=2+1$ Wilson} fermions},}\ }\href@noop {}
  {\bibfield  {journal} {\bibinfo  {journal} {PoS}\ }\textbf {\bibinfo {volume}
  {LATTICE2018}},\ \bibinfo {pages} {129} (\bibinfo {year} {2018})},\ \Eprint
  {http://arxiv.org/abs/1809.10638} {arXiv:1809.10638 [hep-lat]} \BibitemShut
  {NoStop}%
\bibitem [{\citenamefont {Bali}\ \emph {et~al.}(2019)\citenamefont {Bali},
  \citenamefont {Collins}, \citenamefont {Gruber}, \citenamefont {Schäfer},
  \citenamefont {Wein},\ and\ \citenamefont {Wurm}}]{Bali:2018qus}%
  \BibitemOpen
  \bibfield  {author} {\bibinfo {author} {\bibfnamefont {G.~S.}\ \bibnamefont
  {Bali}}, \bibinfo {author} {\bibfnamefont {S.}~\bibnamefont {Collins}},
  \bibinfo {author} {\bibfnamefont {M.}~\bibnamefont {Gruber}}, \bibinfo
  {author} {\bibfnamefont {A.}~\bibnamefont {Schäfer}}, \bibinfo {author}
  {\bibfnamefont {P.}~\bibnamefont {Wein}}, \ and\ \bibinfo {author}
  {\bibfnamefont {T.}~\bibnamefont {Wurm}} (\bibinfo {collaboration} {RQCD}),\
  }\bibfield  {title} {\enquote {\bibinfo {title} {Solving the {PCAC} puzzle
  for nucleon axial and pseudoscalar form factors},}\ }\href {\doibase
  10.1016/j.physletb.2018.12.053} {\bibfield  {journal} {\bibinfo  {journal}
  {Phys. Lett.}\ }\textbf {\bibinfo {volume} {B789}},\ \bibinfo {pages}
  {666--674} (\bibinfo {year} {2019})},\ \Eprint
  {http://arxiv.org/abs/1810.05569} {arXiv:1810.05569 [hep-lat]} \BibitemShut
  {NoStop}%
\bibitem [{\citenamefont {Shintani}\ \emph {et~al.}(2019)\citenamefont
  {Shintani}, \citenamefont {Ishikawa}, \citenamefont {Kuramashi},
  \citenamefont {Sasaki},\ and\ \citenamefont {Yamazaki}}]{Shintani:2018ozy}%
  \BibitemOpen
  \bibfield  {author} {\bibinfo {author} {\bibfnamefont {Eigo}\ \bibnamefont
  {Shintani}}, \bibinfo {author} {\bibfnamefont {Ken-Ichi}\ \bibnamefont
  {Ishikawa}}, \bibinfo {author} {\bibfnamefont {Yoshinobu}\ \bibnamefont
  {Kuramashi}}, \bibinfo {author} {\bibfnamefont {Shoichi}\ \bibnamefont
  {Sasaki}}, \ and\ \bibinfo {author} {\bibfnamefont {Takeshi}\ \bibnamefont
  {Yamazaki}} (\bibinfo {collaboration} {PACS}),\ }\bibfield  {title} {\enquote
  {\bibinfo {title} {Nucleon form factors and root-mean-square radii on a
  $(10.8~\text{fm})^4$ lattice at the physical point},}\ }\href {\doibase
  10.1103/PhysRevD.99.014510} {\bibfield  {journal} {\bibinfo  {journal} {Phys.
  Rev.}\ }\textbf {\bibinfo {volume} {D99}},\ \bibinfo {pages} {014510}
  (\bibinfo {year} {2019})},\ \Eprint {http://arxiv.org/abs/1811.07292}
  {arXiv:1811.07292 [hep-lat]} \BibitemShut {NoStop}%
\bibitem [{\citenamefont {Constantinou}\ \emph {et~al.}(2018)\citenamefont
  {Constantinou}, \citenamefont {Alexandrou}, \citenamefont {Bacchio},
  \citenamefont {Hadjiyiannakou}, \citenamefont {Jansen}, \citenamefont
  {Koutsou},\ and\ \citenamefont {Vaquero Aviles-Casco}}]{Martha:2018lft}%
  \BibitemOpen
  \bibfield  {author} {\bibinfo {author} {\bibfnamefont {Martha}\ \bibnamefont
  {Constantinou}}, \bibinfo {author} {\bibfnamefont {Constantia}\ \bibnamefont
  {Alexandrou}}, \bibinfo {author} {\bibfnamefont {Simone}\ \bibnamefont
  {Bacchio}}, \bibinfo {author} {\bibfnamefont {Kyriakos}\ \bibnamefont
  {Hadjiyiannakou}}, \bibinfo {author} {\bibfnamefont {Karl}\ \bibnamefont
  {Jansen}}, \bibinfo {author} {\bibfnamefont {Giannis}\ \bibnamefont
  {Koutsou}}, \ and\ \bibinfo {author} {\bibfnamefont {Alejandro}\ \bibnamefont
  {Vaquero Aviles-Casco}},\ }\bibfield  {title} {\enquote {\bibinfo {title}
  {\href{https://indico.fnal.gov/event/15949/session/8/contribution/199}{Nucleon
  form factors from {$N_f=2+1+1$} twisted-mass fermions at the physical
  point}},}\ }\href@noop {} {\bibfield  {journal} {\bibinfo  {journal} {PoS}\
  }\textbf {\bibinfo {volume} {LATTICE2018}},\ \bibinfo {pages} {142} (\bibinfo
  {year} {2018})}\BibitemShut {NoStop}%
\bibitem [{\citenamefont {Jang}\ \emph
  {et~al.}(2018{\natexlab{b}})\citenamefont {Jang}, \citenamefont
  {Bhattacharya}, \citenamefont {Gupta}, \citenamefont {Lin},\ and\
  \citenamefont {Yoon}}]{Jang:2018djx}%
  \BibitemOpen
  \bibfield  {author} {\bibinfo {author} {\bibfnamefont {Yong-Chull}\
  \bibnamefont {Jang}}, \bibinfo {author} {\bibfnamefont {Tanmoy}\ \bibnamefont
  {Bhattacharya}}, \bibinfo {author} {\bibfnamefont {Rajan}\ \bibnamefont
  {Gupta}}, \bibinfo {author} {\bibfnamefont {Huey-Wen}\ \bibnamefont {Lin}}, \
  and\ \bibinfo {author} {\bibfnamefont {Boram}\ \bibnamefont {Yoon}} (\bibinfo
  {collaboration} {PNDME}),\ }\bibfield  {title} {\enquote {\bibinfo {title}
  {Updates on nucleon form factors from clover-on-{HISQ} lattice
  formulation},}\ }\href@noop {} {\bibfield  {journal} {\bibinfo  {journal}
  {PoS}\ }\textbf {\bibinfo {volume} {LATTICE2018}},\ \bibinfo {pages} {123}
  (\bibinfo {year} {2018}{\natexlab{b}})},\ \Eprint
  {http://arxiv.org/abs/1901.00060} {arXiv:1901.00060 [hep-lat]} \BibitemShut
  {NoStop}%
\bibitem [{\citenamefont {Bernard}\ \emph {et~al.}(2002)\citenamefont
  {Bernard}, \citenamefont {Elouadrhiri},\ and\ \citenamefont
  {Meissner}}]{Bernard:2001rs}%
  \BibitemOpen
  \bibfield  {author} {\bibinfo {author} {\bibfnamefont {Veronique}\
  \bibnamefont {Bernard}}, \bibinfo {author} {\bibfnamefont {Latifa}\
  \bibnamefont {Elouadrhiri}}, \ and\ \bibinfo {author} {\bibfnamefont
  {Ulf-G.}\ \bibnamefont {Meissner}},\ }\bibfield  {title} {\enquote {\bibinfo
  {title} {Axial structure of the nucleon},}\ }\href {\doibase
  10.1088/0954-3899/28/1/201} {\bibfield  {journal} {\bibinfo  {journal} {J.
  Phys.}\ }\textbf {\bibinfo {volume} {G28}},\ \bibinfo {pages} {R1--R35}
  (\bibinfo {year} {2002})},\ \Eprint {http://arxiv.org/abs/hep-ph/0107088}
  {arXiv:hep-ph/0107088 [hep-ph]} \BibitemShut {NoStop}%
\bibitem [{\citenamefont {Gupta}\ \emph {et~al.}(2018)\citenamefont {Gupta},
  \citenamefont {Jang}, \citenamefont {Yoon}, \citenamefont {Lin},
  \citenamefont {Cirigliano},\ and\ \citenamefont
  {Bhattacharya}}]{Gupta:2018qil}%
  \BibitemOpen
  \bibfield  {author} {\bibinfo {author} {\bibfnamefont {Rajan}\ \bibnamefont
  {Gupta}}, \bibinfo {author} {\bibfnamefont {Yong-Chull}\ \bibnamefont
  {Jang}}, \bibinfo {author} {\bibfnamefont {Boram}\ \bibnamefont {Yoon}},
  \bibinfo {author} {\bibfnamefont {Huey-Wen}\ \bibnamefont {Lin}}, \bibinfo
  {author} {\bibfnamefont {Vincenzo}\ \bibnamefont {Cirigliano}}, \ and\
  \bibinfo {author} {\bibfnamefont {Tanmoy}\ \bibnamefont {Bhattacharya}}
  (\bibinfo {collaboration} {PNDME}),\ }\bibfield  {title} {\enquote {\bibinfo
  {title} {Isovector charges of the nucleon from 2+1+1-flavor lattice {QCD}},}\
  }\href {\doibase 10.1103/PhysRevD.98.034503} {\bibfield  {journal} {\bibinfo
  {journal} {Phys. Rev.}\ }\textbf {\bibinfo {volume} {D98}},\ \bibinfo {pages}
  {034503} (\bibinfo {year} {2018})},\ \Eprint
  {http://arxiv.org/abs/1806.09006} {arXiv:1806.09006 [hep-lat]} \BibitemShut
  {NoStop}%
\bibitem [{\citenamefont {Aoki}\ \emph {et~al.}(2017)\citenamefont {Aoki} \emph
  {et~al.}}]{Aoki:2016frl}%
  \BibitemOpen
  \bibfield  {author} {\bibinfo {author} {\bibfnamefont {S.}~\bibnamefont
  {Aoki}} \emph {et~al.} (\bibinfo {collaboration} {Flavor Lattice Averaging
  Group}),\ }\bibfield  {title} {\enquote {\bibinfo {title} {Review of lattice
  results concerning low-energy particle physics},}\ }\href {\doibase
  10.1140/epjc/s10052-016-4509-7} {\bibfield  {journal} {\bibinfo  {journal}
  {Eur. Phys. J.}\ }\textbf {\bibinfo {volume} {C77}},\ \bibinfo {pages} {112}
  (\bibinfo {year} {2017})},\ \Eprint {http://arxiv.org/abs/1607.00299}
  {arXiv:1607.00299 [hep-lat]} \BibitemShut {NoStop}%
\bibitem [{\citenamefont {Aoki}\ \emph {et~al.}()\citenamefont {Aoki} \emph
  {et~al.}}]{Aoki:2019cca}%
  \BibitemOpen
  \bibfield  {author} {\bibinfo {author} {\bibfnamefont {S.}~\bibnamefont
  {Aoki}} \emph {et~al.} (\bibinfo {collaboration} {Flavor Lattice Averaging
  Group}),\ }\bibfield  {title} {\enquote {\bibinfo {title} {{FLAG} review
  2019},}\ }\href@noop {} {\ }\Eprint {http://arxiv.org/abs/1902.08191}
  {arXiv:1902.08191 [hep-lat]} \BibitemShut {NoStop}%
\bibitem [{FLA()}]{FLAGweb}%
  \BibitemOpen
  \href@noop {} {}\bibinfo {note} {{updates} can be found on the
  \href{http://flag.unibe.ch}{FLAG website}}\BibitemShut {NoStop}%
\bibitem [{\citenamefont {Bhattacharya}\ \emph {et~al.}(2015)\citenamefont
  {Bhattacharya}, \citenamefont {Cirigliano}, \citenamefont {Cohen},
  \citenamefont {Gupta}, \citenamefont {Joseph}, \citenamefont {Lin},\ and\
  \citenamefont {Yoon}}]{Bhattacharya:2015wna}%
  \BibitemOpen
  \bibfield  {author} {\bibinfo {author} {\bibfnamefont {Tanmoy}\ \bibnamefont
  {Bhattacharya}}, \bibinfo {author} {\bibfnamefont {Vincenzo}\ \bibnamefont
  {Cirigliano}}, \bibinfo {author} {\bibfnamefont {Saul}\ \bibnamefont
  {Cohen}}, \bibinfo {author} {\bibfnamefont {Rajan}\ \bibnamefont {Gupta}},
  \bibinfo {author} {\bibfnamefont {Anosh}\ \bibnamefont {Joseph}}, \bibinfo
  {author} {\bibfnamefont {Huey-Wen}\ \bibnamefont {Lin}}, \ and\ \bibinfo
  {author} {\bibfnamefont {Boram}\ \bibnamefont {Yoon}} (\bibinfo
  {collaboration} {PNDME}),\ }\bibfield  {title} {\enquote {\bibinfo {title}
  {Isovector and isoscalar tensor charges of the nucleon from lattice {QCD}},}\
  }\href {\doibase 10.1103/PhysRevD.92.094511} {\bibfield  {journal} {\bibinfo
  {journal} {Phys. Rev.}\ }\textbf {\bibinfo {volume} {D92}},\ \bibinfo {pages}
  {094511} (\bibinfo {year} {2015})},\ \Eprint
  {http://arxiv.org/abs/1506.06411} {arXiv:1506.06411 [hep-lat]} \BibitemShut
  {NoStop}%
\bibitem [{\citenamefont {Arzumanov}\ \emph {et~al.}(2015)\citenamefont
  {Arzumanov}, \citenamefont {Bondarenko}, \citenamefont {Chernyavsky},
  \citenamefont {Geltenbort}, \citenamefont {Morozov}, \citenamefont
  {Nesvizhevsky}, \citenamefont {Panin},\ and\ \citenamefont
  {Strepetov}}]{Arzumanov:2015tea}%
  \BibitemOpen
  \bibfield  {author} {\bibinfo {author} {\bibfnamefont {S.}~\bibnamefont
  {Arzumanov}}, \bibinfo {author} {\bibfnamefont {L.}~\bibnamefont
  {Bondarenko}}, \bibinfo {author} {\bibfnamefont {S.}~\bibnamefont
  {Chernyavsky}}, \bibinfo {author} {\bibfnamefont {P.}~\bibnamefont
  {Geltenbort}}, \bibinfo {author} {\bibfnamefont {V.}~\bibnamefont {Morozov}},
  \bibinfo {author} {\bibfnamefont {V.~V.}\ \bibnamefont {Nesvizhevsky}},
  \bibinfo {author} {\bibfnamefont {{\relax Yu}.}~\bibnamefont {Panin}}, \ and\
  \bibinfo {author} {\bibfnamefont {A.}~\bibnamefont {Strepetov}},\ }\bibfield
  {title} {\enquote {\bibinfo {title} {{A measurement of the neutron lifetime
  using the method of storage of ultracold neutrons and detection of
  inelastically up-scattered neutrons}},}\ }\href {\doibase
  10.1016/j.physletb.2015.04.021} {\bibfield  {journal} {\bibinfo  {journal}
  {Phys. Lett.}\ }\textbf {\bibinfo {volume} {B745}},\ \bibinfo {pages}
  {79--89} (\bibinfo {year} {2015})}\BibitemShut {NoStop}%
\bibitem [{\citenamefont {Yue}\ \emph {et~al.}(2013)\citenamefont {Yue},
  \citenamefont {Dewey}, \citenamefont {Gilliam}, \citenamefont {Greene},
  \citenamefont {Laptev}, \citenamefont {Nico}, \citenamefont {Snow},\ and\
  \citenamefont {Wietfeldt}}]{Yue:2013qrc}%
  \BibitemOpen
  \bibfield  {author} {\bibinfo {author} {\bibfnamefont {A.~T.}\ \bibnamefont
  {Yue}}, \bibinfo {author} {\bibfnamefont {M.~S.}\ \bibnamefont {Dewey}},
  \bibinfo {author} {\bibfnamefont {D.~M.}\ \bibnamefont {Gilliam}}, \bibinfo
  {author} {\bibfnamefont {G.~L.}\ \bibnamefont {Greene}}, \bibinfo {author}
  {\bibfnamefont {A.~B.}\ \bibnamefont {Laptev}}, \bibinfo {author}
  {\bibfnamefont {J.~S.}\ \bibnamefont {Nico}}, \bibinfo {author}
  {\bibfnamefont {W.~M.}\ \bibnamefont {Snow}}, \ and\ \bibinfo {author}
  {\bibfnamefont {F.~E.}\ \bibnamefont {Wietfeldt}},\ }\bibfield  {title}
  {\enquote {\bibinfo {title} {Improved determination of the neutron
  lifetime},}\ }\href {\doibase 10.1103/PhysRevLett.111.222501} {\bibfield
  {journal} {\bibinfo  {journal} {Phys. Rev. Lett.}\ }\textbf {\bibinfo
  {volume} {111}},\ \bibinfo {pages} {222501} (\bibinfo {year} {2013})},\
  \Eprint {http://arxiv.org/abs/1309.2623} {arXiv:1309.2623 [nucl-ex]}
  \BibitemShut {NoStop}%
\bibitem [{\citenamefont {Meyer}\ \emph
  {et~al.}(2016{\natexlab{b}})\citenamefont {Meyer}, \citenamefont {Hill},
  \citenamefont {Kronfeld}, \citenamefont {Li},\ and\ \citenamefont
  {Simone}}]{Meyer:2016kwb}%
  \BibitemOpen
  \bibfield  {author} {\bibinfo {author} {\bibfnamefont {Aaron~S.}\
  \bibnamefont {Meyer}}, \bibinfo {author} {\bibfnamefont {Richard~J.}\
  \bibnamefont {Hill}}, \bibinfo {author} {\bibfnamefont {Andreas~S.}\
  \bibnamefont {Kronfeld}}, \bibinfo {author} {\bibfnamefont {Ruizi}\
  \bibnamefont {Li}}, \ and\ \bibinfo {author} {\bibfnamefont {James~N.}\
  \bibnamefont {Simone}},\ }\bibfield  {title} {\enquote {\bibinfo {title}
  {Calculation of the nucleon axial form factor using staggered lattice
  {QCD}},}\ }\href {\doibase 10.22323/1.256.0179} {\bibfield  {journal}
  {\bibinfo  {journal} {PoS}\ }\textbf {\bibinfo {volume} {LATTICE2016}},\
  \bibinfo {pages} {179} (\bibinfo {year} {2016}{\natexlab{b}})},\ \Eprint
  {http://arxiv.org/abs/1610.04593} {arXiv:1610.04593 [hep-lat]} \BibitemShut
  {NoStop}%
\bibitem [{\citenamefont {Meiman}(1963)}]{Meiman:1963:zex}%
  \BibitemOpen
  \bibfield  {author} {\bibinfo {author} {\bibfnamefont {N.~N.}\ \bibnamefont
  {Meiman}},\ }\bibfield  {title} {\enquote {\bibinfo {title} {Analytic
  expressions for upper limits of coupling constants in quantum field
  theory},}\ }\href@noop {} {\bibfield  {journal} {\bibinfo  {journal} {Sov.
  Phys. JETP}\ }\textbf {\bibinfo {volume} {17}},\ \bibinfo {pages} {830}
  (\bibinfo {year} {1963})},\ \bibinfo {note}
  {[\href{http://www.jetp.ac.ru/cgi-bin/e/index/e/17/4/p830?a=list}{Zh. Eksp.
  Teor. Fiz.~\textbf{44}, 1228 (1963)}]}\BibitemShut {NoStop}%
\bibitem [{\citenamefont {Hill}\ and\ \citenamefont {Paz}(2010)}]{Hill:2010yb}%
  \BibitemOpen
  \bibfield  {author} {\bibinfo {author} {\bibfnamefont {Richard~J.}\
  \bibnamefont {Hill}}\ and\ \bibinfo {author} {\bibfnamefont {Gil}\
  \bibnamefont {Paz}},\ }\bibfield  {title} {\enquote {\bibinfo {title} {Model
  independent extraction of the proton charge radius from electron
  scattering},}\ }\href {\doibase 10.1103/PhysRevD.82.113005} {\bibfield
  {journal} {\bibinfo  {journal} {Phys. Rev.}\ }\textbf {\bibinfo {volume}
  {D82}},\ \bibinfo {pages} {113005} (\bibinfo {year} {2010})},\ \Eprint
  {http://arxiv.org/abs/1008.4619} {arXiv:1008.4619 [hep-ph]} \BibitemShut
  {NoStop}%
\bibitem [{\citenamefont {Hill}(2006)}]{Hill:2006ub}%
  \BibitemOpen
  \bibfield  {author} {\bibinfo {author} {\bibfnamefont {Richard~J.}\
  \bibnamefont {Hill}},\ }\bibfield  {title} {\enquote {\bibinfo {title} {The
  modern description of semileptonic meson form factors},}\ }\href@noop {}
  {\bibfield  {journal} {\bibinfo  {journal} {eConf}\ }\textbf {\bibinfo
  {volume} {C060409}},\ \bibinfo {pages} {027} (\bibinfo {year} {2006})},\
  \Eprint {http://arxiv.org/abs/hep-ph/0606023} {arXiv:hep-ph/0606023 [hep-ph]}
  \BibitemShut {NoStop}%
\bibitem [{\citenamefont {Bernard}\ \emph {et~al.}(2009)\citenamefont {Bernard}
  \emph {et~al.}}]{Bernard:2008dn}%
  \BibitemOpen
  \bibfield  {author} {\bibinfo {author} {\bibfnamefont {C.}~\bibnamefont
  {Bernard}} \emph {et~al.} (\bibinfo {collaboration} {Fermilab Lattice,
  MILC}),\ }\bibfield  {title} {\enquote {\bibinfo {title} {The {$\bar{B}\to
  D^{*}\ell \bar{\nu}$} form factor at zero recoil from three-flavor lattice
  {QCD}: A~model-independent determination of {$|V_{cb}|$}},}\ }\href {\doibase
  10.1103/PhysRevD.79.014506} {\bibfield  {journal} {\bibinfo  {journal} {Phys.
  Rev.}\ }\textbf {\bibinfo {volume} {D79}},\ \bibinfo {pages} {014506}
  (\bibinfo {year} {2009})},\ \Eprint {http://arxiv.org/abs/0808.2519}
  {arXiv:0808.2519 [hep-lat]} \BibitemShut {NoStop}%
\bibitem [{\citenamefont {Bailey}\ \emph {et~al.}(2009)\citenamefont {Bailey}
  \emph {et~al.}}]{Bailey:2008wp}%
  \BibitemOpen
  \bibfield  {author} {\bibinfo {author} {\bibfnamefont {Jon~A.}\ \bibnamefont
  {Bailey}} \emph {et~al.} (\bibinfo {collaboration} {Fermilab Lattice,
  MILC}),\ }\bibfield  {title} {\enquote {\bibinfo {title} {The
  {$B\to\pi\ell\nu$} semileptonic form factor from three-flavor lattice {QCD}:
  A~model-independent determination of {$|V_{ub}|$}},}\ }\href {\doibase
  10.1103/PhysRevD.79.054507} {\bibfield  {journal} {\bibinfo  {journal} {Phys.
  Rev.}\ }\textbf {\bibinfo {volume} {D79}},\ \bibinfo {pages} {054507}
  (\bibinfo {year} {2009})},\ \Eprint {http://arxiv.org/abs/0811.3640}
  {arXiv:0811.3640 [hep-lat]} \BibitemShut {NoStop}%
\bibitem [{\citenamefont {de~Gouvêa}\ and\ \citenamefont
  {Kelly}(2016)}]{deGouvea:2015ndi}%
  \BibitemOpen
  \bibfield  {author} {\bibinfo {author} {\bibfnamefont {André}\ \bibnamefont
  {de~Gouvêa}}\ and\ \bibinfo {author} {\bibfnamefont {Kevin~J.}\ \bibnamefont
  {Kelly}},\ }\bibfield  {title} {\enquote {\bibinfo {title} {Non-standard
  neutrino interactions at {DUNE}},}\ }\href {\doibase
  10.1016/j.nuclphysb.2016.03.013} {\bibfield  {journal} {\bibinfo  {journal}
  {Nucl. Phys.}\ }\textbf {\bibinfo {volume} {B908}},\ \bibinfo {pages}
  {318--335} (\bibinfo {year} {2016})},\ \Eprint
  {http://arxiv.org/abs/1511.05562} {arXiv:1511.05562 [hep-ph]} \BibitemShut
  {NoStop}%
\bibitem [{\citenamefont {Heeck}\ \emph {et~al.}(2018)\citenamefont {Heeck},
  \citenamefont {Lindner}, \citenamefont {Rodejohann},\ and\ \citenamefont
  {Vogl}}]{Heeck:2018nzc}%
  \BibitemOpen
  \bibfield  {author} {\bibinfo {author} {\bibfnamefont {Julian}\ \bibnamefont
  {Heeck}}, \bibinfo {author} {\bibfnamefont {Manfred}\ \bibnamefont
  {Lindner}}, \bibinfo {author} {\bibfnamefont {Werner}\ \bibnamefont
  {Rodejohann}}, \ and\ \bibinfo {author} {\bibfnamefont {Stefan}\ \bibnamefont
  {Vogl}},\ }\bibfield  {title} {\enquote {\bibinfo {title} {Non-standard
  neutrino interactions and neutral gauge bosons},}\ }\href@noop {} {\
  (\bibinfo {year} {2018})},\ \Eprint {http://arxiv.org/abs/1812.04067}
  {arXiv:1812.04067 [hep-ph]} \BibitemShut {NoStop}%
\bibitem [{\citenamefont {Armstrong}\ and\ \citenamefont
  {McKeown}(2012)}]{Armstrong:2012bi}%
  \BibitemOpen
  \bibfield  {author} {\bibinfo {author} {\bibfnamefont {D.~S.}\ \bibnamefont
  {Armstrong}}\ and\ \bibinfo {author} {\bibfnamefont {R.~D.}\ \bibnamefont
  {McKeown}},\ }\bibfield  {title} {\enquote {\bibinfo {title}
  {Parity-violating electron scattering and the electric and magnetic strange
  form factors of the nucleon},}\ }\href {\doibase
  10.1146/annurev-nucl-102010-130419} {\bibfield  {journal} {\bibinfo
  {journal} {Annu. Rev. Nucl. Part. Sci.}\ }\textbf {\bibinfo {volume} {62}},\
  \bibinfo {pages} {337--359} (\bibinfo {year} {2012})},\ \Eprint
  {http://arxiv.org/abs/1207.5238} {arXiv:1207.5238 [nucl-ex]} \BibitemShut
  {NoStop}%
\bibitem [{\citenamefont {Ye}\ \emph {et~al.}(2018)\citenamefont {Ye},
  \citenamefont {Arrington}, \citenamefont {Hill},\ and\ \citenamefont
  {Lee}}]{Ye:2017gyb}%
  \BibitemOpen
  \bibfield  {author} {\bibinfo {author} {\bibfnamefont {Zhihong}\ \bibnamefont
  {Ye}}, \bibinfo {author} {\bibfnamefont {John}\ \bibnamefont {Arrington}},
  \bibinfo {author} {\bibfnamefont {Richard~J.}\ \bibnamefont {Hill}}, \ and\
  \bibinfo {author} {\bibfnamefont {Gabriel}\ \bibnamefont {Lee}},\ }\bibfield
  {title} {\enquote {\bibinfo {title} {Proton and neutron electromagnetic form
  factors and uncertainties},}\ }\href {\doibase
  10.1016/j.physletb.2017.11.023} {\bibfield  {journal} {\bibinfo  {journal}
  {Phys. Lett.}\ }\textbf {\bibinfo {volume} {B777}},\ \bibinfo {pages} {8--15}
  (\bibinfo {year} {2018})},\ \Eprint {http://arxiv.org/abs/1707.09063}
  {arXiv:1707.09063 [nucl-ex]} \BibitemShut {NoStop}%
\bibitem [{\citenamefont {Sufian}(2017)}]{Sufian:2016vso}%
  \BibitemOpen
  \bibfield  {author} {\bibinfo {author} {\bibfnamefont {Raza~Sabbir}\
  \bibnamefont {Sufian}},\ }\bibfield  {title} {\enquote {\bibinfo {title}
  {Neutral weak form factors of proton and neutron},}\ }\href {\doibase
  10.1103/PhysRevD.96.093007} {\bibfield  {journal} {\bibinfo  {journal} {Phys.
  Rev.}\ }\textbf {\bibinfo {volume} {D96}},\ \bibinfo {pages} {093007}
  (\bibinfo {year} {2017})},\ \Eprint {http://arxiv.org/abs/1611.07031}
  {arXiv:1611.07031 [hep-ph]} \BibitemShut {NoStop}%
\bibitem [{\citenamefont {Garvey}\ \emph
  {et~al.}(1993{\natexlab{a}})\citenamefont {Garvey}, \citenamefont {Kolbe},
  \citenamefont {Langanke},\ and\ \citenamefont {Krewald}}]{Garvey:1993sg}%
  \BibitemOpen
  \bibfield  {author} {\bibinfo {author} {\bibfnamefont {G.}~\bibnamefont
  {Garvey}}, \bibinfo {author} {\bibfnamefont {E.}~\bibnamefont {Kolbe}},
  \bibinfo {author} {\bibfnamefont {K.}~\bibnamefont {Langanke}}, \ and\
  \bibinfo {author} {\bibfnamefont {S.}~\bibnamefont {Krewald}},\ }\bibfield
  {title} {\enquote {\bibinfo {title} {{Role of strange quarks in quasielastic
  neutrino scattering}},}\ }\href {\doibase 10.1103/PhysRevC.48.1919}
  {\bibfield  {journal} {\bibinfo  {journal} {Phys. Rev.}\ }\textbf {\bibinfo
  {volume} {C48}},\ \bibinfo {pages} {1919--1925} (\bibinfo {year}
  {1993}{\natexlab{a}})}\BibitemShut {NoStop}%
\bibitem [{\citenamefont {Garvey}\ \emph
  {et~al.}(1993{\natexlab{b}})\citenamefont {Garvey}, \citenamefont {Louis},\
  and\ \citenamefont {White}}]{Garvey:1992cg}%
  \BibitemOpen
  \bibfield  {author} {\bibinfo {author} {\bibfnamefont {G.~T.}\ \bibnamefont
  {Garvey}}, \bibinfo {author} {\bibfnamefont {W.~C.}\ \bibnamefont {Louis}}, \
  and\ \bibinfo {author} {\bibfnamefont {D.~H.}\ \bibnamefont {White}},\
  }\bibfield  {title} {\enquote {\bibinfo {title} {Determination of proton
  strange form-factors from $\nu p$ elastic scattering},}\ }\href {\doibase
  10.1103/PhysRevC.48.761} {\bibfield  {journal} {\bibinfo  {journal} {Phys.
  Rev.}\ }\textbf {\bibinfo {volume} {C48}},\ \bibinfo {pages} {761--765}
  (\bibinfo {year} {1993}{\natexlab{b}})}\BibitemShut {NoStop}%
\bibitem [{\citenamefont {Aguilar-Arevalo}\ \emph
  {et~al.}(2010{\natexlab{b}})\citenamefont {Aguilar-Arevalo} \emph
  {et~al.}}]{AguilarArevalo:2010cx}%
  \BibitemOpen
  \bibfield  {author} {\bibinfo {author} {\bibfnamefont {A.~A.}\ \bibnamefont
  {Aguilar-Arevalo}} \emph {et~al.} (\bibinfo {collaboration} {MiniBooNE}),\
  }\bibfield  {title} {\enquote {\bibinfo {title} {Measurement of the neutrino
  neutral-current elastic differential cross section on mineral oil at
  {$E_\nu\sim1$~GeV}},}\ }\href {\doibase 10.1103/PhysRevD.82.092005}
  {\bibfield  {journal} {\bibinfo  {journal} {Phys. Rev.}\ }\textbf {\bibinfo
  {volume} {D82}},\ \bibinfo {pages} {092005} (\bibinfo {year}
  {2010}{\natexlab{b}})},\ \Eprint {http://arxiv.org/abs/1007.4730}
  {arXiv:1007.4730 [hep-ex]} \BibitemShut {NoStop}%
\bibitem [{\citenamefont {Aguilar-Arevalo}\ \emph {et~al.}(2015)\citenamefont
  {Aguilar-Arevalo} \emph {et~al.}}]{Aguilar-Arevalo:2013nkf}%
  \BibitemOpen
  \bibfield  {author} {\bibinfo {author} {\bibfnamefont {A.~A.}\ \bibnamefont
  {Aguilar-Arevalo}} \emph {et~al.} (\bibinfo {collaboration} {MiniBooNE}),\
  }\bibfield  {title} {\enquote {\bibinfo {title} {Measurement of the
  antineutrino neutral-current elastic differential cross section},}\ }\href
  {\doibase 10.1103/PhysRevD.91.012004} {\bibfield  {journal} {\bibinfo
  {journal} {Phys. Rev.}\ }\textbf {\bibinfo {volume} {D91}},\ \bibinfo {pages}
  {012004} (\bibinfo {year} {2015})},\ \Eprint {http://arxiv.org/abs/1309.7257}
  {arXiv:1309.7257 [hep-ex]} \BibitemShut {NoStop}%
\bibitem [{\citenamefont {Fatima}\ \emph {et~al.}(2018)\citenamefont {Fatima},
  \citenamefont {Sajjad~Athar},\ and\ \citenamefont {Singh}}]{Fatima:2018tzs}%
  \BibitemOpen
  \bibfield  {author} {\bibinfo {author} {\bibfnamefont {A.}~\bibnamefont
  {Fatima}}, \bibinfo {author} {\bibfnamefont {M.}~\bibnamefont
  {Sajjad~Athar}}, \ and\ \bibinfo {author} {\bibfnamefont {S.~K.}\
  \bibnamefont {Singh}},\ }\bibfield  {title} {\enquote {\bibinfo {title}
  {Second class currents and {$T$} violation in quasielastic neutrino and
  antineutrino scattering from nucleons},}\ }\href {\doibase
  10.1103/PhysRevD.98.033005} {\bibfield  {journal} {\bibinfo  {journal} {Phys.
  Rev.}\ }\textbf {\bibinfo {volume} {D98}},\ \bibinfo {pages} {033005}
  (\bibinfo {year} {2018})},\ \Eprint {http://arxiv.org/abs/1806.08597}
  {arXiv:1806.08597 [hep-ph]} \BibitemShut {NoStop}%
\bibitem [{\citenamefont {Lin}(2009)}]{Lin:2008rb}%
  \BibitemOpen
  \bibfield  {author} {\bibinfo {author} {\bibfnamefont {Huey-Wen}\
  \bibnamefont {Lin}},\ }\bibfield  {title} {\enquote {\bibinfo {title}
  {Hyperon physics from lattice {QCD}},}\ }\href {\doibase
  10.1016/j.nuclphysbps.2009.01.029} {\bibfield  {journal} {\bibinfo  {journal}
  {Nucl. Phys. Proc. Suppl.}\ }\textbf {\bibinfo {volume} {187}},\ \bibinfo
  {pages} {200--207} (\bibinfo {year} {2009})},\ \Eprint
  {http://arxiv.org/abs/0812.0411} {arXiv:0812.0411 [hep-lat]} \BibitemShut
  {NoStop}%
\bibitem [{\citenamefont {Sasaki}(2014)}]{Sasaki:2014osa}%
  \BibitemOpen
  \bibfield  {author} {\bibinfo {author} {\bibfnamefont {Shoichi}\ \bibnamefont
  {Sasaki}},\ }\bibfield  {title} {\enquote {\bibinfo {title} {Status of
  semileptonic hyperon decays from lattice {QCD} using 2+1 flavor domain wall
  fermions},}\ }\href {\doibase 10.22323/1.187.0388} {\bibfield  {journal}
  {\bibinfo  {journal} {PoS}\ }\textbf {\bibinfo {volume} {LATTICE2013}},\
  \bibinfo {pages} {388} (\bibinfo {year} {2014})}\BibitemShut {NoStop}%
\bibitem [{\citenamefont {Zeller}\ \emph {et~al.}(2002)\citenamefont {Zeller}
  \emph {et~al.}}]{Zeller:2001hh}%
  \BibitemOpen
  \bibfield  {author} {\bibinfo {author} {\bibfnamefont {G.~P.}\ \bibnamefont
  {Zeller}} \emph {et~al.} (\bibinfo {collaboration} {NuTeV}),\ }\bibfield
  {title} {\enquote {\bibinfo {title} {A precise determination of electroweak
  parameters in neutrino nucleon scattering},}\ }\href {\doibase
  10.1103/PhysRevLett.88.091802} {\bibfield  {journal} {\bibinfo  {journal}
  {Phys. Rev. Lett.}\ }\textbf {\bibinfo {volume} {88}},\ \bibinfo {pages}
  {091802} (\bibinfo {year} {2002})},\ \bibinfo {note}
  {\href{http://doi.org/10.1103/PhysRevLett.90.239902}{\textbf{90}, 239902E
  (2003)}},\ \Eprint {http://arxiv.org/abs/hep-ex/0110059}
  {arXiv:hep-ex/0110059 [hep-ex]} \BibitemShut {NoStop}%
\bibitem [{\citenamefont {Londergan}\ and\ \citenamefont
  {Thomas}(2003)}]{Londergan:2003ij}%
  \BibitemOpen
  \bibfield  {author} {\bibinfo {author} {\bibfnamefont {J.~T.}\ \bibnamefont
  {Londergan}}\ and\ \bibinfo {author} {\bibfnamefont {Anthony~William}\
  \bibnamefont {Thomas}},\ }\bibfield  {title} {\enquote {\bibinfo {title}
  {Charge symmetry violation corrections to determination of the {Weinberg}
  angle in neutrino reactions},}\ }\href {\doibase 10.1103/PhysRevD.67.111901}
  {\bibfield  {journal} {\bibinfo  {journal} {Phys. Rev.}\ }\textbf {\bibinfo
  {volume} {D67}},\ \bibinfo {pages} {111901} (\bibinfo {year} {2003})},\
  \Eprint {http://arxiv.org/abs/hep-ph/0303155} {arXiv:hep-ph/0303155 [hep-ph]}
  \BibitemShut {NoStop}%
\bibitem [{\citenamefont {Ding}\ \emph {et~al.}(2005)\citenamefont {Ding},
  \citenamefont {Xu},\ and\ \citenamefont {Ma}}]{Ding:2004dv}%
  \BibitemOpen
  \bibfield  {author} {\bibinfo {author} {\bibfnamefont {Yong}\ \bibnamefont
  {Ding}}, \bibinfo {author} {\bibfnamefont {Rong-Guang}\ \bibnamefont {Xu}}, \
  and\ \bibinfo {author} {\bibfnamefont {Bo-Qiang}\ \bibnamefont {Ma}},\
  }\bibfield  {title} {\enquote {\bibinfo {title} {Effect of asymmetric
  strange-antistrange sea to the {NuTeV} anomaly},}\ }\href {\doibase
  10.1016/j.physletb.2004.12.054} {\bibfield  {journal} {\bibinfo  {journal}
  {Phys. Lett.}\ }\textbf {\bibinfo {volume} {B607}},\ \bibinfo {pages}
  {101--106} (\bibinfo {year} {2005})},\ \Eprint
  {http://arxiv.org/abs/hep-ph/0408292} {arXiv:hep-ph/0408292 [hep-ph]}
  \BibitemShut {NoStop}%
\bibitem [{\citenamefont {Gluck}\ \emph {et~al.}(2005)\citenamefont {Gluck},
  \citenamefont {Jimenez-Delgado},\ and\ \citenamefont {Reya}}]{Gluck:2005xh}%
  \BibitemOpen
  \bibfield  {author} {\bibinfo {author} {\bibfnamefont {M.}~\bibnamefont
  {Gluck}}, \bibinfo {author} {\bibfnamefont {P.}~\bibnamefont
  {Jimenez-Delgado}}, \ and\ \bibinfo {author} {\bibfnamefont {E.}~\bibnamefont
  {Reya}},\ }\bibfield  {title} {\enquote {\bibinfo {title} {Radiatively
  generated isospin violations in the nucleon and the {NuTeV} anomaly},}\
  }\href {\doibase 10.1103/PhysRevLett.95.022002} {\bibfield  {journal}
  {\bibinfo  {journal} {Phys. Rev. Lett.}\ }\textbf {\bibinfo {volume} {95}},\
  \bibinfo {pages} {022002} (\bibinfo {year} {2005})},\ \Eprint
  {http://arxiv.org/abs/hep-ph/0503103} {arXiv:hep-ph/0503103 [hep-ph]}
  \BibitemShut {NoStop}%
\bibitem [{\citenamefont {Eskola}\ and\ \citenamefont
  {Paukkunen}(2006)}]{Eskola:2006ux}%
  \BibitemOpen
  \bibfield  {author} {\bibinfo {author} {\bibfnamefont {K.~J.}\ \bibnamefont
  {Eskola}}\ and\ \bibinfo {author} {\bibfnamefont {H.}~\bibnamefont
  {Paukkunen}},\ }\bibfield  {title} {\enquote {\bibinfo {title} {{NuTeV}
  $\sin^2\theta_w$ anomaly and nuclear parton distributions revisited},}\
  }\href {\doibase 10.1088/1126-6708/2006/06/008} {\bibfield  {journal}
  {\bibinfo  {journal} {JHEP}\ }\textbf {\bibinfo {volume} {06}},\ \bibinfo
  {pages} {008} (\bibinfo {year} {2006})},\ \Eprint
  {http://arxiv.org/abs/hep-ph/0603155} {arXiv:hep-ph/0603155 [hep-ph]}
  \BibitemShut {NoStop}%
\bibitem [{\citenamefont {Cloet}\ \emph {et~al.}(2009)\citenamefont {Cloet},
  \citenamefont {Bentz},\ and\ \citenamefont {Thomas}}]{Cloet:2009qs}%
  \BibitemOpen
  \bibfield  {author} {\bibinfo {author} {\bibfnamefont {I.~C.}\ \bibnamefont
  {Cloet}}, \bibinfo {author} {\bibfnamefont {W.}~\bibnamefont {Bentz}}, \ and\
  \bibinfo {author} {\bibfnamefont {A.~W.}\ \bibnamefont {Thomas}},\ }\bibfield
   {title} {\enquote {\bibinfo {title} {Isovector {EMC} effect explains the
  {NuTeV} anomaly},}\ }\href {\doibase 10.1103/PhysRevLett.102.252301}
  {\bibfield  {journal} {\bibinfo  {journal} {Phys. Rev. Lett.}\ }\textbf
  {\bibinfo {volume} {102}},\ \bibinfo {pages} {252301} (\bibinfo {year}
  {2009})},\ \Eprint {http://arxiv.org/abs/0901.3559} {arXiv:0901.3559
  [nucl-th]} \BibitemShut {NoStop}%
\bibitem [{\citenamefont {Bentz}\ \emph {et~al.}(2010)\citenamefont {Bentz},
  \citenamefont {Cloet}, \citenamefont {Londergan},\ and\ \citenamefont
  {Thomas}}]{Bentz:2009yy}%
  \BibitemOpen
  \bibfield  {author} {\bibinfo {author} {\bibfnamefont {W.}~\bibnamefont
  {Bentz}}, \bibinfo {author} {\bibfnamefont {I.~C.}\ \bibnamefont {Cloet}},
  \bibinfo {author} {\bibfnamefont {J.~T.}\ \bibnamefont {Londergan}}, \ and\
  \bibinfo {author} {\bibfnamefont {A.~W.}\ \bibnamefont {Thomas}},\ }\bibfield
   {title} {\enquote {\bibinfo {title} {Reassessment of the {NuTeV}
  determination of the weak mixing angle},}\ }\href {\doibase
  10.1016/j.physletb.2010.09.001} {\bibfield  {journal} {\bibinfo  {journal}
  {Phys. Lett.}\ }\textbf {\bibinfo {volume} {B693}},\ \bibinfo {pages}
  {462--466} (\bibinfo {year} {2010})},\ \Eprint
  {http://arxiv.org/abs/0908.3198} {arXiv:0908.3198 [nucl-th]} \BibitemShut
  {NoStop}%
\bibitem [{\citenamefont {Davidson}\ \emph {et~al.}(2002)\citenamefont
  {Davidson}, \citenamefont {Forte}, \citenamefont {Gambino}, \citenamefont
  {Rius},\ and\ \citenamefont {Strumia}}]{Davidson:2001ji}%
  \BibitemOpen
  \bibfield  {author} {\bibinfo {author} {\bibfnamefont {S.}~\bibnamefont
  {Davidson}}, \bibinfo {author} {\bibfnamefont {S.}~\bibnamefont {Forte}},
  \bibinfo {author} {\bibfnamefont {P.}~\bibnamefont {Gambino}}, \bibinfo
  {author} {\bibfnamefont {N.}~\bibnamefont {Rius}}, \ and\ \bibinfo {author}
  {\bibfnamefont {A.}~\bibnamefont {Strumia}},\ }\bibfield  {title} {\enquote
  {\bibinfo {title} {Old and new physics interpretations of the {NuTeV}
  anomaly},}\ }\href {\doibase 10.1088/1126-6708/2002/02/037} {\bibfield
  {journal} {\bibinfo  {journal} {JHEP}\ }\textbf {\bibinfo {volume} {02}},\
  \bibinfo {pages} {037} (\bibinfo {year} {2002})},\ \Eprint
  {http://arxiv.org/abs/hep-ph/0112302} {arXiv:hep-ph/0112302 [hep-ph]}
  \BibitemShut {NoStop}%
\bibitem [{\citenamefont {Kretzer}\ \emph {et~al.}(2004)\citenamefont
  {Kretzer}, \citenamefont {Olness}, \citenamefont {Pumplin}, \citenamefont
  {Stump}, \citenamefont {Tung},\ and\ \citenamefont {Reno}}]{Kretzer:2003wy}%
  \BibitemOpen
  \bibfield  {author} {\bibinfo {author} {\bibfnamefont {Stefan}\ \bibnamefont
  {Kretzer}}, \bibinfo {author} {\bibfnamefont {Fredrick}\ \bibnamefont
  {Olness}}, \bibinfo {author} {\bibfnamefont {Jon}\ \bibnamefont {Pumplin}},
  \bibinfo {author} {\bibfnamefont {Daniel}\ \bibnamefont {Stump}}, \bibinfo
  {author} {\bibfnamefont {Wu-Ki}\ \bibnamefont {Tung}}, \ and\ \bibinfo
  {author} {\bibfnamefont {Mary~Hall}\ \bibnamefont {Reno}},\ }\bibfield
  {title} {\enquote {\bibinfo {title} {The parton structure of the nucleon and
  precision determination of the {Weinberg} angle in neutrino scattering},}\
  }\href {\doibase 10.1103/PhysRevLett.93.041802} {\bibfield  {journal}
  {\bibinfo  {journal} {Phys. Rev. Lett.}\ }\textbf {\bibinfo {volume} {93}},\
  \bibinfo {pages} {041802} (\bibinfo {year} {2004})},\ \Eprint
  {http://arxiv.org/abs/hep-ph/0312322} {arXiv:hep-ph/0312322 [hep-ph]}
  \BibitemShut {NoStop}%
\bibitem [{\citenamefont {Lai}\ \emph {et~al.}(2007)\citenamefont {Lai},
  \citenamefont {Nadolsky}, \citenamefont {Pumplin}, \citenamefont {Stump},
  \citenamefont {Tung},\ and\ \citenamefont {Yuan}}]{Lai:2007dq}%
  \BibitemOpen
  \bibfield  {author} {\bibinfo {author} {\bibfnamefont {H.~L.}\ \bibnamefont
  {Lai}}, \bibinfo {author} {\bibfnamefont {Pavel~M.}\ \bibnamefont
  {Nadolsky}}, \bibinfo {author} {\bibfnamefont {J.}~\bibnamefont {Pumplin}},
  \bibinfo {author} {\bibfnamefont {D.}~\bibnamefont {Stump}}, \bibinfo
  {author} {\bibfnamefont {W.~K.}\ \bibnamefont {Tung}}, \ and\ \bibinfo
  {author} {\bibfnamefont {C.-P.}\ \bibnamefont {Yuan}},\ }\bibfield  {title}
  {\enquote {\bibinfo {title} {The strange parton distribution of the nucleon:
  Global analysis and applications},}\ }\href {\doibase
  10.1088/1126-6708/2007/04/089} {\bibfield  {journal} {\bibinfo  {journal}
  {JHEP}\ }\textbf {\bibinfo {volume} {04}},\ \bibinfo {pages} {089} (\bibinfo
  {year} {2007})},\ \Eprint {http://arxiv.org/abs/hep-ph/0702268}
  {arXiv:hep-ph/0702268 [hep-ph]} \BibitemShut {NoStop}%
\bibitem [{\citenamefont {Mason}(2006)}]{Mason:2006qa}%
  \BibitemOpen
  \bibfield  {author} {\bibinfo {author} {\bibfnamefont {David~Alexander}\
  \bibnamefont {Mason}},\ }\emph {\bibinfo {title} {Measurement of the
  strange-antistrange asymmetry at {NLO} in {QCD} from {NuTeV} dimuon data}},\
  \href {\doibase 10.2172/879078} {Ph.D. thesis},\ \bibinfo  {school}
  {University of Oregon} (\bibinfo {year} {2006})\BibitemShut {NoStop}%
\bibitem [{\citenamefont {Leinweber}\ \emph {et~al.}(1993)\citenamefont
  {Leinweber}, \citenamefont {Draper},\ and\ \citenamefont
  {Woloshyn}}]{Leinweber:1992pv}%
  \BibitemOpen
  \bibfield  {author} {\bibinfo {author} {\bibfnamefont {Derek~B.}\
  \bibnamefont {Leinweber}}, \bibinfo {author} {\bibfnamefont {Terrence}\
  \bibnamefont {Draper}}, \ and\ \bibinfo {author} {\bibfnamefont {R.~M.}\
  \bibnamefont {Woloshyn}},\ }\bibfield  {title} {\enquote {\bibinfo {title}
  {Baryon octet to decuplet electromagnetic transitions},}\ }\href {\doibase
  10.1103/PhysRevD.48.2230} {\bibfield  {journal} {\bibinfo  {journal} {Phys.
  Rev.}\ }\textbf {\bibinfo {volume} {D48}},\ \bibinfo {pages} {2230--2249}
  (\bibinfo {year} {1993})},\ \Eprint {http://arxiv.org/abs/hep-lat/9212016}
  {arXiv:hep-lat/9212016 [hep-lat]} \BibitemShut {NoStop}%
\bibitem [{\citenamefont {Alexandrou}\ \emph {et~al.}(2008)\citenamefont
  {Alexandrou}, \citenamefont {Koutsou}, \citenamefont {Neff}, \citenamefont
  {Negele}, \citenamefont {Schroers},\ and\ \citenamefont
  {Tsapalis}}]{Alexandrou:2007dt}%
  \BibitemOpen
  \bibfield  {author} {\bibinfo {author} {\bibfnamefont {C.}~\bibnamefont
  {Alexandrou}}, \bibinfo {author} {\bibfnamefont {G.}~\bibnamefont {Koutsou}},
  \bibinfo {author} {\bibfnamefont {H.}~\bibnamefont {Neff}}, \bibinfo {author}
  {\bibfnamefont {John~W.}\ \bibnamefont {Negele}}, \bibinfo {author}
  {\bibfnamefont {W.}~\bibnamefont {Schroers}}, \ and\ \bibinfo {author}
  {\bibfnamefont {A.}~\bibnamefont {Tsapalis}},\ }\bibfield  {title} {\enquote
  {\bibinfo {title} {Nucleon to {$\Delta$} electromagnetic transition form
  factors in lattice {QCD}},}\ }\href {\doibase 10.1103/PhysRevD.77.085012}
  {\bibfield  {journal} {\bibinfo  {journal} {Phys. Rev.}\ }\textbf {\bibinfo
  {volume} {D77}},\ \bibinfo {pages} {085012} (\bibinfo {year} {2008})},\
  \Eprint {http://arxiv.org/abs/0710.4621} {arXiv:0710.4621 [hep-lat]}
  \BibitemShut {NoStop}%
\bibitem [{\citenamefont {Alexandrou}\ \emph {et~al.}(2007)\citenamefont
  {Alexandrou}, \citenamefont {Koutsou}, \citenamefont {Leontiou},
  \citenamefont {Negele},\ and\ \citenamefont {Tsapalis}}]{Alexandrou:2009vqd}%
  \BibitemOpen
  \bibfield  {author} {\bibinfo {author} {\bibfnamefont {C.}~\bibnamefont
  {Alexandrou}}, \bibinfo {author} {\bibfnamefont {G.}~\bibnamefont {Koutsou}},
  \bibinfo {author} {\bibfnamefont {Th.}\ \bibnamefont {Leontiou}}, \bibinfo
  {author} {\bibfnamefont {John~W.}\ \bibnamefont {Negele}}, \ and\ \bibinfo
  {author} {\bibfnamefont {A.}~\bibnamefont {Tsapalis}},\ }\bibfield  {title}
  {\enquote {\bibinfo {title} {Axial nucleon and nucleon to {$\Delta$} form
  factors and the {Goldberger-Treiman} relations from lattice {QCD}},}\ }\href
  {\doibase 10.1103/PhysRevD.76.094511} {\bibfield  {journal} {\bibinfo
  {journal} {Phys. Rev.}\ }\textbf {\bibinfo {volume} {D76}},\ \bibinfo {pages}
  {094511} (\bibinfo {year} {2007})},\ \bibinfo {note}
  {[\href{http://doi.org/10.1103/PhysRevD.80.099901}{\textbf{D80}, 099901
  (2009)}]},\ \Eprint {http://arxiv.org/abs/0706.3011} {arXiv:0706.3011
  [hep-lat]} \BibitemShut {NoStop}%
\bibitem [{\citenamefont {Aguilar-Arevalo}\ \emph {et~al.}(2018)\citenamefont
  {Aguilar-Arevalo} \emph {et~al.}}]{Aguilar-Arevalo:2018gpe}%
  \BibitemOpen
  \bibfield  {author} {\bibinfo {author} {\bibfnamefont {A.~A.}\ \bibnamefont
  {Aguilar-Arevalo}} \emph {et~al.} (\bibinfo {collaboration} {MiniBooNE}),\
  }\bibfield  {title} {\enquote {\bibinfo {title} {Significant excess of
  electron-like events in the {MiniBooNE} short-baseline neutrino
  experiment},}\ }\href {\doibase 10.1103/PhysRevLett.121.221801} {\bibfield
  {journal} {\bibinfo  {journal} {Phys. Rev. Lett.}\ }\textbf {\bibinfo
  {volume} {121}},\ \bibinfo {pages} {221801} (\bibinfo {year} {2018})},\
  \Eprint {http://arxiv.org/abs/1805.12028} {arXiv:1805.12028 [hep-ex]}
  \BibitemShut {NoStop}%
\bibitem [{\citenamefont {{L\"uscher}}(1986)}]{Luscher:1986pf}%
  \BibitemOpen
  \bibfield  {author} {\bibinfo {author} {\bibfnamefont {M.}~\bibnamefont
  {{L\"uscher}}},\ }\bibfield  {title} {\enquote {\bibinfo {title} {Volume
  dependence of the energy spectrum in massive quantum field theories~2:
  Scattering states},}\ }\href {\doibase 10.1007/BF01211097} {\bibfield
  {journal} {\bibinfo  {journal} {Commun. Math. Phys.}\ }\textbf {\bibinfo
  {volume} {105}},\ \bibinfo {pages} {153--188} (\bibinfo {year}
  {1986})}\BibitemShut {NoStop}%
\bibitem [{\citenamefont {{L\"uscher}}(1991)}]{Luscher:1990ux}%
  \BibitemOpen
  \bibfield  {author} {\bibinfo {author} {\bibfnamefont {M.}~\bibnamefont
  {{L\"uscher}}},\ }\bibfield  {title} {\enquote {\bibinfo {title} {Two
  particle states on a torus and their relation to the scattering matrix},}\
  }\href {\doibase 10.1016/0550-3213(91)90366-6} {\bibfield  {journal}
  {\bibinfo  {journal} {Nucl. Phys.}\ }\textbf {\bibinfo {volume} {B354}},\
  \bibinfo {pages} {531--578} (\bibinfo {year} {1991})}\BibitemShut {NoStop}%
\bibitem [{\citenamefont {Lellouch}\ and\ \citenamefont
  {{L\"uscher}}(2001)}]{Lellouch:2000pv}%
  \BibitemOpen
  \bibfield  {author} {\bibinfo {author} {\bibfnamefont {Laurent}\ \bibnamefont
  {Lellouch}}\ and\ \bibinfo {author} {\bibfnamefont {Martin}\ \bibnamefont
  {{L\"uscher}}},\ }\bibfield  {title} {\enquote {\bibinfo {title} {Weak
  transition matrix elements from finite volume correlation functions},}\
  }\href {\doibase 10.1007/s002200100410} {\bibfield  {journal} {\bibinfo
  {journal} {Commun. Math. Phys.}\ }\textbf {\bibinfo {volume} {219}},\
  \bibinfo {pages} {31--44} (\bibinfo {year} {2001})},\ \Eprint
  {http://arxiv.org/abs/hep-lat/0003023} {arXiv:hep-lat/0003023 [hep-lat]}
  \BibitemShut {NoStop}%
\bibitem [{\citenamefont {Aoki}\ \emph {et~al.}(2007)\citenamefont {Aoki} \emph
  {et~al.}}]{Aoki:2007rd}%
  \BibitemOpen
  \bibfield  {author} {\bibinfo {author} {\bibfnamefont {S.}~\bibnamefont
  {Aoki}} \emph {et~al.} (\bibinfo {collaboration} {CP-PACS}),\ }\bibfield
  {title} {\enquote {\bibinfo {title} {Lattice {QCD} calculation of the $\rho$
  meson decay width},}\ }\href {\doibase 10.1103/PhysRevD.76.094506} {\bibfield
   {journal} {\bibinfo  {journal} {Phys. Rev.}\ }\textbf {\bibinfo {volume}
  {D76}},\ \bibinfo {pages} {094506} (\bibinfo {year} {2007})},\ \Eprint
  {http://arxiv.org/abs/0708.3705} {arXiv:0708.3705 [hep-lat]} \BibitemShut
  {NoStop}%
\bibitem [{\citenamefont {Feng}\ \emph {et~al.}(2011)\citenamefont {Feng},
  \citenamefont {Jansen},\ and\ \citenamefont {Renner}}]{Feng:2010es}%
  \BibitemOpen
  \bibfield  {author} {\bibinfo {author} {\bibfnamefont {Xu}~\bibnamefont
  {Feng}}, \bibinfo {author} {\bibfnamefont {Karl}\ \bibnamefont {Jansen}}, \
  and\ \bibinfo {author} {\bibfnamefont {Dru~B.}\ \bibnamefont {Renner}},\
  }\bibfield  {title} {\enquote {\bibinfo {title} {Resonance parameters of the
  rho-meson from lattice {QCD}},}\ }\href {\doibase 10.1103/PhysRevD.83.094505}
  {\bibfield  {journal} {\bibinfo  {journal} {Phys. Rev.}\ }\textbf {\bibinfo
  {volume} {D83}},\ \bibinfo {pages} {094505} (\bibinfo {year} {2011})},\
  \Eprint {http://arxiv.org/abs/1011.5288} {arXiv:1011.5288 [hep-lat]}
  \BibitemShut {NoStop}%
\bibitem [{\citenamefont {Dudek}\ \emph {et~al.}(2013)\citenamefont {Dudek},
  \citenamefont {Edwards},\ and\ \citenamefont {Thomas}}]{Dudek:2012xn}%
  \BibitemOpen
  \bibfield  {author} {\bibinfo {author} {\bibfnamefont {Jozef~J.}\
  \bibnamefont {Dudek}}, \bibinfo {author} {\bibfnamefont {Robert~G.}\
  \bibnamefont {Edwards}}, \ and\ \bibinfo {author} {\bibfnamefont
  {Christopher~E.}\ \bibnamefont {Thomas}} (\bibinfo {collaboration} {Hadron
  Spectrum}),\ }\bibfield  {title} {\enquote {\bibinfo {title} {Energy
  dependence of the $\rho$ resonance in $\pi\pi$ elastic scattering from
  lattice {QCD}},}\ }\href {\doibase 10.1103/PhysRevD.87.034505,
  10.1103/PhysRevD.90.099902} {\bibfield  {journal} {\bibinfo  {journal} {Phys.
  Rev.}\ }\textbf {\bibinfo {volume} {D87}},\ \bibinfo {pages} {034505}
  (\bibinfo {year} {2013})},\ \bibinfo {note} {[Erratum: Phys.
  Rev.D90,no.9,099902(2014)]},\ \Eprint {http://arxiv.org/abs/1212.0830}
  {arXiv:1212.0830 [hep-ph]} \BibitemShut {NoStop}%
\bibitem [{\citenamefont {Wilson}\ \emph {et~al.}(2015)\citenamefont {Wilson},
  \citenamefont {Briceño}, \citenamefont {Dudek}, \citenamefont {Edwards},\
  and\ \citenamefont {Thomas}}]{Wilson:2015dqa}%
  \BibitemOpen
  \bibfield  {author} {\bibinfo {author} {\bibfnamefont {David~J.}\
  \bibnamefont {Wilson}}, \bibinfo {author} {\bibfnamefont {Raúl~A.}\
  \bibnamefont {Briceño}}, \bibinfo {author} {\bibfnamefont {Jozef~J.}\
  \bibnamefont {Dudek}}, \bibinfo {author} {\bibfnamefont {Robert~G.}\
  \bibnamefont {Edwards}}, \ and\ \bibinfo {author} {\bibfnamefont
  {Christopher~E.}\ \bibnamefont {Thomas}},\ }\bibfield  {title} {\enquote
  {\bibinfo {title} {{Coupled $\pi\pi, K\bar{K}$ scattering in $P$-wave and the
  $\rho$ resonance from lattice QCD}},}\ }\href {\doibase
  10.1103/PhysRevD.92.094502} {\bibfield  {journal} {\bibinfo  {journal} {Phys.
  Rev.}\ }\textbf {\bibinfo {volume} {D92}},\ \bibinfo {pages} {094502}
  (\bibinfo {year} {2015})},\ \Eprint {http://arxiv.org/abs/1507.02599}
  {arXiv:1507.02599 [hep-ph]} \BibitemShut {NoStop}%
\bibitem [{\citenamefont {Alexandrou}\ \emph
  {et~al.}(2017{\natexlab{b}})\citenamefont {Alexandrou}, \citenamefont
  {Leskovec}, \citenamefont {Meinel}, \citenamefont {Negele}, \citenamefont
  {Paul}, \citenamefont {Petschlies}, \citenamefont {Pochinsky}, \citenamefont
  {Rendon},\ and\ \citenamefont {Syritsyn}}]{Alexandrou:2017mpi}%
  \BibitemOpen
  \bibfield  {author} {\bibinfo {author} {\bibfnamefont {Constantia}\
  \bibnamefont {Alexandrou}}, \bibinfo {author} {\bibfnamefont {Luka}\
  \bibnamefont {Leskovec}}, \bibinfo {author} {\bibfnamefont {Stefan}\
  \bibnamefont {Meinel}}, \bibinfo {author} {\bibfnamefont {John}\ \bibnamefont
  {Negele}}, \bibinfo {author} {\bibfnamefont {Srijit}\ \bibnamefont {Paul}},
  \bibinfo {author} {\bibfnamefont {Marcus}\ \bibnamefont {Petschlies}},
  \bibinfo {author} {\bibfnamefont {Andrew}\ \bibnamefont {Pochinsky}},
  \bibinfo {author} {\bibfnamefont {Gumaro}\ \bibnamefont {Rendon}}, \ and\
  \bibinfo {author} {\bibfnamefont {Sergey}\ \bibnamefont {Syritsyn}},\
  }\bibfield  {title} {\enquote {\bibinfo {title} {{$P$-wave $\pi\pi$
  scattering and the $\rho$ resonance from lattice QCD}},}\ }\href {\doibase
  10.1103/PhysRevD.96.034525} {\bibfield  {journal} {\bibinfo  {journal} {Phys.
  Rev.}\ }\textbf {\bibinfo {volume} {D96}},\ \bibinfo {pages} {034525}
  (\bibinfo {year} {2017}{\natexlab{b}})},\ \Eprint
  {http://arxiv.org/abs/1704.05439} {arXiv:1704.05439 [hep-lat]} \BibitemShut
  {NoStop}%
\bibitem [{\citenamefont {Beane}\ \emph
  {et~al.}(2008{\natexlab{a}})\citenamefont {Beane}, \citenamefont {Luu},
  \citenamefont {Orginos}, \citenamefont {Parreno}, \citenamefont {Savage},
  \citenamefont {Torok},\ and\ \citenamefont {Walker-Loud}}]{Beane:2007xs}%
  \BibitemOpen
  \bibfield  {author} {\bibinfo {author} {\bibfnamefont {Silas~R.}\
  \bibnamefont {Beane}}, \bibinfo {author} {\bibfnamefont {Thomas~C.}\
  \bibnamefont {Luu}}, \bibinfo {author} {\bibfnamefont {Kostas}\ \bibnamefont
  {Orginos}}, \bibinfo {author} {\bibfnamefont {Assumpta}\ \bibnamefont
  {Parreno}}, \bibinfo {author} {\bibfnamefont {Martin~J.}\ \bibnamefont
  {Savage}}, \bibinfo {author} {\bibfnamefont {Aaron}\ \bibnamefont {Torok}}, \
  and\ \bibinfo {author} {\bibfnamefont {Andre}\ \bibnamefont {Walker-Loud}},\
  }\bibfield  {title} {\enquote {\bibinfo {title} {{Precise Determination of
  the I=2 pi pi Scattering Length from Mixed-Action Lattice QCD}},}\ }\href
  {\doibase 10.1103/PhysRevD.77.014505} {\bibfield  {journal} {\bibinfo
  {journal} {Phys. Rev.}\ }\textbf {\bibinfo {volume} {D77}},\ \bibinfo {pages}
  {014505} (\bibinfo {year} {2008}{\natexlab{a}})},\ \Eprint
  {http://arxiv.org/abs/0706.3026} {arXiv:0706.3026 [hep-lat]} \BibitemShut
  {NoStop}%
\bibitem [{\citenamefont {Feng}\ \emph {et~al.}(2010)\citenamefont {Feng},
  \citenamefont {Jansen},\ and\ \citenamefont {Renner}}]{Feng:2009ij}%
  \BibitemOpen
  \bibfield  {author} {\bibinfo {author} {\bibfnamefont {Xu}~\bibnamefont
  {Feng}}, \bibinfo {author} {\bibfnamefont {Karl}\ \bibnamefont {Jansen}}, \
  and\ \bibinfo {author} {\bibfnamefont {Dru~B.}\ \bibnamefont {Renner}},\
  }\bibfield  {title} {\enquote {\bibinfo {title} {The $\pi^+\pi^+$ scattering
  length from maximally twisted mass lattice {QCD}},}\ }\href {\doibase
  10.1016/j.physletb.2010.01.018} {\bibfield  {journal} {\bibinfo  {journal}
  {Phys. Lett.}\ }\textbf {\bibinfo {volume} {B684}},\ \bibinfo {pages}
  {268--274} (\bibinfo {year} {2010})},\ \Eprint
  {http://arxiv.org/abs/0909.3255} {arXiv:0909.3255 [hep-lat]} \BibitemShut
  {NoStop}%
\bibitem [{\citenamefont {Beane}\ \emph {et~al.}(2012)\citenamefont {Beane},
  \citenamefont {Chang}, \citenamefont {Detmold}, \citenamefont {Lin},
  \citenamefont {Luu}, \citenamefont {Orginos}, \citenamefont {Parreno},
  \citenamefont {Savage}, \citenamefont {Torok},\ and\ \citenamefont
  {Walker-Loud}}]{Beane:2011sc}%
  \BibitemOpen
  \bibfield  {author} {\bibinfo {author} {\bibfnamefont {S.~R.}\ \bibnamefont
  {Beane}}, \bibinfo {author} {\bibfnamefont {E.}~\bibnamefont {Chang}},
  \bibinfo {author} {\bibfnamefont {W.}~\bibnamefont {Detmold}}, \bibinfo
  {author} {\bibfnamefont {H.~W.}\ \bibnamefont {Lin}}, \bibinfo {author}
  {\bibfnamefont {T.~C.}\ \bibnamefont {Luu}}, \bibinfo {author} {\bibfnamefont
  {K.}~\bibnamefont {Orginos}}, \bibinfo {author} {\bibfnamefont
  {A.}~\bibnamefont {Parreno}}, \bibinfo {author} {\bibfnamefont {M.~J.}\
  \bibnamefont {Savage}}, \bibinfo {author} {\bibfnamefont {A.}~\bibnamefont
  {Torok}}, \ and\ \bibinfo {author} {\bibfnamefont {A.}~\bibnamefont
  {Walker-Loud}} (\bibinfo {collaboration} {NPLQCD}),\ }\bibfield  {title}
  {\enquote {\bibinfo {title} {The {$I=2$} $\pi\pi$ s-wave scattering phase
  shift from lattice {QCD}},}\ }\href {\doibase 10.1103/PhysRevD.85.034505}
  {\bibfield  {journal} {\bibinfo  {journal} {Phys. Rev.}\ }\textbf {\bibinfo
  {volume} {D85}},\ \bibinfo {pages} {034505} (\bibinfo {year} {2012})},\
  \Eprint {http://arxiv.org/abs/1107.5023} {arXiv:1107.5023 [hep-lat]}
  \BibitemShut {NoStop}%
\bibitem [{\citenamefont {Dudek}\ \emph {et~al.}(2011)\citenamefont {Dudek},
  \citenamefont {Edwards}, \citenamefont {Peardon}, \citenamefont {Richards},\
  and\ \citenamefont {Thomas}}]{Dudek:2010ew}%
  \BibitemOpen
  \bibfield  {author} {\bibinfo {author} {\bibfnamefont {Jozef~J.}\
  \bibnamefont {Dudek}}, \bibinfo {author} {\bibfnamefont {Robert~G.}\
  \bibnamefont {Edwards}}, \bibinfo {author} {\bibfnamefont {Michael~J.}\
  \bibnamefont {Peardon}}, \bibinfo {author} {\bibfnamefont {David~G.}\
  \bibnamefont {Richards}}, \ and\ \bibinfo {author} {\bibfnamefont
  {Christopher~E.}\ \bibnamefont {Thomas}},\ }\bibfield  {title} {\enquote
  {\bibinfo {title} {The phase-shift of isospin-2 $\pi$-$\pi$ scattering from
  lattice {QCD}},}\ }\href {\doibase 10.1103/PhysRevD.83.071504} {\bibfield
  {journal} {\bibinfo  {journal} {Phys. Rev.}\ }\textbf {\bibinfo {volume}
  {D83}},\ \bibinfo {pages} {071504} (\bibinfo {year} {2011})},\ \Eprint
  {http://arxiv.org/abs/1011.6352} {arXiv:1011.6352 [hep-ph]} \BibitemShut
  {NoStop}%
\bibitem [{\citenamefont {Dudek}\ \emph {et~al.}(2012)\citenamefont {Dudek},
  \citenamefont {Edwards},\ and\ \citenamefont {Thomas}}]{Dudek:2012gj}%
  \BibitemOpen
  \bibfield  {author} {\bibinfo {author} {\bibfnamefont {Jozef~J.}\
  \bibnamefont {Dudek}}, \bibinfo {author} {\bibfnamefont {Robert~G.}\
  \bibnamefont {Edwards}}, \ and\ \bibinfo {author} {\bibfnamefont
  {Christopher~E.}\ \bibnamefont {Thomas}},\ }\bibfield  {title} {\enquote
  {\bibinfo {title} {{S} and {D}-wave phase shifts in isospin-2 $\pi\pi$
  scattering from lattice {QCD}},}\ }\href {\doibase
  10.1103/PhysRevD.86.034031} {\bibfield  {journal} {\bibinfo  {journal} {Phys.
  Rev.}\ }\textbf {\bibinfo {volume} {D86}},\ \bibinfo {pages} {034031}
  (\bibinfo {year} {2012})},\ \Eprint {http://arxiv.org/abs/1203.6041}
  {arXiv:1203.6041 [hep-ph]} \BibitemShut {NoStop}%
\bibitem [{\citenamefont {Detmold}\ and\ \citenamefont
  {Savage}(2004)}]{Detmold:2004qn}%
  \BibitemOpen
  \bibfield  {author} {\bibinfo {author} {\bibfnamefont {William}\ \bibnamefont
  {Detmold}}\ and\ \bibinfo {author} {\bibfnamefont {Martin~J.}\ \bibnamefont
  {Savage}},\ }\bibfield  {title} {\enquote {\bibinfo {title} {Electroweak
  matrix elements in the two nucleon sector from lattice {QCD}},}\ }\href
  {\doibase 10.1016/j.nuclphysa.2004.07.007} {\bibfield  {journal} {\bibinfo
  {journal} {Nucl. Phys.}\ }\textbf {\bibinfo {volume} {A743}},\ \bibinfo
  {pages} {170--193} (\bibinfo {year} {2004})},\ \Eprint
  {http://arxiv.org/abs/hep-lat/0403005} {arXiv:hep-lat/0403005 [hep-lat]}
  \BibitemShut {NoStop}%
\bibitem [{\citenamefont {He}\ \emph {et~al.}(2005)\citenamefont {He},
  \citenamefont {Feng},\ and\ \citenamefont {Liu}}]{He:2005ey}%
  \BibitemOpen
  \bibfield  {author} {\bibinfo {author} {\bibfnamefont {Song}\ \bibnamefont
  {He}}, \bibinfo {author} {\bibfnamefont {Xu}~\bibnamefont {Feng}}, \ and\
  \bibinfo {author} {\bibfnamefont {Chuan}\ \bibnamefont {Liu}},\ }\bibfield
  {title} {\enquote {\bibinfo {title} {{Two particle states and the S-matrix
  elements in multi-channel scattering}},}\ }\href {\doibase
  10.1088/1126-6708/2005/07/011} {\bibfield  {journal} {\bibinfo  {journal}
  {JHEP}\ }\textbf {\bibinfo {volume} {07}},\ \bibinfo {pages} {011} (\bibinfo
  {year} {2005})},\ \Eprint {http://arxiv.org/abs/hep-lat/0504019}
  {arXiv:hep-lat/0504019 [hep-lat]} \BibitemShut {NoStop}%
\bibitem [{\citenamefont {Hansen}\ and\ \citenamefont
  {Sharpe}(2012)}]{Hansen:2012tf}%
  \BibitemOpen
  \bibfield  {author} {\bibinfo {author} {\bibfnamefont {Maxwell~T.}\
  \bibnamefont {Hansen}}\ and\ \bibinfo {author} {\bibfnamefont {Stephen~R.}\
  \bibnamefont {Sharpe}},\ }\bibfield  {title} {\enquote {\bibinfo {title}
  {Multiple-channel generalization of {Lellouch-L\"uscher} formula},}\ }\href
  {\doibase 10.1103/PhysRevD.86.016007} {\bibfield  {journal} {\bibinfo
  {journal} {Phys. Rev.}\ }\textbf {\bibinfo {volume} {D86}},\ \bibinfo {pages}
  {016007} (\bibinfo {year} {2012})},\ \Eprint {http://arxiv.org/abs/1204.0826}
  {arXiv:1204.0826 [hep-lat]} \BibitemShut {NoStop}%
\bibitem [{\citenamefont {Briceño}\ and\ \citenamefont
  {Davoudi}(2013{\natexlab{a}})}]{Briceno:2012yi}%
  \BibitemOpen
  \bibfield  {author} {\bibinfo {author} {\bibfnamefont {Raúl~A.}\
  \bibnamefont {Briceño}}\ and\ \bibinfo {author} {\bibfnamefont {Zohreh}\
  \bibnamefont {Davoudi}},\ }\bibfield  {title} {\enquote {\bibinfo {title}
  {{Moving multichannel systems in a finite volume with application to
  proton-proton fusion}},}\ }\href {\doibase 10.1103/PhysRevD.88.094507}
  {\bibfield  {journal} {\bibinfo  {journal} {Phys. Rev.}\ }\textbf {\bibinfo
  {volume} {D88}},\ \bibinfo {pages} {094507} (\bibinfo {year}
  {2013}{\natexlab{a}})},\ \Eprint {http://arxiv.org/abs/1204.1110}
  {arXiv:1204.1110 [hep-lat]} \BibitemShut {NoStop}%
\bibitem [{\citenamefont {Guo}\ \emph {et~al.}(2013)\citenamefont {Guo},
  \citenamefont {Dudek}, \citenamefont {Edwards},\ and\ \citenamefont
  {Szczepaniak}}]{Guo:2012hv}%
  \BibitemOpen
  \bibfield  {author} {\bibinfo {author} {\bibfnamefont {Peng}\ \bibnamefont
  {Guo}}, \bibinfo {author} {\bibfnamefont {Jozef}\ \bibnamefont {Dudek}},
  \bibinfo {author} {\bibfnamefont {Robert}\ \bibnamefont {Edwards}}, \ and\
  \bibinfo {author} {\bibfnamefont {Adam~P.}\ \bibnamefont {Szczepaniak}},\
  }\bibfield  {title} {\enquote {\bibinfo {title} {{Coupled-channel scattering
  on a torus}},}\ }\href {\doibase 10.1103/PhysRevD.88.014501} {\bibfield
  {journal} {\bibinfo  {journal} {Phys. Rev.}\ }\textbf {\bibinfo {volume}
  {D88}},\ \bibinfo {pages} {014501} (\bibinfo {year} {2013})},\ \Eprint
  {http://arxiv.org/abs/1211.0929} {arXiv:1211.0929 [hep-lat]} \BibitemShut
  {NoStop}%
\bibitem [{\citenamefont {Kreuzer}\ and\ \citenamefont
  {Hammer}(2011)}]{Kreuzer:2010ti}%
  \BibitemOpen
  \bibfield  {author} {\bibinfo {author} {\bibfnamefont {Simon}\ \bibnamefont
  {Kreuzer}}\ and\ \bibinfo {author} {\bibfnamefont {H.-W.}\ \bibnamefont
  {Hammer}},\ }\bibfield  {title} {\enquote {\bibinfo {title} {The triton in a
  finite volume},}\ }\href {\doibase 10.1016/j.physletb.2010.10.003} {\bibfield
   {journal} {\bibinfo  {journal} {Phys. Lett.}\ }\textbf {\bibinfo {volume}
  {B694}},\ \bibinfo {pages} {424--429} (\bibinfo {year} {2011})},\ \Eprint
  {http://arxiv.org/abs/1008.4499} {arXiv:1008.4499 [hep-lat]} \BibitemShut
  {NoStop}%
\bibitem [{\citenamefont {Briceño}\ and\ \citenamefont
  {Davoudi}(2013{\natexlab{b}})}]{Briceno:2012rv}%
  \BibitemOpen
  \bibfield  {author} {\bibinfo {author} {\bibfnamefont {Raúl~A.}\
  \bibnamefont {Briceño}}\ and\ \bibinfo {author} {\bibfnamefont {Zohreh}\
  \bibnamefont {Davoudi}},\ }\bibfield  {title} {\enquote {\bibinfo {title}
  {Three-particle scattering amplitudes from a finite volume formalism},}\
  }\href {\doibase 10.1103/PhysRevD.87.094507} {\bibfield  {journal} {\bibinfo
  {journal} {Phys. Rev.}\ }\textbf {\bibinfo {volume} {D87}},\ \bibinfo {pages}
  {094507} (\bibinfo {year} {2013}{\natexlab{b}})},\ \Eprint
  {http://arxiv.org/abs/1212.3398} {arXiv:1212.3398 [hep-lat]} \BibitemShut
  {NoStop}%
\bibitem [{\citenamefont {Meißner}\ \emph {et~al.}(2015)\citenamefont
  {Meißner}, \citenamefont {Ríos},\ and\ \citenamefont
  {Rusetsky}}]{Meissner:2014dea}%
  \BibitemOpen
  \bibfield  {author} {\bibinfo {author} {\bibfnamefont {Ulf-G.}\ \bibnamefont
  {Meißner}}, \bibinfo {author} {\bibfnamefont {Guillermo}\ \bibnamefont
  {Ríos}}, \ and\ \bibinfo {author} {\bibfnamefont {Akaki}\ \bibnamefont
  {Rusetsky}},\ }\bibfield  {title} {\enquote {\bibinfo {title} {Spectrum of
  three-body bound states in a finite volume},}\ }\href {\doibase
  10.1103/PhysRevLett.117.069902, 10.1103/PhysRevLett.114.091602} {\bibfield
  {journal} {\bibinfo  {journal} {Phys. Rev. Lett.}\ }\textbf {\bibinfo
  {volume} {114}},\ \bibinfo {pages} {091602} (\bibinfo {year} {2015})},\
  \bibinfo {note} {[Erratum: Phys. Rev. Lett.117,no.6,069902(2016)]},\ \Eprint
  {http://arxiv.org/abs/1412.4969} {arXiv:1412.4969 [hep-lat]} \BibitemShut
  {NoStop}%
\bibitem [{\citenamefont {Briceño}\ \emph {et~al.}(2017)\citenamefont
  {Briceño}, \citenamefont {Hansen},\ and\ \citenamefont
  {Sharpe}}]{Briceno:2017tce}%
  \BibitemOpen
  \bibfield  {author} {\bibinfo {author} {\bibfnamefont {Raúl~A.}\
  \bibnamefont {Briceño}}, \bibinfo {author} {\bibfnamefont {Maxwell~T.}\
  \bibnamefont {Hansen}}, \ and\ \bibinfo {author} {\bibfnamefont {Stephen~R.}\
  \bibnamefont {Sharpe}},\ }\bibfield  {title} {\enquote {\bibinfo {title}
  {Relating the finite-volume spectrum and the two-and-three-particle {$S$}
  matrix for relativistic systems of identical scalar particles},}\ }\href
  {\doibase 10.1103/PhysRevD.95.074510} {\bibfield  {journal} {\bibinfo
  {journal} {Phys. Rev.}\ }\textbf {\bibinfo {volume} {D95}},\ \bibinfo {pages}
  {074510} (\bibinfo {year} {2017})},\ \Eprint
  {http://arxiv.org/abs/1701.07465} {arXiv:1701.07465 [hep-lat]} \BibitemShut
  {NoStop}%
\bibitem [{\citenamefont {Briceño}\ \emph {et~al.}(2019)\citenamefont
  {Briceño}, \citenamefont {Hansen},\ and\ \citenamefont
  {Sharpe}}]{Briceno:2018aml}%
  \BibitemOpen
  \bibfield  {author} {\bibinfo {author} {\bibfnamefont {Raúl~A.}\
  \bibnamefont {Briceño}}, \bibinfo {author} {\bibfnamefont {Maxwell~T.}\
  \bibnamefont {Hansen}}, \ and\ \bibinfo {author} {\bibfnamefont {Stephen~R.}\
  \bibnamefont {Sharpe}},\ }\bibfield  {title} {\enquote {\bibinfo {title}
  {Three-particle systems with resonant subprocesses in a finite volume},}\
  }\href {\doibase 10.1103/PhysRevD.99.014516} {\bibfield  {journal} {\bibinfo
  {journal} {Phys. Rev.}\ }\textbf {\bibinfo {volume} {D99}},\ \bibinfo {pages}
  {014516} (\bibinfo {year} {2019})},\ \Eprint
  {http://arxiv.org/abs/1810.01429} {arXiv:1810.01429 [hep-lat]} \BibitemShut
  {NoStop}%
\bibitem [{\citenamefont {Döring}\ \emph {et~al.}(2018)\citenamefont
  {Döring}, \citenamefont {Hammer}, \citenamefont {Mai}, \citenamefont {Pang},
  \citenamefont {Rusetsky},\ and\ \citenamefont {Wu}}]{Doring:2018xxx}%
  \BibitemOpen
  \bibfield  {author} {\bibinfo {author} {\bibfnamefont {M.}~\bibnamefont
  {Döring}}, \bibinfo {author} {\bibfnamefont {H.-W.}\ \bibnamefont {Hammer}},
  \bibinfo {author} {\bibfnamefont {M.}~\bibnamefont {Mai}}, \bibinfo {author}
  {\bibfnamefont {J.-Y.}\ \bibnamefont {Pang}}, \bibinfo {author}
  {\bibfnamefont {A.}~\bibnamefont {Rusetsky}}, \ and\ \bibinfo {author}
  {\bibfnamefont {J.}~\bibnamefont {Wu}},\ }\bibfield  {title} {\enquote
  {\bibinfo {title} {Three-body spectrum in a finite volume: the role of cubic
  symmetry},}\ }\href {\doibase 10.1103/PhysRevD.97.114508} {\bibfield
  {journal} {\bibinfo  {journal} {Phys. Rev.}\ }\textbf {\bibinfo {volume}
  {D97}},\ \bibinfo {pages} {114508} (\bibinfo {year} {2018})},\ \Eprint
  {http://arxiv.org/abs/1802.03362} {arXiv:1802.03362 [hep-lat]} \BibitemShut
  {NoStop}%
\bibitem [{\citenamefont {Beane}\ \emph
  {et~al.}(2008{\natexlab{b}})\citenamefont {Beane}, \citenamefont {Detmold},
  \citenamefont {Luu}, \citenamefont {Orginos}, \citenamefont {Savage},\ and\
  \citenamefont {Torok}}]{Beane:2007es}%
  \BibitemOpen
  \bibfield  {author} {\bibinfo {author} {\bibfnamefont {Silas~R.}\
  \bibnamefont {Beane}}, \bibinfo {author} {\bibfnamefont {William}\
  \bibnamefont {Detmold}}, \bibinfo {author} {\bibfnamefont {Thomas~C.}\
  \bibnamefont {Luu}}, \bibinfo {author} {\bibfnamefont {Kostas}\ \bibnamefont
  {Orginos}}, \bibinfo {author} {\bibfnamefont {Martin~J.}\ \bibnamefont
  {Savage}}, \ and\ \bibinfo {author} {\bibfnamefont {Aaron}\ \bibnamefont
  {Torok}},\ }\bibfield  {title} {\enquote {\bibinfo {title} {Multi-pion
  systems in lattice {QCD} and the three-pion interaction},}\ }\href {\doibase
  10.1103/PhysRevLett.100.082004} {\bibfield  {journal} {\bibinfo  {journal}
  {Phys. Rev. Lett.}\ }\textbf {\bibinfo {volume} {100}},\ \bibinfo {pages}
  {082004} (\bibinfo {year} {2008}{\natexlab{b}})},\ \Eprint
  {http://arxiv.org/abs/0710.1827} {arXiv:0710.1827 [hep-lat]} \BibitemShut
  {NoStop}%
\bibitem [{\citenamefont {Dudek}\ \emph {et~al.}(2016)\citenamefont {Dudek},
  \citenamefont {Edwards},\ and\ \citenamefont {Wilson}}]{Dudek:2016cru}%
  \BibitemOpen
  \bibfield  {author} {\bibinfo {author} {\bibfnamefont {Jozef~J.}\
  \bibnamefont {Dudek}}, \bibinfo {author} {\bibfnamefont {Robert~G.}\
  \bibnamefont {Edwards}}, \ and\ \bibinfo {author} {\bibfnamefont {David~J.}\
  \bibnamefont {Wilson}} (\bibinfo {collaboration} {Hadron Spectrum}),\
  }\bibfield  {title} {\enquote {\bibinfo {title} {An $a_0$ resonance in
  strongly coupled $\pi \eta$, {$K\bar{K}$} scattering from lattice {QCD}},}\
  }\href {\doibase 10.1103/PhysRevD.93.094506} {\bibfield  {journal} {\bibinfo
  {journal} {Phys. Rev.}\ }\textbf {\bibinfo {volume} {D93}},\ \bibinfo {pages}
  {094506} (\bibinfo {year} {2016})},\ \Eprint
  {http://arxiv.org/abs/1602.05122} {arXiv:1602.05122 [hep-ph]} \BibitemShut
  {NoStop}%
\bibitem [{\citenamefont {Briceño}\ \emph
  {et~al.}(2018{\natexlab{a}})\citenamefont {Briceño}, \citenamefont {Dudek},
  \citenamefont {Edwards},\ and\ \citenamefont {Wilson}}]{Briceno:2017qmb}%
  \BibitemOpen
  \bibfield  {author} {\bibinfo {author} {\bibfnamefont {Raúl~A.}\
  \bibnamefont {Briceño}}, \bibinfo {author} {\bibfnamefont {Jozef~J.}\
  \bibnamefont {Dudek}}, \bibinfo {author} {\bibfnamefont {Robert~G.}\
  \bibnamefont {Edwards}}, \ and\ \bibinfo {author} {\bibfnamefont {David~J.}\
  \bibnamefont {Wilson}} (\bibinfo {collaboration} {Hadron Spectrum}),\
  }\bibfield  {title} {\enquote {\bibinfo {title} {{Isoscalar $\pi\pi$,
  $K\bar{K}$, $\eta\eta$ scattering and the $\sigma$, $f_0$, $f_2$ mesons from
  QCD}},}\ }\href {\doibase 10.1103/PhysRevD.97.054513} {\bibfield  {journal}
  {\bibinfo  {journal} {Phys. Rev.}\ }\textbf {\bibinfo {volume} {D97}},\
  \bibinfo {pages} {054513} (\bibinfo {year} {2018}{\natexlab{a}})},\ \Eprint
  {http://arxiv.org/abs/1708.06667} {arXiv:1708.06667 [hep-lat]} \BibitemShut
  {NoStop}%
\bibitem [{\citenamefont {Woss}\ \emph {et~al.}(2018)\citenamefont {Woss},
  \citenamefont {Thomas}, \citenamefont {Dudek}, \citenamefont {Edwards},\ and\
  \citenamefont {Wilson}}]{Woss:2018irj}%
  \BibitemOpen
  \bibfield  {author} {\bibinfo {author} {\bibfnamefont {Antoni}\ \bibnamefont
  {Woss}}, \bibinfo {author} {\bibfnamefont {Christopher~E.}\ \bibnamefont
  {Thomas}}, \bibinfo {author} {\bibfnamefont {Jozef~J.}\ \bibnamefont
  {Dudek}}, \bibinfo {author} {\bibfnamefont {Robert~G.}\ \bibnamefont
  {Edwards}}, \ and\ \bibinfo {author} {\bibfnamefont {David~J.}\ \bibnamefont
  {Wilson}} (\bibinfo {collaboration} {Hadron Spectrum}),\ }\bibfield  {title}
  {\enquote {\bibinfo {title} {Dynamically-coupled partial-waves in $\rho\pi$
  isospin-2 scattering from lattice {QCD}},}\ }\href {\doibase
  10.1007/JHEP07(2018)043} {\bibfield  {journal} {\bibinfo  {journal} {JHEP}\
  }\textbf {\bibinfo {volume} {07}},\ \bibinfo {pages} {043} (\bibinfo {year}
  {2018})},\ \Eprint {http://arxiv.org/abs/1802.05580} {arXiv:1802.05580
  [hep-lat]} \BibitemShut {NoStop}%
\bibitem [{\citenamefont {Briceño}\ and\ \citenamefont
  {Hansen}(2015)}]{Briceno:2015csa}%
  \BibitemOpen
  \bibfield  {author} {\bibinfo {author} {\bibfnamefont {Raúl~A.}\
  \bibnamefont {Briceño}}\ and\ \bibinfo {author} {\bibfnamefont {Maxwell~T.}\
  \bibnamefont {Hansen}},\ }\bibfield  {title} {\enquote {\bibinfo {title}
  {Multichannel $0\to 2$ and $1\to 2$ transition amplitudes for arbitrary spin
  particles in a finite volume},}\ }\href {\doibase 10.1103/PhysRevD.92.074509}
  {\bibfield  {journal} {\bibinfo  {journal} {Phys. Rev.}\ }\textbf {\bibinfo
  {volume} {D92}},\ \bibinfo {pages} {074509} (\bibinfo {year} {2015})},\
  \Eprint {http://arxiv.org/abs/1502.04314} {arXiv:1502.04314 [hep-lat]}
  \BibitemShut {NoStop}%
\bibitem [{\citenamefont {Briceño}\ \emph {et~al.}(2016)\citenamefont
  {Briceño}, \citenamefont {Dudek}, \citenamefont {Edwards}, \citenamefont
  {Shultz}, \citenamefont {Thomas},\ and\ \citenamefont
  {Wilson}}]{Briceno:2016kkp}%
  \BibitemOpen
  \bibfield  {author} {\bibinfo {author} {\bibfnamefont {Raúl~A.}\
  \bibnamefont {Briceño}}, \bibinfo {author} {\bibfnamefont {Jozef~J.}\
  \bibnamefont {Dudek}}, \bibinfo {author} {\bibfnamefont {Robert~G.}\
  \bibnamefont {Edwards}}, \bibinfo {author} {\bibfnamefont {Christian~J.}\
  \bibnamefont {Shultz}}, \bibinfo {author} {\bibfnamefont {Christopher~E.}\
  \bibnamefont {Thomas}}, \ and\ \bibinfo {author} {\bibfnamefont {David~J.}\
  \bibnamefont {Wilson}},\ }\bibfield  {title} {\enquote {\bibinfo {title} {The
  $\pi\pi\to\pi\gamma^\star$ amplitude and the resonant
  $\rho\to\pi\gamma^\star$ transition from lattice {QCD}},}\ }\href {\doibase
  10.1103/PhysRevD.93.114508} {\bibfield  {journal} {\bibinfo  {journal} {Phys.
  Rev.}\ }\textbf {\bibinfo {volume} {D93}},\ \bibinfo {pages} {114508}
  (\bibinfo {year} {2016})},\ \Eprint {http://arxiv.org/abs/1604.03530}
  {arXiv:1604.03530 [hep-ph]} \BibitemShut {NoStop}%
\bibitem [{\citenamefont {Alexandrou}\ \emph {et~al.}(2018)\citenamefont
  {Alexandrou}, \citenamefont {Leskovec}, \citenamefont {Meinel}, \citenamefont
  {Negele}, \citenamefont {Paul}, \citenamefont {Petschlies}, \citenamefont
  {Pochinsky}, \citenamefont {Rendon},\ and\ \citenamefont
  {Syritsyn}}]{Alexandrou:2018jbt}%
  \BibitemOpen
  \bibfield  {author} {\bibinfo {author} {\bibfnamefont {Constantia}\
  \bibnamefont {Alexandrou}}, \bibinfo {author} {\bibfnamefont {Luka}\
  \bibnamefont {Leskovec}}, \bibinfo {author} {\bibfnamefont {Stefan}\
  \bibnamefont {Meinel}}, \bibinfo {author} {\bibfnamefont {John}\ \bibnamefont
  {Negele}}, \bibinfo {author} {\bibfnamefont {Srijit}\ \bibnamefont {Paul}},
  \bibinfo {author} {\bibfnamefont {Marcus}\ \bibnamefont {Petschlies}},
  \bibinfo {author} {\bibfnamefont {Andrew}\ \bibnamefont {Pochinsky}},
  \bibinfo {author} {\bibfnamefont {Gumaro}\ \bibnamefont {Rendon}}, \ and\
  \bibinfo {author} {\bibfnamefont {Sergey}\ \bibnamefont {Syritsyn}},\
  }\bibfield  {title} {\enquote {\bibinfo {title} {$\pi\gamma \to \pi\pi$
  transition and the $\rho$ radiative decay width from lattice {QCD}},}\ }\href
  {\doibase 10.1103/PhysRevD.98.074502} {\bibfield  {journal} {\bibinfo
  {journal} {Phys. Rev.}\ }\textbf {\bibinfo {volume} {D98}},\ \bibinfo {pages}
  {074502} (\bibinfo {year} {2018})},\ \Eprint
  {http://arxiv.org/abs/1807.08357} {arXiv:1807.08357 [hep-lat]} \BibitemShut
  {NoStop}%
\bibitem [{\citenamefont {Briceño}\ and\ \citenamefont
  {Hansen}(2016)}]{Briceno:2015tza}%
  \BibitemOpen
  \bibfield  {author} {\bibinfo {author} {\bibfnamefont {Raúl~A.}\
  \bibnamefont {Briceño}}\ and\ \bibinfo {author} {\bibfnamefont {Maxwell~T.}\
  \bibnamefont {Hansen}},\ }\bibfield  {title} {\enquote {\bibinfo {title}
  {Relativistic, model-independent, multichannel $2\to 2$ transition amplitudes
  in a finite volume},}\ }\href {\doibase 10.1103/PhysRevD.94.013008}
  {\bibfield  {journal} {\bibinfo  {journal} {Phys. Rev.}\ }\textbf {\bibinfo
  {volume} {D94}},\ \bibinfo {pages} {013008} (\bibinfo {year} {2016})},\
  \Eprint {http://arxiv.org/abs/1509.08507} {arXiv:1509.08507 [hep-lat]}
  \BibitemShut {NoStop}%
\bibitem [{\citenamefont {Baroni}\ \emph {et~al.}(2018)\citenamefont {Baroni},
  \citenamefont {Briceño}, \citenamefont {Hansen},\ and\ \citenamefont
  {Ortega-Gama}}]{Baroni:2018iau}%
  \BibitemOpen
  \bibfield  {author} {\bibinfo {author} {\bibfnamefont {Alessandro}\
  \bibnamefont {Baroni}}, \bibinfo {author} {\bibfnamefont {Raúl~A.}\
  \bibnamefont {Briceño}}, \bibinfo {author} {\bibfnamefont {Maxwell~T.}\
  \bibnamefont {Hansen}}, \ and\ \bibinfo {author} {\bibfnamefont {Felipe~G.}\
  \bibnamefont {Ortega-Gama}},\ }\bibfield  {title} {\enquote {\bibinfo {title}
  {{Form factors of two-hadron states from a covariant finite-volume
  formalism}},}\ }\href@noop {} {\  (\bibinfo {year} {2018})},\ \Eprint
  {http://arxiv.org/abs/1812.10504} {arXiv:1812.10504 [hep-lat]} \BibitemShut
  {NoStop}%
\bibitem [{\citenamefont {Andersen}\ \emph {et~al.}(2018)\citenamefont
  {Andersen}, \citenamefont {Bulava}, \citenamefont {Hörz},\ and\
  \citenamefont {Morningstar}}]{Andersen:2017una}%
  \BibitemOpen
  \bibfield  {author} {\bibinfo {author} {\bibfnamefont {Christian~W.}\
  \bibnamefont {Andersen}}, \bibinfo {author} {\bibfnamefont {John}\
  \bibnamefont {Bulava}}, \bibinfo {author} {\bibfnamefont {Ben}\ \bibnamefont
  {Hörz}}, \ and\ \bibinfo {author} {\bibfnamefont {Colin}\ \bibnamefont
  {Morningstar}},\ }\bibfield  {title} {\enquote {\bibinfo {title} {Elastic
  {$I=3/2$} {$P$}-wave nucleon-pion scattering amplitude and the
  {$\Delta$(1232)} resonance from {$N_f=2+1$} lattice {QCD}},}\ }\href
  {\doibase 10.1103/PhysRevD.97.014506} {\bibfield  {journal} {\bibinfo
  {journal} {Phys. Rev.}\ }\textbf {\bibinfo {volume} {D97}},\ \bibinfo {pages}
  {014506} (\bibinfo {year} {2018})},\ \Eprint
  {http://arxiv.org/abs/1710.01557} {arXiv:1710.01557 [hep-lat]} \BibitemShut
  {NoStop}%
\bibitem [{\citenamefont {Rocco}\ \emph {et~al.}(2016)\citenamefont {Rocco},
  \citenamefont {Lovato},\ and\ \citenamefont {Benhar}}]{Rocco:2015cil}%
  \BibitemOpen
  \bibfield  {author} {\bibinfo {author} {\bibfnamefont {Noemi}\ \bibnamefont
  {Rocco}}, \bibinfo {author} {\bibfnamefont {Alessandro}\ \bibnamefont
  {Lovato}}, \ and\ \bibinfo {author} {\bibfnamefont {Omar}\ \bibnamefont
  {Benhar}},\ }\bibfield  {title} {\enquote {\bibinfo {title} {{Unified
  description of electron-nucleus scattering within the spectral function
  formalism}},}\ }\href {\doibase 10.1103/PhysRevLett.116.192501} {\bibfield
  {journal} {\bibinfo  {journal} {Phys. Rev. Lett.}\ }\textbf {\bibinfo
  {volume} {116}},\ \bibinfo {pages} {192501} (\bibinfo {year} {2016})},\
  \Eprint {http://arxiv.org/abs/1512.07426} {arXiv:1512.07426 [nucl-th]}
  \BibitemShut {NoStop}%
\bibitem [{\citenamefont {Liu}\ and\ \citenamefont {Dong}(1994)}]{Liu:1993cv}%
  \BibitemOpen
  \bibfield  {author} {\bibinfo {author} {\bibfnamefont {Keh-Fei}\ \bibnamefont
  {Liu}}\ and\ \bibinfo {author} {\bibfnamefont {Shao-Jing}\ \bibnamefont
  {Dong}},\ }\bibfield  {title} {\enquote {\bibinfo {title} {Origin of
  difference between $\bar d$ and $\bar u$ partons in the nucleon},}\ }\href
  {\doibase 10.1103/PhysRevLett.72.1790} {\bibfield  {journal} {\bibinfo
  {journal} {Phys. Rev. Lett.}\ }\textbf {\bibinfo {volume} {72}},\ \bibinfo
  {pages} {1790--1793} (\bibinfo {year} {1994})},\ \Eprint
  {http://arxiv.org/abs/hep-ph/9306299} {arXiv:hep-ph/9306299 [hep-ph]}
  \BibitemShut {NoStop}%
\bibitem [{\citenamefont {Liu}\ \emph {et~al.}(1999)\citenamefont {Liu},
  \citenamefont {Dong}, \citenamefont {Draper}, \citenamefont {Leinweber},
  \citenamefont {Sloan}, \citenamefont {Wilcox},\ and\ \citenamefont
  {Woloshyn}}]{Liu:1998um}%
  \BibitemOpen
  \bibfield  {author} {\bibinfo {author} {\bibfnamefont {K.~F.}\ \bibnamefont
  {Liu}}, \bibinfo {author} {\bibfnamefont {S.~J.}\ \bibnamefont {Dong}},
  \bibinfo {author} {\bibfnamefont {Terrence}\ \bibnamefont {Draper}}, \bibinfo
  {author} {\bibfnamefont {D.}~\bibnamefont {Leinweber}}, \bibinfo {author}
  {\bibfnamefont {J.~H.}\ \bibnamefont {Sloan}}, \bibinfo {author}
  {\bibfnamefont {W.}~\bibnamefont {Wilcox}}, \ and\ \bibinfo {author}
  {\bibfnamefont {R.~M.}\ \bibnamefont {Woloshyn}},\ }\bibfield  {title}
  {\enquote {\bibinfo {title} {Valence {QCD}: connecting {QCD} to the quark
  model},}\ }\href {\doibase 10.1103/PhysRevD.59.112001} {\bibfield  {journal}
  {\bibinfo  {journal} {Phys. Rev.}\ }\textbf {\bibinfo {volume} {D59}},\
  \bibinfo {pages} {112001} (\bibinfo {year} {1999})},\ \Eprint
  {http://arxiv.org/abs/hep-ph/9806491} {arXiv:hep-ph/9806491 [hep-ph]}
  \BibitemShut {NoStop}%
\bibitem [{\citenamefont {Liu}(2000)}]{Liu:1999ak}%
  \BibitemOpen
  \bibfield  {author} {\bibinfo {author} {\bibfnamefont {Keh-Fei}\ \bibnamefont
  {Liu}},\ }\bibfield  {title} {\enquote {\bibinfo {title} {{Parton degrees of
  freedom from the path-integral formalism}},}\ }\href {\doibase
  10.1103/PhysRevD.62.074501} {\bibfield  {journal} {\bibinfo  {journal} {Phys.
  Rev.}\ }\textbf {\bibinfo {volume} {D62}},\ \bibinfo {pages} {074501}
  (\bibinfo {year} {2000})},\ \Eprint {http://arxiv.org/abs/hep-ph/9910306}
  {arXiv:hep-ph/9910306 [hep-ph]} \BibitemShut {NoStop}%
\bibitem [{\citenamefont {Aglietti}\ \emph {et~al.}(1998)\citenamefont
  {Aglietti}, \citenamefont {Ciuchini}, \citenamefont {Corbo}, \citenamefont
  {Franco}, \citenamefont {Martinelli},\ and\ \citenamefont
  {Silvestrini}}]{Aglietti:1998mz}%
  \BibitemOpen
  \bibfield  {author} {\bibinfo {author} {\bibfnamefont {U.}~\bibnamefont
  {Aglietti}}, \bibinfo {author} {\bibfnamefont {Marco}\ \bibnamefont
  {Ciuchini}}, \bibinfo {author} {\bibfnamefont {G.}~\bibnamefont {Corbo}},
  \bibinfo {author} {\bibfnamefont {E.}~\bibnamefont {Franco}}, \bibinfo
  {author} {\bibfnamefont {G.}~\bibnamefont {Martinelli}}, \ and\ \bibinfo
  {author} {\bibfnamefont {L.}~\bibnamefont {Silvestrini}},\ }\bibfield
  {title} {\enquote {\bibinfo {title} {Model independent determination of the
  shape function for inclusive $b$ decays and of the structure functions in
  {DIS}},}\ }\href {\doibase 10.1016/S0370-2693(98)00677-7} {\bibfield
  {journal} {\bibinfo  {journal} {Phys. Lett.}\ }\textbf {\bibinfo {volume}
  {B432}},\ \bibinfo {pages} {411--420} (\bibinfo {year} {1998})},\ \Eprint
  {http://arxiv.org/abs/hep-ph/9804416} {arXiv:hep-ph/9804416 [hep-ph]}
  \BibitemShut {NoStop}%
\bibitem [{\citenamefont {Detmold}\ and\ \citenamefont
  {Lin}(2006)}]{Detmold:2005gg}%
  \BibitemOpen
  \bibfield  {author} {\bibinfo {author} {\bibfnamefont {William}\ \bibnamefont
  {Detmold}}\ and\ \bibinfo {author} {\bibfnamefont {C.~J.~David}\ \bibnamefont
  {Lin}},\ }\bibfield  {title} {\enquote {\bibinfo {title} {Deep-inelastic
  scattering and the operator product expansion in lattice {QCD}},}\ }\href
  {\doibase 10.1103/PhysRevD.73.014501} {\bibfield  {journal} {\bibinfo
  {journal} {Phys. Rev.}\ }\textbf {\bibinfo {volume} {D73}},\ \bibinfo {pages}
  {014501} (\bibinfo {year} {2006})},\ \Eprint
  {http://arxiv.org/abs/hep-lat/0507007} {arXiv:hep-lat/0507007 [hep-lat]}
  \BibitemShut {NoStop}%
\bibitem [{\citenamefont {Liu}(2016)}]{Liu:2016djw}%
  \BibitemOpen
  \bibfield  {author} {\bibinfo {author} {\bibfnamefont {Keh-Fei}\ \bibnamefont
  {Liu}},\ }\bibfield  {title} {\enquote {\bibinfo {title} {{Parton
  distribution function from the hadronic tensor on the lattice}},}\ }\href
  {\doibase 10.22323/1.251.0115} {\bibfield  {journal} {\bibinfo  {journal}
  {PoS}\ }\textbf {\bibinfo {volume} {LATTICE2015}},\ \bibinfo {pages} {115}
  (\bibinfo {year} {2016})},\ \Eprint {http://arxiv.org/abs/1603.07352}
  {arXiv:1603.07352 [hep-ph]} \BibitemShut {NoStop}%
\bibitem [{\citenamefont {Chambers}\ \emph {et~al.}(2017)\citenamefont
  {Chambers}, \citenamefont {Horsley}, \citenamefont {Nakamura}, \citenamefont
  {Perlt}, \citenamefont {Rakow}, \citenamefont {Schierholz}, \citenamefont
  {Schiller}, \citenamefont {Somfleth}, \citenamefont {Young},\ and\
  \citenamefont {Zanotti}}]{Chambers:2017dov}%
  \BibitemOpen
  \bibfield  {author} {\bibinfo {author} {\bibfnamefont {A.~J.}\ \bibnamefont
  {Chambers}}, \bibinfo {author} {\bibfnamefont {R.}~\bibnamefont {Horsley}},
  \bibinfo {author} {\bibfnamefont {Y.}~\bibnamefont {Nakamura}}, \bibinfo
  {author} {\bibfnamefont {H.}~\bibnamefont {Perlt}}, \bibinfo {author}
  {\bibfnamefont {P.~E.~L.}\ \bibnamefont {Rakow}}, \bibinfo {author}
  {\bibfnamefont {G.}~\bibnamefont {Schierholz}}, \bibinfo {author}
  {\bibfnamefont {A.}~\bibnamefont {Schiller}}, \bibinfo {author}
  {\bibfnamefont {K.}~\bibnamefont {Somfleth}}, \bibinfo {author}
  {\bibfnamefont {R.~D.}\ \bibnamefont {Young}}, \ and\ \bibinfo {author}
  {\bibfnamefont {J.~M.}\ \bibnamefont {Zanotti}} (\bibinfo {collaboration}
  {QCDSF}),\ }\bibfield  {title} {\enquote {\bibinfo {title} {Nucleon structure
  functions from operator product expansion on the lattice},}\ }\href {\doibase
  10.1103/PhysRevLett.118.242001} {\bibfield  {journal} {\bibinfo  {journal}
  {Phys. Rev. Lett.}\ }\textbf {\bibinfo {volume} {118}},\ \bibinfo {pages}
  {242001} (\bibinfo {year} {2017})},\ \Eprint
  {http://arxiv.org/abs/1703.01153} {arXiv:1703.01153 [hep-lat]} \BibitemShut
  {NoStop}%
\bibitem [{\citenamefont {Hansen}\ \emph {et~al.}(2017)\citenamefont {Hansen},
  \citenamefont {Meyer},\ and\ \citenamefont {Robaina}}]{Hansen:2017mnd}%
  \BibitemOpen
  \bibfield  {author} {\bibinfo {author} {\bibfnamefont {Maxwell~T.}\
  \bibnamefont {Hansen}}, \bibinfo {author} {\bibfnamefont {Harvey~B.}\
  \bibnamefont {Meyer}}, \ and\ \bibinfo {author} {\bibfnamefont {Daniel}\
  \bibnamefont {Robaina}},\ }\bibfield  {title} {\enquote {\bibinfo {title}
  {From deep inelastic scattering to heavy-flavor semileptonic decays: Total
  rates into multihadron final states from lattice {QCD}},}\ }\href {\doibase
  10.1103/PhysRevD.96.094513} {\bibfield  {journal} {\bibinfo  {journal} {Phys.
  Rev.}\ }\textbf {\bibinfo {volume} {D96}},\ \bibinfo {pages} {094513}
  (\bibinfo {year} {2017})},\ \Eprint {http://arxiv.org/abs/1704.08993}
  {arXiv:1704.08993 [hep-lat]} \BibitemShut {NoStop}%
\bibitem [{\citenamefont {Liang}\ \emph
  {et~al.}(2018{\natexlab{b}})\citenamefont {Liang}, \citenamefont {Liu},\ and\
  \citenamefont {Yang}}]{Liang:2017mye}%
  \BibitemOpen
  \bibfield  {author} {\bibinfo {author} {\bibfnamefont {Jian}\ \bibnamefont
  {Liang}}, \bibinfo {author} {\bibfnamefont {Keh-Fei}\ \bibnamefont {Liu}}, \
  and\ \bibinfo {author} {\bibfnamefont {Yi-Bo}\ \bibnamefont {Yang}},\
  }\bibfield  {title} {\enquote {\bibinfo {title} {{Lattice calculation of
  hadronic tensor of the nucleon}},}\ }\href {\doibase
  10.1051/epjconf/201817514014} {\bibfield  {journal} {\bibinfo  {journal} {EPJ
  Web Conf.}\ }\textbf {\bibinfo {volume} {175}},\ \bibinfo {pages} {14014}
  (\bibinfo {year} {2018}{\natexlab{b}})},\ \Eprint
  {http://arxiv.org/abs/1710.11145} {arXiv:1710.11145 [hep-lat]} \BibitemShut
  {NoStop}%
\bibitem [{\citenamefont {Ji}(2013)}]{Ji:2013dva}%
  \BibitemOpen
  \bibfield  {author} {\bibinfo {author} {\bibfnamefont {Xiangdong}\
  \bibnamefont {Ji}},\ }\bibfield  {title} {\enquote {\bibinfo {title} {Parton
  physics on a {Euclidean} lattice},}\ }\href {\doibase
  10.1103/PhysRevLett.110.262002} {\bibfield  {journal} {\bibinfo  {journal}
  {Phys. Rev. Lett.}\ }\textbf {\bibinfo {volume} {110}},\ \bibinfo {pages}
  {262002} (\bibinfo {year} {2013})},\ \Eprint {http://arxiv.org/abs/1305.1539}
  {arXiv:1305.1539 [hep-ph]} \BibitemShut {NoStop}%
\bibitem [{\citenamefont {Rossi}\ and\ \citenamefont
  {Testa}(2018)}]{Rossi:2018zkn}%
  \BibitemOpen
  \bibfield  {author} {\bibinfo {author} {\bibfnamefont {Giancarlo}\
  \bibnamefont {Rossi}}\ and\ \bibinfo {author} {\bibfnamefont {Massimo}\
  \bibnamefont {Testa}},\ }\bibfield  {title} {\enquote {\bibinfo {title}
  {{Euclidean versus Minkowski short distance}},}\ }\href {\doibase
  10.1103/PhysRevD.98.054028} {\bibfield  {journal} {\bibinfo  {journal} {Phys.
  Rev.}\ }\textbf {\bibinfo {volume} {D98}},\ \bibinfo {pages} {054028}
  (\bibinfo {year} {2018})},\ \Eprint {http://arxiv.org/abs/1806.00808}
  {arXiv:1806.00808 [hep-lat]} \BibitemShut {NoStop}%
\bibitem [{\citenamefont {Cichy}\ and\ \citenamefont
  {Constantinou}(2018)}]{Cichy:2018mum}%
  \BibitemOpen
  \bibfield  {author} {\bibinfo {author} {\bibfnamefont {Krzysztof}\
  \bibnamefont {Cichy}}\ and\ \bibinfo {author} {\bibfnamefont {Martha}\
  \bibnamefont {Constantinou}},\ }\bibfield  {title} {\enquote {\bibinfo
  {title} {A guide to light-cone {PDFs} from lattice {QCD}: an overview of
  approaches, techniques and results},}\ }\href@noop {} {\  (\bibinfo {year}
  {2018})},\ \Eprint {http://arxiv.org/abs/1811.07248} {arXiv:1811.07248
  [hep-lat]} \BibitemShut {NoStop}%
\bibitem [{\citenamefont {Monahan}(2018)}]{Monahan:2018euv}%
  \BibitemOpen
  \bibfield  {author} {\bibinfo {author} {\bibfnamefont {Christopher}\
  \bibnamefont {Monahan}},\ }\bibfield  {title} {\enquote {\bibinfo {title}
  {Recent developments in $x$-dependent structure calculations},}\ }\href@noop
  {} {\bibfield  {journal} {\bibinfo  {journal} {PoS}\ }\textbf {\bibinfo
  {volume} {LATTICE2018}},\ \bibinfo {pages} {018} (\bibinfo {year} {2018})},\
  \Eprint {http://arxiv.org/abs/1811.00678} {arXiv:1811.00678 [hep-lat]}
  \BibitemShut {NoStop}%
\bibitem [{\citenamefont {Lin}\ \emph {et~al.}(2015)\citenamefont {Lin},
  \citenamefont {Chen}, \citenamefont {Cohen},\ and\ \citenamefont
  {Ji}}]{Lin:2014zya}%
  \BibitemOpen
  \bibfield  {author} {\bibinfo {author} {\bibfnamefont {Huey-Wen}\
  \bibnamefont {Lin}}, \bibinfo {author} {\bibfnamefont {Jiunn-Wei}\
  \bibnamefont {Chen}}, \bibinfo {author} {\bibfnamefont {Saul~D.}\
  \bibnamefont {Cohen}}, \ and\ \bibinfo {author} {\bibfnamefont {Xiangdong}\
  \bibnamefont {Ji}},\ }\bibfield  {title} {\enquote {\bibinfo {title} {Flavor
  structure of the nucleon sea from lattice {QCD}},}\ }\href {\doibase
  10.1103/PhysRevD.91.054510} {\bibfield  {journal} {\bibinfo  {journal} {Phys.
  Rev.}\ }\textbf {\bibinfo {volume} {D91}},\ \bibinfo {pages} {054510}
  (\bibinfo {year} {2015})},\ \Eprint {http://arxiv.org/abs/1402.1462}
  {arXiv:1402.1462 [hep-ph]} \BibitemShut {NoStop}%
\bibitem [{\citenamefont {Radyushkin}(2017)}]{Radyushkin:2017cyf}%
  \BibitemOpen
  \bibfield  {author} {\bibinfo {author} {\bibfnamefont {A.~V.}\ \bibnamefont
  {Radyushkin}},\ }\bibfield  {title} {\enquote {\bibinfo {title}
  {{Quasi-parton distribution functions, momentum distributions, and
  pseudo-parton distribution functions}},}\ }\href {\doibase
  10.1103/PhysRevD.96.034025} {\bibfield  {journal} {\bibinfo  {journal} {Phys.
  Rev.}\ }\textbf {\bibinfo {volume} {D96}},\ \bibinfo {pages} {034025}
  (\bibinfo {year} {2017})},\ \Eprint {http://arxiv.org/abs/1705.01488}
  {arXiv:1705.01488 [hep-ph]} \BibitemShut {NoStop}%
\bibitem [{\citenamefont {Ma}\ and\ \citenamefont {Qiu}(2018)}]{Ma:2017pxb}%
  \BibitemOpen
  \bibfield  {author} {\bibinfo {author} {\bibfnamefont {Yan-Qing}\
  \bibnamefont {Ma}}\ and\ \bibinfo {author} {\bibfnamefont {Jian-Wei}\
  \bibnamefont {Qiu}},\ }\bibfield  {title} {\enquote {\bibinfo {title}
  {Exploring partonic structure of hadrons using \emph{ab initio} lattice {QCD}
  calculations},}\ }\href {\doibase 10.1103/PhysRevLett.120.022003} {\bibfield
  {journal} {\bibinfo  {journal} {Phys. Rev. Lett.}\ }\textbf {\bibinfo
  {volume} {120}},\ \bibinfo {pages} {022003} (\bibinfo {year} {2018})},\
  \Eprint {http://arxiv.org/abs/1709.03018} {arXiv:1709.03018 [hep-ph]}
  \BibitemShut {NoStop}%
\bibitem [{\citenamefont {Chen}\ \emph {et~al.}(2018)\citenamefont {Chen},
  \citenamefont {Jin}, \citenamefont {Lin}, \citenamefont {Liu}, \citenamefont
  {Schäfer}, \citenamefont {Yang}, \citenamefont {Zhang},\ and\ \citenamefont
  {Zhao}}]{Chen:2018fwa}%
  \BibitemOpen
  \bibfield  {author} {\bibinfo {author} {\bibfnamefont {Jiunn-Wei}\
  \bibnamefont {Chen}}, \bibinfo {author} {\bibfnamefont {Luchang}\
  \bibnamefont {Jin}}, \bibinfo {author} {\bibfnamefont {Huey-Wen}\
  \bibnamefont {Lin}}, \bibinfo {author} {\bibfnamefont {Yu-Sheng}\
  \bibnamefont {Liu}}, \bibinfo {author} {\bibfnamefont {Andreas}\ \bibnamefont
  {Schäfer}}, \bibinfo {author} {\bibfnamefont {Yi-Bo}\ \bibnamefont {Yang}},
  \bibinfo {author} {\bibfnamefont {Jian-Hui}\ \bibnamefont {Zhang}}, \ and\
  \bibinfo {author} {\bibfnamefont {Yong}\ \bibnamefont {Zhao}} (\bibinfo
  {collaboration} {LP$^3$}),\ }\bibfield  {title} {\enquote {\bibinfo {title}
  {First direct lattice-{QCD} calculation of the $x$-dependence of the pion
  parton distribution function},}\ }\href@noop {} {\  (\bibinfo {year}
  {2018})},\ \Eprint {http://arxiv.org/abs/1804.01483} {arXiv:1804.01483
  [hep-lat]} \BibitemShut {NoStop}%
\bibitem [{\citenamefont {Briceño}\ \emph
  {et~al.}(2018{\natexlab{b}})\citenamefont {Briceño}, \citenamefont
  {Guerrero}, \citenamefont {Hansen},\ and\ \citenamefont
  {Monahan}}]{Briceno:2018lfj}%
  \BibitemOpen
  \bibfield  {author} {\bibinfo {author} {\bibfnamefont {Raúl~A.}\
  \bibnamefont {Briceño}}, \bibinfo {author} {\bibfnamefont {Juan~V.}\
  \bibnamefont {Guerrero}}, \bibinfo {author} {\bibfnamefont {Maxwell~T.}\
  \bibnamefont {Hansen}}, \ and\ \bibinfo {author} {\bibfnamefont
  {Christopher~J.}\ \bibnamefont {Monahan}},\ }\bibfield  {title} {\enquote
  {\bibinfo {title} {{Finite-volume effects due to spatially nonlocal
  operators}},}\ }\href {\doibase 10.1103/PhysRevD.98.014511} {\bibfield
  {journal} {\bibinfo  {journal} {Phys. Rev.}\ }\textbf {\bibinfo {volume}
  {D98}},\ \bibinfo {pages} {014511} (\bibinfo {year} {2018}{\natexlab{b}})},\
  \Eprint {http://arxiv.org/abs/1805.01034} {arXiv:1805.01034 [hep-lat]}
  \BibitemShut {NoStop}%
\bibitem [{\citenamefont {Bali}\ \emph {et~al.}(2016)\citenamefont {Bali},
  \citenamefont {Lang}, \citenamefont {Musch},\ and\ \citenamefont
  {Schäfer}}]{Bali:2016lva}%
  \BibitemOpen
  \bibfield  {author} {\bibinfo {author} {\bibfnamefont {Gunnar~S.}\
  \bibnamefont {Bali}}, \bibinfo {author} {\bibfnamefont {Bernhard}\
  \bibnamefont {Lang}}, \bibinfo {author} {\bibfnamefont {Bernhard~U.}\
  \bibnamefont {Musch}}, \ and\ \bibinfo {author} {\bibfnamefont {Andreas}\
  \bibnamefont {Schäfer}},\ }\bibfield  {title} {\enquote {\bibinfo {title}
  {{Novel quark smearing for hadrons with high momenta in lattice QCD}},}\
  }\href {\doibase 10.1103/PhysRevD.93.094515} {\bibfield  {journal} {\bibinfo
  {journal} {Phys. Rev.}\ }\textbf {\bibinfo {volume} {D93}},\ \bibinfo {pages}
  {094515} (\bibinfo {year} {2016})},\ \Eprint
  {http://arxiv.org/abs/1602.05525} {arXiv:1602.05525 [hep-lat]} \BibitemShut
  {NoStop}%
\bibitem [{\citenamefont {Buck}\ and\ \citenamefont
  {Perez}(1983)}]{Buck:1975ae}%
  \BibitemOpen
  \bibfield  {author} {\bibinfo {author} {\bibfnamefont {B.}~\bibnamefont
  {Buck}}\ and\ \bibinfo {author} {\bibfnamefont {S.~M.}\ \bibnamefont
  {Perez}},\ }\bibfield  {title} {\enquote {\bibinfo {title} {New look at
  magnetic moments and beta decays of mirror nuclei},}\ }\href {\doibase
  10.1103/PhysRevLett.50.1975} {\bibfield  {journal} {\bibinfo  {journal}
  {Phys. Rev. Lett.}\ }\textbf {\bibinfo {volume} {50}},\ \bibinfo {pages}
  {1975--1978} (\bibinfo {year} {1983})}\BibitemShut {NoStop}%
\bibitem [{\citenamefont {Krofcheck}\ \emph {et~al.}(1985)\citenamefont
  {Krofcheck} \emph {et~al.}}]{Krofcheck:1985fg}%
  \BibitemOpen
  \bibfield  {author} {\bibinfo {author} {\bibfnamefont {D.}~\bibnamefont
  {Krofcheck}} \emph {et~al.},\ }\bibfield  {title} {\enquote {\bibinfo {title}
  {{Gamow-Teller} strength function in {$^{71}$Ge} via the $(p, n)$ reaction at
  medium-energies},}\ }\href {\doibase 10.1103/PhysRevLett.55.1051} {\bibfield
  {journal} {\bibinfo  {journal} {Phys. Rev. Lett.}\ }\textbf {\bibinfo
  {volume} {55}},\ \bibinfo {pages} {1051--1054} (\bibinfo {year}
  {1985})}\BibitemShut {NoStop}%
\bibitem [{\citenamefont {Chou}\ \emph {et~al.}(1993)\citenamefont {Chou},
  \citenamefont {Warburton},\ and\ \citenamefont {Brown}}]{Chou:1993zz}%
  \BibitemOpen
  \bibfield  {author} {\bibinfo {author} {\bibfnamefont {W.~T.}\ \bibnamefont
  {Chou}}, \bibinfo {author} {\bibfnamefont {E.~K.}\ \bibnamefont {Warburton}},
  \ and\ \bibinfo {author} {\bibfnamefont {B.~Alex}\ \bibnamefont {Brown}},\
  }\bibfield  {title} {\enquote {\bibinfo {title} {{Gamow-Teller} beta-decay
  rates for {$A\le18$} nuclei},}\ }\href {\doibase 10.1103/PhysRevC.47.163}
  {\bibfield  {journal} {\bibinfo  {journal} {Phys. Rev.}\ }\textbf {\bibinfo
  {volume} {C47}},\ \bibinfo {pages} {163--177} (\bibinfo {year}
  {1993})}\BibitemShut {NoStop}%
\bibitem [{\citenamefont {Baroni}\ \emph {et~al.}(2016)\citenamefont {Baroni},
  \citenamefont {Girlanda}, \citenamefont {Pastore}, \citenamefont
  {Schiavilla},\ and\ \citenamefont {Viviani}}]{Baroni:2015}%
  \BibitemOpen
  \bibfield  {author} {\bibinfo {author} {\bibfnamefont {A.}~\bibnamefont
  {Baroni}}, \bibinfo {author} {\bibfnamefont {L.}~\bibnamefont {Girlanda}},
  \bibinfo {author} {\bibfnamefont {S.}~\bibnamefont {Pastore}}, \bibinfo
  {author} {\bibfnamefont {R.}~\bibnamefont {Schiavilla}}, \ and\ \bibinfo
  {author} {\bibfnamefont {M.}~\bibnamefont {Viviani}},\ }\bibfield  {title}
  {\enquote {\bibinfo {title} {Nuclear axial currents in chiral effective field
  theory},}\ }\href {\doibase 10.1103/PhysRevC.93.049902,
  10.1103/PhysRevC.93.015501, 10.1103/PhysRevC.95.059901} {\bibfield  {journal}
  {\bibinfo  {journal} {Phys. Rev.}\ }\textbf {\bibinfo {volume} {C93}},\
  \bibinfo {pages} {015501} (\bibinfo {year} {2016})},\ \bibinfo {note}
  {[Erratum: Phys. Rev.C95,no.5,059901(2017)]},\ \Eprint
  {http://arxiv.org/abs/1509.07039} {arXiv:1509.07039 [nucl-th]} \BibitemShut
  {NoStop}%
\bibitem [{\citenamefont {Krebs}\ \emph {et~al.}(2017)\citenamefont {Krebs},
  \citenamefont {Epelbaum},\ and\ \citenamefont {Meißner}}]{Krebs:2016rqz}%
  \BibitemOpen
  \bibfield  {author} {\bibinfo {author} {\bibfnamefont {H.}~\bibnamefont
  {Krebs}}, \bibinfo {author} {\bibfnamefont {E.}~\bibnamefont {Epelbaum}}, \
  and\ \bibinfo {author} {\bibfnamefont {U.~G.}\ \bibnamefont {Meißner}},\
  }\bibfield  {title} {\enquote {\bibinfo {title} {{Nuclear axial current
  operators to fourth order in chiral effective field theory}},}\ }\href
  {\doibase 10.1016/j.aop.2017.01.021} {\bibfield  {journal} {\bibinfo
  {journal} {Ann. Phys.}\ }\textbf {\bibinfo {volume} {378}},\ \bibinfo {pages}
  {317--395} (\bibinfo {year} {2017})},\ \Eprint
  {http://arxiv.org/abs/1610.03569} {arXiv:1610.03569 [nucl-th]} \BibitemShut
  {NoStop}%
\bibitem [{\citenamefont {Pastore}\ \emph {et~al.}(2018)\citenamefont
  {Pastore}, \citenamefont {Baroni}, \citenamefont {Carlson}, \citenamefont
  {Gandolfi}, \citenamefont {Pieper}, \citenamefont {Schiavilla},\ and\
  \citenamefont {Wiringa}}]{Pastore:2017uwc}%
  \BibitemOpen
  \bibfield  {author} {\bibinfo {author} {\bibfnamefont {S.}~\bibnamefont
  {Pastore}}, \bibinfo {author} {\bibfnamefont {A.}~\bibnamefont {Baroni}},
  \bibinfo {author} {\bibfnamefont {J.}~\bibnamefont {Carlson}}, \bibinfo
  {author} {\bibfnamefont {S.}~\bibnamefont {Gandolfi}}, \bibinfo {author}
  {\bibfnamefont {Steven~C.}\ \bibnamefont {Pieper}}, \bibinfo {author}
  {\bibfnamefont {R.}~\bibnamefont {Schiavilla}}, \ and\ \bibinfo {author}
  {\bibfnamefont {R.~B.}\ \bibnamefont {Wiringa}},\ }\bibfield  {title}
  {\enquote {\bibinfo {title} {Quantum {Monte Carlo} calculations of weak
  transitions in {$A=6$-$10$} nuclei},}\ }\href {\doibase
  10.1103/PhysRevC.97.022501} {\bibfield  {journal} {\bibinfo  {journal} {Phys.
  Rev.}\ }\textbf {\bibinfo {volume} {C97}},\ \bibinfo {pages} {022501}
  (\bibinfo {year} {2018})},\ \Eprint {http://arxiv.org/abs/1709.03592}
  {arXiv:1709.03592 [nucl-th]} \BibitemShut {NoStop}%
\bibitem [{\citenamefont {Lovato}\ \emph {et~al.}(2015)\citenamefont {Lovato},
  \citenamefont {Gandolfi}, \citenamefont {Carlson}, \citenamefont {Pieper},\
  and\ \citenamefont {Schiavilla}}]{Lovato:2015qka}%
  \BibitemOpen
  \bibfield  {author} {\bibinfo {author} {\bibfnamefont {A.}~\bibnamefont
  {Lovato}}, \bibinfo {author} {\bibfnamefont {S.}~\bibnamefont {Gandolfi}},
  \bibinfo {author} {\bibfnamefont {J.}~\bibnamefont {Carlson}}, \bibinfo
  {author} {\bibfnamefont {Steven~C.}\ \bibnamefont {Pieper}}, \ and\ \bibinfo
  {author} {\bibfnamefont {R.}~\bibnamefont {Schiavilla}},\ }\bibfield  {title}
  {\enquote {\bibinfo {title} {Electromagnetic and neutral-weak response
  functions of {$^4$He} and {$^{12}$C}},}\ }\href {\doibase
  10.1103/PhysRevC.91.062501} {\bibfield  {journal} {\bibinfo  {journal} {Phys.
  Rev.}\ }\textbf {\bibinfo {volume} {C91}},\ \bibinfo {pages} {062501}
  (\bibinfo {year} {2015})},\ \Eprint {http://arxiv.org/abs/1501.01981}
  {arXiv:1501.01981 [nucl-th]} \BibitemShut {NoStop}%
\bibitem [{\citenamefont {Lovato}\ \emph {et~al.}(2018)\citenamefont {Lovato},
  \citenamefont {Gandolfi}, \citenamefont {Carlson}, \citenamefont {Lusk},
  \citenamefont {Pieper},\ and\ \citenamefont {Schiavilla}}]{Lovato:2017cux}%
  \BibitemOpen
  \bibfield  {author} {\bibinfo {author} {\bibfnamefont {A.}~\bibnamefont
  {Lovato}}, \bibinfo {author} {\bibfnamefont {S.}~\bibnamefont {Gandolfi}},
  \bibinfo {author} {\bibfnamefont {J.}~\bibnamefont {Carlson}}, \bibinfo
  {author} {\bibfnamefont {Ewing}\ \bibnamefont {Lusk}}, \bibinfo {author}
  {\bibfnamefont {Steven~C.}\ \bibnamefont {Pieper}}, \ and\ \bibinfo {author}
  {\bibfnamefont {R.}~\bibnamefont {Schiavilla}},\ }\bibfield  {title}
  {\enquote {\bibinfo {title} {Quantum {Monte Carlo} calculation of
  neutral-current $\nu$-{$^{12}$C} inclusive quasielastic scattering},}\ }\href
  {\doibase 10.1103/PhysRevC.97.022502} {\bibfield  {journal} {\bibinfo
  {journal} {Phys. Rev.}\ }\textbf {\bibinfo {volume} {C97}},\ \bibinfo {pages}
  {022502} (\bibinfo {year} {2018})},\ \Eprint
  {http://arxiv.org/abs/1711.02047} {arXiv:1711.02047 [nucl-th]} \BibitemShut
  {NoStop}%
\bibitem [{\citenamefont {Savage}\ \emph {et~al.}(2017)\citenamefont {Savage},
  \citenamefont {Shanahan}, \citenamefont {Tiburzi}, \citenamefont {Wagman},
  \citenamefont {Winter}, \citenamefont {Beane}, \citenamefont {Chang},
  \citenamefont {Davoudi}, \citenamefont {Detmold},\ and\ \citenamefont
  {Orginos}}]{Savage:2016kon}%
  \BibitemOpen
  \bibfield  {author} {\bibinfo {author} {\bibfnamefont {Martin~J.}\
  \bibnamefont {Savage}}, \bibinfo {author} {\bibfnamefont {Phiala~E.}\
  \bibnamefont {Shanahan}}, \bibinfo {author} {\bibfnamefont {Brian~C.}\
  \bibnamefont {Tiburzi}}, \bibinfo {author} {\bibfnamefont {Michael~L.}\
  \bibnamefont {Wagman}}, \bibinfo {author} {\bibfnamefont {Frank}\
  \bibnamefont {Winter}}, \bibinfo {author} {\bibfnamefont {Silas~R.}\
  \bibnamefont {Beane}}, \bibinfo {author} {\bibfnamefont {Emmanuel}\
  \bibnamefont {Chang}}, \bibinfo {author} {\bibfnamefont {Zohreh}\
  \bibnamefont {Davoudi}}, \bibinfo {author} {\bibfnamefont {William}\
  \bibnamefont {Detmold}}, \ and\ \bibinfo {author} {\bibfnamefont {Kostas}\
  \bibnamefont {Orginos}} (\bibinfo {collaboration} {NPLQCD}),\ }\bibfield
  {title} {\enquote {\bibinfo {title} {Proton-proton fusion and tritium $\beta$
  decay from lattice quantum chromodynamics},}\ }\href {\doibase
  10.1103/PhysRevLett.119.062002} {\bibfield  {journal} {\bibinfo  {journal}
  {Phys. Rev. Lett.}\ }\textbf {\bibinfo {volume} {119}},\ \bibinfo {pages}
  {062002} (\bibinfo {year} {2017})},\ \Eprint
  {http://arxiv.org/abs/1610.04545} {arXiv:1610.04545 [hep-lat]} \BibitemShut
  {NoStop}%
\bibitem [{\citenamefont {Chang}\ \emph
  {et~al.}(2018{\natexlab{b}})\citenamefont {Chang}, \citenamefont {Davoudi},
  \citenamefont {Detmold}, \citenamefont {Gambhir}, \citenamefont {Orginos},
  \citenamefont {Savage}, \citenamefont {Shanahan}, \citenamefont {Wagman},\
  and\ \citenamefont {Winter}}]{Chang:2017eiq}%
  \BibitemOpen
  \bibfield  {author} {\bibinfo {author} {\bibfnamefont {Emmanuel}\
  \bibnamefont {Chang}}, \bibinfo {author} {\bibfnamefont {Zohreh}\
  \bibnamefont {Davoudi}}, \bibinfo {author} {\bibfnamefont {William}\
  \bibnamefont {Detmold}}, \bibinfo {author} {\bibfnamefont {Arjun~S.}\
  \bibnamefont {Gambhir}}, \bibinfo {author} {\bibfnamefont {Kostas}\
  \bibnamefont {Orginos}}, \bibinfo {author} {\bibfnamefont {Martin~J.}\
  \bibnamefont {Savage}}, \bibinfo {author} {\bibfnamefont {Phiala~E.}\
  \bibnamefont {Shanahan}}, \bibinfo {author} {\bibfnamefont {Michael~L.}\
  \bibnamefont {Wagman}}, \ and\ \bibinfo {author} {\bibfnamefont {Frank}\
  \bibnamefont {Winter}} (\bibinfo {collaboration} {NPLQCD}),\ }\bibfield
  {title} {\enquote {\bibinfo {title} {Scalar, axial, and tensor interactions
  of light nuclei from lattice {QCD}},}\ }\href {\doibase
  10.1103/PhysRevLett.120.152002} {\bibfield  {journal} {\bibinfo  {journal}
  {Phys. Rev. Lett.}\ }\textbf {\bibinfo {volume} {120}},\ \bibinfo {pages}
  {152002} (\bibinfo {year} {2018}{\natexlab{b}})},\ \Eprint
  {http://arxiv.org/abs/1712.03221} {arXiv:1712.03221 [hep-lat]} \BibitemShut
  {NoStop}%
\bibitem [{\citenamefont {Tiburzi}\ \emph {et~al.}(2017)\citenamefont
  {Tiburzi}, \citenamefont {Wagman}, \citenamefont {Winter}, \citenamefont
  {Chang}, \citenamefont {Davoudi}, \citenamefont {Detmold}, \citenamefont
  {Orginos}, \citenamefont {Savage},\ and\ \citenamefont
  {Shanahan}}]{Tiburzi:2017iux}%
  \BibitemOpen
  \bibfield  {author} {\bibinfo {author} {\bibfnamefont {Brian~C.}\
  \bibnamefont {Tiburzi}}, \bibinfo {author} {\bibfnamefont {Michael~L.}\
  \bibnamefont {Wagman}}, \bibinfo {author} {\bibfnamefont {Frank}\
  \bibnamefont {Winter}}, \bibinfo {author} {\bibfnamefont {Emmanuel}\
  \bibnamefont {Chang}}, \bibinfo {author} {\bibfnamefont {Zohreh}\
  \bibnamefont {Davoudi}}, \bibinfo {author} {\bibfnamefont {William}\
  \bibnamefont {Detmold}}, \bibinfo {author} {\bibfnamefont {Kostas}\
  \bibnamefont {Orginos}}, \bibinfo {author} {\bibfnamefont {Martin~J.}\
  \bibnamefont {Savage}}, \ and\ \bibinfo {author} {\bibfnamefont {Phiala~E.}\
  \bibnamefont {Shanahan}},\ }\bibfield  {title} {\enquote {\bibinfo {title}
  {{Double-$\beta$ decay matrix elements from lattice quantum
  chromodynamics}},}\ }\href {\doibase 10.1103/PhysRevD.96.054505} {\bibfield
  {journal} {\bibinfo  {journal} {Phys. Rev.}\ }\textbf {\bibinfo {volume}
  {D96}},\ \bibinfo {pages} {054505} (\bibinfo {year} {2017})},\ \Eprint
  {http://arxiv.org/abs/1702.02929} {arXiv:1702.02929 [hep-lat]} \BibitemShut
  {NoStop}%
\bibitem [{\citenamefont {Barnea}\ \emph {et~al.}(2015)\citenamefont {Barnea},
  \citenamefont {Contessi}, \citenamefont {Gazit}, \citenamefont {Pederiva},\
  and\ \citenamefont {van Kolck}}]{Barnea:2013uqa}%
  \BibitemOpen
  \bibfield  {author} {\bibinfo {author} {\bibfnamefont {N.}~\bibnamefont
  {Barnea}}, \bibinfo {author} {\bibfnamefont {L.}~\bibnamefont {Contessi}},
  \bibinfo {author} {\bibfnamefont {D.}~\bibnamefont {Gazit}}, \bibinfo
  {author} {\bibfnamefont {F.}~\bibnamefont {Pederiva}}, \ and\ \bibinfo
  {author} {\bibfnamefont {U.}~\bibnamefont {van Kolck}},\ }\bibfield  {title}
  {\enquote {\bibinfo {title} {Effective field theory for lattice nuclei},}\
  }\href {\doibase 10.1103/PhysRevLett.114.052501} {\bibfield  {journal}
  {\bibinfo  {journal} {Phys. Rev. Lett.}\ }\textbf {\bibinfo {volume} {114}},\
  \bibinfo {pages} {052501} (\bibinfo {year} {2015})},\ \Eprint
  {http://arxiv.org/abs/1311.4966} {arXiv:1311.4966 [nucl-th]} \BibitemShut
  {NoStop}%
\bibitem [{\citenamefont {Bansal}\ \emph {et~al.}(2018)\citenamefont {Bansal},
  \citenamefont {Binder}, \citenamefont {Ekström}, \citenamefont {Hagen},
  \citenamefont {Jansen},\ and\ \citenamefont {Papenbrock}}]{Bansal:2017pwn}%
  \BibitemOpen
  \bibfield  {author} {\bibinfo {author} {\bibfnamefont {A.}~\bibnamefont
  {Bansal}}, \bibinfo {author} {\bibfnamefont {S.}~\bibnamefont {Binder}},
  \bibinfo {author} {\bibfnamefont {A.}~\bibnamefont {Ekström}}, \bibinfo
  {author} {\bibfnamefont {G.}~\bibnamefont {Hagen}}, \bibinfo {author}
  {\bibfnamefont {G.~R.}\ \bibnamefont {Jansen}}, \ and\ \bibinfo {author}
  {\bibfnamefont {T.}~\bibnamefont {Papenbrock}},\ }\bibfield  {title}
  {\enquote {\bibinfo {title} {Pionless effective field theory for atomic
  nuclei and lattice nuclei},}\ }\href {\doibase 10.1103/PhysRevC.98.054301}
  {\bibfield  {journal} {\bibinfo  {journal} {Phys. Rev.}\ }\textbf {\bibinfo
  {volume} {C98}},\ \bibinfo {pages} {054301} (\bibinfo {year} {2018})},\
  \Eprint {http://arxiv.org/abs/1712.10246} {arXiv:1712.10246 [nucl-th]}
  \BibitemShut {NoStop}%
\bibitem [{\citenamefont {Contessi}\ \emph {et~al.}(2017)\citenamefont
  {Contessi}, \citenamefont {Lovato}, \citenamefont {Pederiva}, \citenamefont
  {Roggero}, \citenamefont {Kirscher},\ and\ \citenamefont {{van
  Kolck}}}]{Contessi:2017rww}%
  \BibitemOpen
  \bibfield  {author} {\bibinfo {author} {\bibfnamefont {L.}~\bibnamefont
  {Contessi}}, \bibinfo {author} {\bibfnamefont {A.}~\bibnamefont {Lovato}},
  \bibinfo {author} {\bibfnamefont {F.}~\bibnamefont {Pederiva}}, \bibinfo
  {author} {\bibfnamefont {A.}~\bibnamefont {Roggero}}, \bibinfo {author}
  {\bibfnamefont {J.}~\bibnamefont {Kirscher}}, \ and\ \bibinfo {author}
  {\bibfnamefont {U.}~\bibnamefont {{van Kolck}}},\ }\bibfield  {title}
  {\enquote {\bibinfo {title} {Ground-state properties of {$^{4}$He} and
  {$^{16}$O} extrapolated from lattice {QCD} with pionless {EFT}},}\ }\href
  {\doibase 10.1016/j.physletb.2017.07.048} {\bibfield  {journal} {\bibinfo
  {journal} {Phys. Lett.}\ }\textbf {\bibinfo {volume} {B772}},\ \bibinfo
  {pages} {839--848} (\bibinfo {year} {2017})},\ \Eprint
  {http://arxiv.org/abs/1701.06516} {arXiv:1701.06516 [nucl-th]} \BibitemShut
  {NoStop}%
\bibitem [{\citenamefont {Kirscher}\ \emph {et~al.}(2017)\citenamefont
  {Kirscher}, \citenamefont {Pazy}, \citenamefont {Drachman},\ and\
  \citenamefont {Barnea}}]{Kirscher:2017fqc}%
  \BibitemOpen
  \bibfield  {author} {\bibinfo {author} {\bibfnamefont {Johannes}\
  \bibnamefont {Kirscher}}, \bibinfo {author} {\bibfnamefont {Ehoud}\
  \bibnamefont {Pazy}}, \bibinfo {author} {\bibfnamefont {Jonathan}\
  \bibnamefont {Drachman}}, \ and\ \bibinfo {author} {\bibfnamefont {Nir}\
  \bibnamefont {Barnea}},\ }\bibfield  {title} {\enquote {\bibinfo {title}
  {{Electromagnetic characteristics of $A \leq 3$ physical and lattice
  nuclei}},}\ }\href {\doibase 10.1103/PhysRevC.96.024001} {\bibfield
  {journal} {\bibinfo  {journal} {Phys. Rev.}\ }\textbf {\bibinfo {volume}
  {C96}},\ \bibinfo {pages} {024001} (\bibinfo {year} {2017})},\ \Eprint
  {http://arxiv.org/abs/1702.07268} {arXiv:1702.07268 [nucl-th]} \BibitemShut
  {NoStop}%
\bibitem [{\citenamefont {Roggero}\ and\ \citenamefont
  {Carlson}(2018)}]{Roggero:2018hrn}%
  \BibitemOpen
  \bibfield  {author} {\bibinfo {author} {\bibfnamefont {Alessandro}\
  \bibnamefont {Roggero}}\ and\ \bibinfo {author} {\bibfnamefont {Joseph}\
  \bibnamefont {Carlson}},\ }\bibfield  {title} {\enquote {\bibinfo {title}
  {{Linear response on a quantum computer}},}\ }\href@noop {} {\  (\bibinfo
  {year} {2018})},\ \Eprint {http://arxiv.org/abs/1804.01505} {arXiv:1804.01505
  [quant-ph]} \BibitemShut {NoStop}%
\bibitem [{\citenamefont {Detmold}\ and\ \citenamefont
  {Endres}(2014)}]{Detmold:2014hla}%
  \BibitemOpen
  \bibfield  {author} {\bibinfo {author} {\bibfnamefont {William}\ \bibnamefont
  {Detmold}}\ and\ \bibinfo {author} {\bibfnamefont {Michael~G.}\ \bibnamefont
  {Endres}},\ }\bibfield  {title} {\enquote {\bibinfo {title} {{Signal/noise
  enhancement strategies for stochastically estimated correlation
  functions}},}\ }\href {\doibase 10.1103/PhysRevD.90.034503} {\bibfield
  {journal} {\bibinfo  {journal} {Phys. Rev.}\ }\textbf {\bibinfo {volume}
  {D90}},\ \bibinfo {pages} {034503} (\bibinfo {year} {2014})},\ \Eprint
  {http://arxiv.org/abs/1404.6816} {arXiv:1404.6816 [hep-lat]} \BibitemShut
  {NoStop}%
\bibitem [{\citenamefont {Beane}\ \emph {et~al.}(2015)\citenamefont {Beane},
  \citenamefont {Detmold}, \citenamefont {Orginos},\ and\ \citenamefont
  {Savage}}]{Beane:2014oea}%
  \BibitemOpen
  \bibfield  {author} {\bibinfo {author} {\bibfnamefont {Silas~R.}\
  \bibnamefont {Beane}}, \bibinfo {author} {\bibfnamefont {William}\
  \bibnamefont {Detmold}}, \bibinfo {author} {\bibfnamefont {Kostas}\
  \bibnamefont {Orginos}}, \ and\ \bibinfo {author} {\bibfnamefont {Martin~J.}\
  \bibnamefont {Savage}},\ }\bibfield  {title} {\enquote {\bibinfo {title}
  {Uncertainty quantification in lattice {QCD} calculations for nuclear
  physics},}\ }\href {\doibase 10.1088/0954-3899/42/3/034022} {\bibfield
  {journal} {\bibinfo  {journal} {J. Phys.}\ }\textbf {\bibinfo {volume}
  {G42}},\ \bibinfo {pages} {034022} (\bibinfo {year} {2015})},\ \Eprint
  {http://arxiv.org/abs/1410.2937} {arXiv:1410.2937 [nucl-th]} \BibitemShut
  {NoStop}%
\bibitem [{\citenamefont {Wagman}\ and\ \citenamefont
  {Savage}(2017{\natexlab{a}})}]{Wagman:2017xfh}%
  \BibitemOpen
  \bibfield  {author} {\bibinfo {author} {\bibfnamefont {Michael~L.}\
  \bibnamefont {Wagman}}\ and\ \bibinfo {author} {\bibfnamefont {Martin~J.}\
  \bibnamefont {Savage}},\ }\bibfield  {title} {\enquote {\bibinfo {title}
  {Taming the signal-to-noise problem in lattice {QCD} by phase reweighting},}\
  }\href@noop {} {\  (\bibinfo {year} {2017}{\natexlab{a}})},\ \Eprint
  {http://arxiv.org/abs/1704.07356} {arXiv:1704.07356 [hep-lat]} \BibitemShut
  {NoStop}%
\bibitem [{\citenamefont {Wagman}\ and\ \citenamefont
  {Savage}(2017{\natexlab{b}})}]{Wagman:2016bam}%
  \BibitemOpen
  \bibfield  {author} {\bibinfo {author} {\bibfnamefont {Michael~L.}\
  \bibnamefont {Wagman}}\ and\ \bibinfo {author} {\bibfnamefont {Martin~J.}\
  \bibnamefont {Savage}},\ }\bibfield  {title} {\enquote {\bibinfo {title}
  {Statistics of baryon correlation functions in lattice {QCD}},}\ }\href
  {\doibase 10.1103/PhysRevD.96.114508} {\bibfield  {journal} {\bibinfo
  {journal} {Phys. Rev.}\ }\textbf {\bibinfo {volume} {D96}},\ \bibinfo {pages}
  {114508} (\bibinfo {year} {2017}{\natexlab{b}})},\ \Eprint
  {http://arxiv.org/abs/1611.07643} {arXiv:1611.07643 [hep-lat]} \BibitemShut
  {NoStop}%
\bibitem [{\citenamefont {Detmold}\ \emph {et~al.}(2018)\citenamefont
  {Detmold}, \citenamefont {Kanwar},\ and\ \citenamefont
  {Wagman}}]{Detmold:2018eqd}%
  \BibitemOpen
  \bibfield  {author} {\bibinfo {author} {\bibfnamefont {William}\ \bibnamefont
  {Detmold}}, \bibinfo {author} {\bibfnamefont {Gurtej}\ \bibnamefont
  {Kanwar}}, \ and\ \bibinfo {author} {\bibfnamefont {Michael~L.}\ \bibnamefont
  {Wagman}},\ }\bibfield  {title} {\enquote {\bibinfo {title} {Phase unwrapping
  and one-dimensional sign problems},}\ }\href {\doibase
  10.1103/PhysRevD.98.074511} {\bibfield  {journal} {\bibinfo  {journal} {Phys.
  Rev.}\ }\textbf {\bibinfo {volume} {D98}},\ \bibinfo {pages} {074511}
  (\bibinfo {year} {2018})},\ \Eprint {http://arxiv.org/abs/1806.01832}
  {arXiv:1806.01832 [hep-lat]} \BibitemShut {NoStop}%
\end{thebibliography}%

\end{document}